\documentclass[11pt,letterpaper]{article}

\usepackage[utf8x]{inputenc}%
\usepackage{tcolorbox,fancybox}%
\usepackage{physics}
\usepackage{cite}
\usepackage{moresize}
\usepackage{ntheorem}
\usepackage{mathtools}

\usepackage[english]{babel}
\usepackage{amsmath,amssymb,amsfonts}
\numberwithin{equation}{section}
\usepackage{graphicx}
\usepackage{booktabs}
\usepackage{color,xcolor}
\usepackage[numbers,sort&compress]{natbib}

\usepackage{tikz}
\usepackage[compat=1.1.0]{tikz-feynman}
\usetikzlibrary{positioning}
\usetikzlibrary{decorations.text}
\usetikzlibrary{decorations.pathmorphing}

\usetikzlibrary{calc}
\usetikzlibrary{shapes.misc}
\tikzset{
        >=latex,
    photon/.style={decorate, decoration={snake}, draw=black, thick},
    fermionnoarrow/.style={draw=black, postaction={decorate}, thick},
    scalar/.style={draw=black, postaction={decorate}, decoration={markings,mark=at position .55 with {\arrow{>}}}, thick, dashed},
    scalarnoarrow/.style={draw=black, postaction={decorate},  thick, dashed},
    fermion/.style={draw=black, postaction={decorate},decoration={markings,mark=at position .55 with {\arrow{>}}}, thick},
    gluon/.style={decorate, draw=black, decoration={coil,amplitude=4pt, segment length=5pt}, thick},
    vertex/.style={draw,shape=circle,fill=black,minimum size=3pt,inner sep=0pt},
    fillvertex/.style={draw,shape=circle,fill=black,minimum size=5pt,inner sep=0pt},
    openvertex/.style={draw,shape=circle,minimum size=5pt,inner sep=0pt},
    blob/.style={draw=red,shape=circle,fill=red,minimum size=6pt,inner sep=0pt},
    redvertex/.style={draw=red,shape=circle,fill=red,minimum size=3pt,inner sep=0pt},
    cross/.style={cross out, draw=black,thick, minimum size=5pt, inner sep=0pt, outer sep=0pt}
}

\usepackage{physics}
\usepackage{orcidlink}
\usepackage{slashed}
\usepackage{mathrsfs}
\usepackage{enumerate}
\usepackage{mdframed}
\usepackage{url}
\usepackage[makeroom]{cancel}
\usepackage{geometry}

\usepackage{pifont}
\newcommand{\cmark}{\ding{51}}%
\newcommand{\xmark}{\ding{55}}%

\newtheorem*{thm-non}{Theorem}

\usepackage{hyperref} 
\hypersetup{
    colorlinks=true,       
    linkcolor=red,          
    citecolor=blue,        
    filecolor=magenta,      
    urlcolor=purple           
}
\usepackage[all]{hypcap} 

\def\beqn{\begin{eqnarray}}
\def\eeqn{\end{eqnarray}}
\def\beqs{\begin{subequations}}
\def\eeqs{\end{subequations}}
\def\beq{\begin{equation}}
\def\eeq{\end{equation}}
\def\ba{\begin{array}}
\def\ea{\end{array}}

\def\non{\nonumber\\}

\def\hf{\frac{1}{2}}
\def\[{\left[}
\def\]{\right]}
\def\({\left(}
\def\){\right)}
\newcommand\para{\paragraph{}}

\def\gSU{\rm SU}

\def\gG{\rm G}

\newcommand{\rep}[1]{\mathbf{#1}}
\newcommand{\repb}[1]{\mathbf{\overline{#1}}}

\def\Bc{\mathcal{B}}

\def\Dc{\mathcal{D}}
\def\Ec{\mathcal{E}}
\def\Fc{\mathcal{F}}
\def\Gc{\mathcal{G}}

\def\Lc{\mathcal{L}}

\def\Nc{\mathcal{N}}
\def\Oc{\mathcal{O}}
\def\Pc{\mathcal{P}}
\def\Qc{\mathcal{Q}}
\def\Rc{\mathcal{R}}
\def\Sc{\mathcal{S}}
\def\Tc{\mathcal{T}}

\def\Xc{\mathcal{X}}
\def\Yc{\mathcal{Y}}

\def\DG{\mathfrak{D}}  \def\dG{\mathfrak{d}}
\def\EG{\mathfrak{E}}  \def\eG{\mathfrak{e}}
  
  \def\gG{\mathfrak{g}}

  \def\nG{\mathfrak{n}}
  \def\oG{\mathfrak{o}}

  \def\sG{\mathfrak{s}}
  
\def\UG{\mathfrak{U}}  \def\uG{\mathfrak{u}}

\title{
{\bf Nonmaximal symmetry breaking patterns in the supersymmetric $\widehat {\sG\uG}(8)_{k_U =1}$ theory} \\
\author{\large Ning Chen$^1$\,\orcidlink{0000-0002-0032-9012}, Jianan Tian$^2$\,\orcidlink{0009-0001-4677-1906}, Bin Wang$^3$\,\orcidlink{0009-0001-8556-7115} }
\date{\small \it
$^1\, ^2 \, ^3$School of Physics, Nankai University, Tianjin, 300071, China \\
}
}

\begin{document}

\maketitle
\setlength{\parskip}{0.2ex}

\begin{abstract}
\bigskip
We study the nonmaximal symmetry breaking patterns of ${\mathfrak{s} \mathfrak{u}}(8) \to \mathfrak{g}_{531}/\mathfrak{g}_{351}$ allowed in a supersymmetric theory with the affine Lie algebra of $\widehat {\sG\uG}(8)_{k_U=1}$, which was recently proposed to explain the three-generational standard model quark/lepton mass hierarchies as well as the Cabbibo-Kobayashi-Maskawa mixing pattern.
Three gauge couplings are found to achieve the unification below the Planck scales, according to the unification relations determined by the affine Lie algebra at level of $1$.
We also prove that a separate nonmaximal symmetry breaking pattern of ${\mathfrak{s} \mathfrak{u}}(8) \to \mathfrak{g}_{621}$ is unrealistic due to the additional massless vectorlike quarks in the spectrum.
\end{abstract}

\vspace{10.5cm}
{\emph{Emails:}\\  
$^{\,1}$\url{chenning_symmetry@nankai.edu.cn}\\
$^{\,2}$\url{tianjianan@mail.nankai.edu.cn}\\
$^{\,3}$\url{wb@mail.nankai.edu.cn}\\
 }

\thispagestyle{empty}  
\newpage  
 
\setcounter{page}{1}  

\vspace{1.0cm}
\eject
\tableofcontents

\section{INTRODUCTION}

\para
Grand unified theories (GUTs) were proposed to not only unify all three Standard Model (SM) symmetries into one simple Lie algebra of either $\sG\uG(5)$~\cite{Georgi:1974sy} or $\sG\oG(10)$~\cite{Fritzsch:1974nn}, but to also unify all SM fermions into some anomaly-free irreducible representations (IRs) of some simple Lie algebra $\gG_U$.
The Georgi-Glashow $\sG\uG(5)$ GUT~\cite{Georgi:1974sy} and the Fritzsch-Minkowski $\sG\oG(10)$ GUT~\cite{Fritzsch:1974nn} contain the chiral fermions of $ 3\times \[ \repb{5_F} \oplus \rep{10_F} \] $ and $ 3 \times \rep{16_F} $, respectively.
Neither framework can fully explain the three-generational SM fermion mass hierarchies and the Cabibbo-Kobayashi-Maskawa (CKM) mixing pattern~\cite{Cabibbo:1963yz,Kobayashi:1973fv} due to three repetitive chiral irreducible anomaly-free fermion sets (IRAFFSs) \footnote{The definition of the chiral IRAFFS was previously given in Refs.~\cite{Chen:2023qxi,Chen:2024cht}.} in the flavor sector.

\para
In a seminal paper by Georgi~\cite{Georgi:1979md}, he first proposed to avoid the repetitive structure in the flavor sector in a unified ${\sG \uG}(11)$ theory.
This was recently relaxed with two following chiral IRAFFSs at the GUT scale:
\beqn\label{eq:SU8_3gen_fermions}
\{ f_L \}_{  {\sG\uG}(8) }^{n_g=3}&=& \Big[ \repb{8_F}^\omega \oplus \rep{28_F} \Big] \bigoplus \Big[ \repb{8_F}^{ \dot \omega } \oplus \rep{56_F} \Big] \,,~ {\rm dim}_{ \mathbf{F}}= 156\,, \non
&& \Omega \equiv ( \omega \,, \dot \omega ) \,, ~ \omega = ( 3\,, {\rm IV}\,, {\rm V}\,, {\rm VI}) \,, ~  \dot \omega = (\dot 1\,, \dot 2\,, \dot {\rm VII}\,, \dot {\rm VIII}\,, \dot {\rm IX} ) \,,
\eeqn
starting from a nonsupersymmetric ${\sG\uG}(8)$ theory~\cite{Barr:2008pn,Chen:2023qxi,Chen:2024cht}.
Here, the undotted/dotted indices represent the $\repb{8_F}$'s in the rank-two chiral IRAFFS and the rank-three chiral IRAFFS, \footnote{The rank-two and the rank-three chiral IRAFFSs are named after the ${\sG\uG}(8)$ rank-two and rank-three antisymmetric fermions of $\rep{28_F}$ and $\rep{56_F}$, respectively.} respectively.
We also distinguish the heavy partner fermions and the SM fermions in terms of the Roman numbers and the Arabic numbers.
The global symmetries of the chiral fermions in Eq.~\eqref{eq:SU8_3gen_fermions} are
\beqn\label{eq:global_SU8}
\widetilde{ \Gc}_{\rm global} \[{\sG\uG}(8)\,, n_g=3 \]&=& \Big[ \widetilde{ {\rm SU}}(4)_\omega  \otimes \widetilde{ {\rm U}}(1)_{T_2}  \otimes \widetilde{ {\rm U}}(1)_{{\rm PQ}_2} \Big]  \bigotimes \Big[ \widetilde{ {\rm SU}}(5)_{\dot \omega } \otimes \widetilde{ {\rm U}}(1)_{T_3} \otimes \widetilde{ {\rm U}}(1)_{{\rm PQ}_3 }  \Big]  \,,
\eeqn
with the $\widetilde{ {\rm U}}(1)_{T_i}$ being the nonanomalous global symmetries, and the $\widetilde{ {\rm U}}(1)_{{\rm PQ}_i}$ being the anomalous global Peccei-Quinn symmetries~\cite{Peccei:1977hh}.
To count the SM generations by decomposing the ${\sG\uG}(8)$ fermion IRs into the ${\sG\uG}(5)$ IRs~\cite{Georgi:1979md}, one finds three identical SM $\repb{5_F}$'s and three distinctive $\rep{10_F}$'s~\cite{Barr:2008pn}.
It is sufficient to obtain three distinctive SM generations in the UV setup of the ${\sG\uG}(8)$ theory.

\para
The nonsupersymmetric ${\sG\uG}(8)$ theory was analyzed according to its maximal symmetry breaking pattern of ${\sG\uG}(8) \to \gG_{441} \equiv {\sG\uG}(4)_s \oplus {\sG\uG}(4)_W \oplus {\uG}(1)_{X_0}$~\cite{Chen:2024cht,Chen:2024deo}, which was favored by the minimization condition of the adjoint Higgs potential~\cite{Li:1973mq}.
It was also found that the renormalization group equations (RGEs) of the gauge couplings in the nonsupersymmetric ${\sG\uG}(8)$ theory cannot achieve the unification in the field theoretical interpretation of $\alpha_{4s}(v_U) = \alpha_{4 W} (v_U)= \alpha_{1} (v_U)$~\cite{Barr:2008pn}, even if the one-loop threshold effects~\cite{Hall:1980kf,Weinberg:1980wa,Langacker:1992rq} and the higher-dimension operators were taken into account.
The main difficulties of achieving the gauge coupling unification are (i) the Cartan discontinuities of different Abelian gauge couplings through the symmetry breaking stage of ${\sG\uG}(N) \oplus {\uG}(1)_X  \to  {\sG\uG}(N-1) \oplus  {\uG}(1)_{X^\prime }$ given by
\beqn\label{eq:Cartan_discont}
&& \alpha_{ X^\prime }^{-1} =  \frac{2}{ N(N-1)} \alpha_{N}^{-1} +  \alpha_{X}^{-1} \,,
\eeqn
where $(\alpha_N \,, \alpha_{ X^\prime } \,, \alpha_{ X } )$ are the gauge couplings of the $({\sG\uG}(N)\,, {\uG}(1)_X\,, {\uG}(1)_{X^\prime })$ symmetries, and (ii) the suggested intermediate symmetry breaking scales, which were previously determined according to the observed SM quark/lepton masses and the CKM mixing patterns~\cite{Chen:2024cht}.
In the context of string theory, it was shown by Ginsparg~\cite{Ginsparg:1987ee} that the gauge coupling unification can be achieved as
\beqn\label{eq:affine_unification}
&& k_s \alpha_{ s} (v_U) = k_W \alpha_{ W} (v_U) = k_1 \alpha_{ 1 }  (v_U) \,,
\eeqn
with $(k_s\,, k_W\,, k_1)$ being the affine levels of the corresponding strong, weak, and the Abelian gauge couplings in the physical basis. \footnote{Some early studies of the higher affine levels in the string-inspired unified theories include Refs.~\cite{Font:1990uw,Dienes:1996du}.}
In a recent paper~\cite{Chen:2024wcj}, we have proved that the reasonable conformal embedding based on the affine Lie algebra of
\beqn\label{eq:SU8_441embedding}
&& \widehat{ \sG \uG}(4)_{s\,, k_s = 1} \oplus \widehat{ \sG \uG}(4)_{W\,, k_W = 1} \oplus \hat{  \uG}(1)_{ 1 \,, k_{1\,,{\tt phys}} =  \frac{1}{4 } }  \subset \widehat{ \sG \uG}(8)_{ k_U = 1} \,
\eeqn
can achieve the gauge coupling unification with the $\Nc=1$ supersymmetric (SUSY) extension between the scale of $v_{441} \leq \mu \leq v_U$.
All chiral superfields are tabulated in Table~\ref{tab:SUSY_SU8_setup} so that (i) the $[\sG \uG(8)]^3$ gauge anomaly is vanishing, and (ii) a nonanomalous global symmetry of $[\sG \uG(8)]^2 \cdot \widetilde {\rm U}(1)_T=0$ is well defined. \footnote{The global $\widetilde {\rm U}(1)_T$ will evolve to the global $\widetilde {\rm U}(1)_{ B-L}$ symmetry in the SM~\cite{Chen:2023qxi}.}
Correspondingly, we expect the following Yukawa coupling terms from the holomorphic superpotential:
\beqn\label{eq:SUSYSU8_Yukawa}
W_Y &=& Y_\Bc \repb{8_F}^\omega \rep{28_F} \repb{8_H}_{\,, \omega} + Y_\Dc \repb{8_F}^{\dot \omega } \rep{56_F} \repb{28_H}_{\,, \dot \omega}  + Y_\Tc \rep{28_F} \rep{28_F} \rep{70_H}  \non
&& + \frac{c_4 }{ M_{\rm pl}} \rep{56_F} \rep{56_F} \rep{63_H} \rep{28_H}^{\dot \omega} \,,
\eeqn
with the reduced Planck scale of $ M_{\rm pl} = (8 \pi G_N)^{-\frac{1}{2}} = 2.4 \times 10^{18}\,{\rm GeV}$.

\begin{table}[htp]
\begin{center}
\begin{tabular}{c|cccc}
\hline \hline
$\sG\uG(8)$ &    $\widetilde{ {\rm SU}}(4)_\omega $ &  $\widetilde{  {\rm SU}}(5)_{ \dot \omega }$  &  $\widetilde {\rm U}(1)_{\rm PQ}$  &  $\widetilde {\rm U}(1)_T$  \\
\hline
$\repb{8_F}^{\omega }$    & $\rep{4}$  & $\rep{1}$  &  $p$ &  $-3t$  \\ 
$\repb{8_F}^{\dot \omega }$   & $\rep{1}$  & $\rep{5}$  &  $p$  &  $-3t$  \\ 
$ \rep{28_F} $    &  $\rep{1}$ & $\rep{1}$  &  $q_2$ & $+2t$  \\ 
$\rep{56_F}$   & $\rep{1}$  & $\rep{1}$  &  $q_3$ & $+t$  \\ 
\hline
$\repb{8_H}_{\,,\omega }/\rep{8_H}^{\omega }$   &  $\repb{4}/\rep{4}$  & $\rep{1}$  & $-( p+q_2) /p+q_2$  &  $+t/-t$  \\
$\repb{28_H}_{\,,\dot\omega  }/\rep{28_H}^{\dot\omega  }$   & $\rep{1}$  & $\repb{5}/\rep{5}$  & $-(p+q_3 )/p+q_3$  &  $+2t/-2t $  \\
$\rep{70_H}/\repb{70_H}$   & $\rep{1}$  & $\rep{1}$  &  $-2q_2/+2q_2$ & $-4t/+4t$ \\
$\rep{63_H}$   & $\rep{1}$  & $\rep{1}$  &  $0$ & $0$ \\
\hline\hline
\end{tabular}
\end{center}
\caption{
The representations and charges of all $\widehat {\sG\uG}(8)_{k_U =1}$ chiral superfields under the global symmetries in Eq.~\eqref{eq:global_SU8}.
}
\label{tab:SUSY_SU8_setup}
\end{table}%

\para
It turns out the Yukawa coupling terms of $\repb{8_F}^\omega \rep{28_F} \repb{8_H}_{\,, \omega}$, $\repb{8_F}^{\dot \omega } \rep{56_F} \repb{28_H}_{\,, \dot \omega}$, and $\rep{56_F} \rep{56_F} \rep{63_H} \rep{28_H}^{\dot \omega}$ give vectorlike fermion masses through the sequential stages after the breaking of ${ \sG \uG}(8) \to \gG_{441}$~\cite{Chen:2023qxi}, leaving three generational massless SM quarks/leptons plus twenty-three left-handed sterile neutrinos before the electroweak symmetry breaking (EWSB) stage according to the `t Hooft anomaly matching conditions~\cite{tHooft:1979rat}. 
Furthermore, the previous analyses of the maximal symmetry breaking patterns~\cite{Chen:2023qxi,Chen:2024cht,Chen:2024yhb} reveal the origins of three-generational SM quark/lepton masses and the observed CKM mixing pattern~\cite{Cabibbo:1963yz,Kobayashi:1973fv} in the quark sector through the Yukawa coupling to one unique SM Higgs boson within the $\rep{70_H}$ in the spectrum.
Specifically, the Yukawa coupling of $\rep{28_F} \rep{28_F} \rep{70_H}$ in Eq.~\eqref{eq:SUSYSU8_Yukawa} leads to the top quark mass, and other SM up-type quark masses are due to two $d=5$ operators of
\beqs\label{eqs:d5_ffHH}
\beqn
c_4\, \Oc_\Fc^{ (4\,,1)}  &=& c_4\, \rep{56_F}  \rep{56_F} \cdot \repb{28_H}_{\,, \dot \omega }  \cdot  \rep{70_H}  \,,\label{eq:d5_ffHH_41} \\[1mm]
c_5\, \Oc_\Fc^{ (5\,,1)}  &=& c_5\, \rep{28_F}  \rep{56_F} \cdot  \repb{8_H}_{\,,\omega }  \cdot  \rep{70_H} \,.\label{eq:d5_ffHH_51}
\eeqn
\eeqs
To generate the masses for all down-type quarks and charged leptons, we conjecture two sets of Higgs mixing terms,
\beqs\label{eqs:d5_Hmixings}
\beqn
d_{\mathscr A}\, \Oc_{\mathscr A}^{d=5} &\equiv&d_{\mathscr A}\,  \epsilon_{ \omega_1 \omega_2 \omega_3  \omega_4 } \repb{8_{H}}_{\,, \omega_1}^\dag   \repb{8_{H}}_{\,, \omega_2}^\dag \repb{8_{H}}_{\,, \omega_3}^\dag  \repb{8_{H}}_{\,, \omega_4 }^\dag  \rep{70_H}^\dag  \,, \quad \Pc\Qc = 2 ( 2p + 3 q_2 ) \neq 0 \,, \label{eq:d5_Hmixing_A}\\[1mm]
d_{\mathscr B}\, \Oc_{\mathscr B}^{d=5} &\equiv&d_{\mathscr B}\,   ( \repb{28_H}_{\,,\dot \kappa_1 }^\dag \repb{28_H}_{\,,\dot \kappa_2 } ) \cdot  \repb{28_H}_{\,,\dot \omega_1}^\dag \repb{28_H}_{\,,\dot \omega_2}^\dag   \rep{70_H}^\dag  \,, \quad  \Pc\Qc =  2 ( p + q_2 + q_3) \,, \non
&& {\rm with} ~~ \dot \kappa_2 \neq ( \dot \kappa_1 \,, \dot \omega_1 \,, \dot \omega_2 )  \,, \label{eq:d5_Hmixing_B}
\eeqn
\eeqs
where one of the Higgs fields will mix with two renormalizable Yukawa couplings of $Y_\Bc  \repb{8_F}^\omega  \rep{28_F}  \repb{8_{H}}_{\,,\omega }+ Y_\Dc \repb{8_F}^{\dot \omega  } \rep{56_F}  \repb{28_{H}}_{\,,\dot \omega }   +{\rm H.c.}$ in Eq.~\eqref{eq:SUSYSU8_Yukawa}.
Both $d=5$ Yukawa couplings in Eqs.~\eqref{eqs:d5_ffHH} and Higgs mixing terms in Eqs.~\eqref{eqs:d5_Hmixings} violate the global symmetries in Eq.~\eqref{eq:global_SU8} explicitly, and they were conjectured to be suppressed by the reduced Planck scale of $ M_{\rm pl} = (8 \pi G_N)^{-\frac{1}{2}} = 2.4 \times 10^{18}\,{\rm GeV}$~\cite{Chen:2024cht}.
Three intermediate symmetry breaking scales beyond the EW scale enter into the SM quark/lepton mass matrices, and hence are set by requiring a reasonable matching with the experimental measurements in the flavor sector.

\para
It was previously pointed out by Witten that other nonmaximal symmetry breaking patterns are possible in a SUSY theory~\cite{Witten:1981nf}.
The superpotential for the adjoint field $\rep{63_H}$ takes the form of
\beqn\label{eq:SU8_SUSYpotential}
W&\supset&  \hf \mu_U \Tr( \rep{63_H} )^2 + \frac{1}{ 3 } \lambda_U \Tr( \rep{63_H} )^3 \,,
\eeqn
and the corresponding field equation for the unbroken SUSY reads
\beqn
&& \lambda_U ( \rep{63_H}^2  -  \frac{1}{8} \mathbb{I}_{8} \Tr( \rep{63_H} )^2  ) + \mu_U \rep{63_H} =0  \,.
\eeqn
The vacuum expectation values (VEVs) of the adjoint Higgs field and the corresponding symmetry breaking patterns are
\beqs\label{eqs:SUSYSU8_Patterns}
\beqn
{\sG \uG}(8) \to \gG_{441}~&:&~ \langle \rep{63_H} \rangle = \frac{1}{ \sqrt{2} } v_U {\rm diag} ( \frac{1}{4} \mathbb{I}_4 \,, -  \frac{1}{4}  \mathbb{I}_4 ) \,,\label{eq:SUSYSU8_441Pattern} \\[1mm]
{\sG \uG}(8) \to  \gG_{531}/\gG_{351}~&:&~ \langle \rep{63_H} \rangle = \frac{ 3 \mu_U }{ 2 \lambda_U } {\rm diag} (\mathbb{I}_5 \,, -  \frac{5}{3} \mathbb{I}_3 ) / \langle \rep{63_H} \rangle = \frac{ 5 \mu_U }{ 2 \lambda_U } {\rm diag} (\mathbb{I}_3 \,, -  \frac{3}{5 } \mathbb{I}_5 ) \,, \label{eq:SUSYSU8_531Pattern} \\[1mm]
{\sG \uG}(8) \to  \gG_{621}~&:&~ \langle \rep{63_H} \rangle = \frac{ \mu_U }{ 2 \lambda_U } {\rm diag} (\mathbb{I}_6 \,, -  3 \mathbb{I}_2 ) \,.\label{eq:SUSYSU8_621Pattern}
\eeqn
\eeqs
The maximal symmetry breaking pattern of ${\sG \uG}(8)  \to  \gG_{441}$ can be achieved with the vanishing mass term of $\mu_U$ in Eq.~\eqref{eq:SU8_SUSYpotential}, and have been recently studied in Refs.~\cite{Chen:2024cht,Chen:2024deo,Chen:2024yhb,Chen:2024wcj}.
As for the nonmaximal symmetry breaking patterns in Eqs.~\eqref{eq:SUSYSU8_531Pattern} and \eqref{eq:SUSYSU8_621Pattern}, we have also proved in Ref.~\cite{Chen:2024wcj} that the corresponding conformal embeddings are
\beqs
\beqn
&& \widehat{ \sG \uG}( 5)_{ k_{s/W} = 1} \oplus \widehat{ \sG \uG}(3)_{ k_{W/s }= 1} \oplus \hat{  \uG}(1)_{ k_{1\,,{\tt phys}} =  \frac{1}{4 } }  \subset \widehat{ \sG \uG}(8)_{ k_U = 1} \,,\label{eq:SU8_531embedding} \\[1mm]
&& \widehat{ \sG \uG}( 6)_{ k_{s} = 1} \oplus \widehat{ \sG \uG}(2 )_{ k_{W }= 1} \oplus \hat{  \uG}(1)_{ k_{1\,,{\tt phys}} =  \frac{1}{4 } }  \subset \widehat{ \sG \uG}(8)_{ k_U = 1} \,.\label{eq:SU8_621embedding}
\eeqn
\eeqs
Given the possible nonmaximal symmetry breaking patterns in the SUSY extension, as well as the conformal embeddings in Eqs.~\eqref{eq:SU8_531embedding} and \eqref{eq:SU8_621embedding}, it is a central question to check whether the gauge coupling unification can be achieved according to Eq.~\eqref{eq:affine_unification}.

\para
If one considers the nonmaximal symmetry breaking patterns of ${\sG\uG}(8) \to \gG_{531}$ or ${\sG\uG}(8) \to \gG_{351}$ at the GUT scale, there can be four possible sequential symmetry breaking patterns, \footnote{The acronyms stand for the strong-strong-weak (SSW), strong-weak-strong (SWS), weak-strong-strong (WSS), and weak-weak-weak (WWW) symmetry breaking patterns, respectively.} namely,
\beqs\label{eqs:SU8_531351Patterns}
\beqn
{\rm SSW}~&:&~ {\sG\uG}(8) \to \gG_{531} \to  \gG_{431} \to \gG_{331}  \to \gG_{\rm SM} \,, \non
&& \gG_{531} \equiv {\sG\uG}(5)_{s} \oplus  {\sG\uG}(3)_W \oplus  {\uG}(1)_{X_0 }  \,, \,  \gG_{331} \equiv {\sG\uG}(3)_{c} \oplus  {\sG\uG}(3)_W \oplus  {\uG}(1)_{X_2 } \,, \label{eq:SUSYSU8_531_SSWPattern} \\
{\rm SWS}~&:&~ {\sG\uG}(8) \to \gG_{531} \to \gG_{431} \to \gG_{421} \to \gG_{\rm SM} \,, \non
&& \gG_{431} \equiv {\sG\uG}(4)_{s} \oplus  {\sG\uG}(3)_W \oplus  {\uG}(1)_{X_1 }  \,, \,  \gG_{421} \equiv {\sG\uG}(4)_{s}  \oplus  {\sG\uG}(2)_W \oplus  {\uG}(1)_{X_2 } \,, \label{eq:SUSYSU8_531_SWSPattern} \\
{\rm WSS}~&:&~ {\sG\uG}(8) \to \gG_{531} \to \gG_{521} \to \gG_{421} \to \gG_{\rm SM} \,, \non
&& \gG_{521} \equiv {\sG\uG}(5)_{s} \oplus {\sG\uG}(2)_W \oplus   {\uG}(1)_{X_1 }  \,, \label{eq:SUSYSU8_531_WSSPattern} \\
{\rm WWW}~&:&~ {\sG\uG}(8) \to \gG_{351} \to \gG_{341} \to \gG_{331} \to \gG_{\rm SM} \,, \non
&& \gG_{351} \equiv {\sG\uG}(3)_{c} \oplus  {\sG\uG}(5)_W \oplus  {\uG}(1)_{X_0 }  \,, \,  \gG_{341} \equiv {\sG\uG}(3)_{c}  \oplus  {\sG\uG}(4)_W \oplus  {\uG}(1)_{X_1 } \,, \label{eq:SUSYSU8_351_WWWPattern}
\eeqn
\eeqs
with $\gG_{\rm SM} \equiv  {\sG\uG}(3)_{c} \oplus {\sG\uG}(2)_W \oplus  {\uG}(1)_{Y }$.
In contrast to the maximal symmetry breaking pattern of $ {\sG \uG}(8) \to \gG_{441}$, the ${\uG}(1)_1$ gauge coupling is normalized as $\alpha_1 = \frac{16}{15 } \alpha_{X_0}$ in the above nonmaximal symmetry breaking patterns of $ {\sG \uG}(8) \to  \gG_{531}/\gG_{351} $.
Additionally, there can also be a possible symmetry breaking pattern of ${\sG\uG}(8) \to  \gG_{621} \equiv {\sG\uG}(6)_{s} \oplus  {\sG\uG}(2 )_W \oplus  {\uG}(1)_{X_0 } $, with the sequential breaking of ${\sG\uG}(6)_{s} \oplus {\uG}(1)_{X_0 }  \to ... \to  {\sG\uG}(3)_{c} \oplus {\uG}(1)_{ Y }$ in the extended strong sector.
However, this turns out to be a no-go pattern, since a pair of vectorlike quarks of $(\uG\,, \dG)$ in the spectrum remains massless through this symmetry breaking pattern.

\para
The possible $d = 5$ gravity-induced operator of
\beqn\label{eq:HSW_op}
&& \Oc_{\rm HSW} = - \frac{1}{2} \frac{c_{\rm HSW}}{M_{\rm pl}} \Tr \[ \rep{63_H} U_{\mu \nu} U^{\mu \nu} \]  \,
\eeqn
suggested by Hill-Shafi-Wetterich (HSW)~\cite{Hill:1983xh,Shafi:1983gz} could have effect on the gauge coupling unification.
Here, the $U_{\mu \nu}$ represents the $ {\sG\uG}(8) $ field strength tensor. 
After the GUT-scale symmetry breaking with the VEVs in Eq.~\eqref{eqs:SUSYSU8_Patterns}, this HSW operator shifts the ${\sG\uG}( N_s)_s$, ${\sG\uG}( N_W)_W$, and ${\uG}(1)_{X_0 }$ field strengths for different symmetry breaking patterns as follows:
\beqs\label{eqs:SU8_531351_HSW}
\beqn
{\sG\uG}(8) \to \gG_{531} &:& -\frac{1}{2} \left( 1 - \frac{3}{4} \epsilon \right) \Tr \[ G_{\mu \nu} G^{\mu \nu} \] -\frac{1}{2} \left( 1 + \frac{5}{4} \epsilon \right) \Tr \[ W_{\mu \nu} W^{\mu \nu} \]  \non 
&& - \frac{1}{4} \left( 1 + \frac{1}{2} \epsilon  \right)  \frac{15}{16} X_{0 \mu \nu} X_0^{\mu \nu} \,,    \label{eq:SUSYSU8_531_HSW} \\[1mm]
{\sG\uG}(8) \to \gG_{351} &:& -\frac{1}{2} \left( 1 - \frac{5}{4} \epsilon  \right) \Tr \[ G_{\mu \nu} G^{\mu \nu} \] -\frac{1}{2} \left( 1 + \frac{3}{4 } \epsilon \right) \Tr \[ W_{\mu \nu} W^{\mu \nu} \]  \non 
&& -\frac{1}{4} \left( 1 - \frac{1}{2}  \epsilon \right) \frac{15}{16} X_{0 \mu \nu} X_0^{\mu \nu} \,,   \label{eq:SUSYSU8_351_HSW}
\eeqn
\eeqs
where $ \epsilon \equiv \frac{1}{\sqrt{15}}  c_{\rm HSW} \zeta_0$ with $\zeta_0 \equiv \frac{v_U}{M_{\rm pl}}$. 
Consequently, the gauge coupling unification is modified into
\beqs\label{eqs:SU8_531351_GCU}
\beqn
{\sG\uG}(8) \to \gG_{531} &:&  \left( 1 - \frac{3}{4} \epsilon \right) \alpha_{5s} (v_U) = \left( 1 + \frac{5}{4} \epsilon \right) \alpha_{3W} (v_U) = \left( 1 + \frac{1}{2} \epsilon  \right) \frac{1}{4} \alpha_{1} (v_U) = \alpha_{U} (v_U)  \,, \label{eq:SUSYSU8_531_SSW_GCU} \\[1mm]
{\sG\uG}(8) \to \gG_{351} &:&  \left( 1 - \frac{5}{4} \epsilon  \right) \alpha_{3s} (v_U) = \left( 1 + \frac{3}{4 } \epsilon \right) \alpha_{5W} (v_U) = \left( 1 - \frac{1}{2}  \epsilon \right) \frac{1}{4} \alpha_{1} (v_U) = \alpha_{U} (v_U)  \,,\label{eq:SUSYSU8_351_WWW_GCU} 
\label{eq:SUSYSU8_621_GCU}
\eeqn
\eeqs
with the affine levels in Eq.~\eqref{eq:SU8_531embedding} and the gravity-induced operator in Eq.~\eqref{eq:HSW_op} taken into account.

\para
The rest of the paper will be devoted to analyzing the symmetry breaking patterns in Eqs.~\eqref{eq:SUSYSU8_531Pattern} and \eqref{eq:SUSYSU8_621Pattern}, with the focus on the gauge coupling evolutions in terms of the corresponding RGEs.
In Secs.~\ref{section:SSW_pattern}, \ref{section:SWS_pattern}, \ref{section:WSS_pattern}, and \ref{section:WWW_pattern}, we analyze the sequential symmetry breaking patterns following the nonminimal breaking patterns of ${\sG \uG}(8) \to  \gG_{531}$ and ${\sG \uG}(8) \to  \gG_{351}$ at the GUT scale.
Particularly, we derive the RGEs along each symmetry breaking pattern and find that the gauge coupling unification can be achieved in the context of the affine Lie algebra of $\widehat {\sG \uG}(8)_{k_U=1}$.
In Sec.~\ref{section:SSS_pattern}, we prove that the symmetry breaking pattern in Eq.~\eqref{eq:SUSYSU8_621Pattern} is unrealistic, since the Yukawa coupling of $\rep{56_F} \rep{56_F} \rep{63_H} \rep{28_H}^{\dot \omega}$ in Eq.~\eqref{eq:SUSYSU8_Yukawa} leads to vanishing mass terms for vectorlike fermions of $(\uG\,, \dG)$ within the $\rep{56_F}$.
We summarize the results in Sec.~\ref{section:conclusion}. 
In Appendix~\ref{section:RGEs}, we list the generic results of the one- and two-loop $\beta$ coefficients.
We present the details of all possible symmetry breaking patterns after the GUT-scale symmetry breaking of ${\sG \uG}(8)\to \gG_{531}/\gG_{351}$ in Ref.~\cite{Wang:2023gut}.
By reproducing the SM quark/lepton mass hierarchies and the CKM mixing pattern in Ref.~\cite{Chen:2024cht}, we identify the SM flavors as well as determine the intermediate symmetry breaking scales for each pattern.
Accordingly, the intermediate symmetry breaking scales are used for the RGEs in Secs.~\ref{section:SSW_pattern}, \ref{section:SWS_pattern}, \ref{section:WSS_pattern}, and \ref{section:WWW_pattern}.

\section{THE SSW SYMMETRY BREAKING PATTERN}
\label{section:SSW_pattern}

\subsection{Decompositions of the ${\sG \uG}$(8) fermions}

\begin{table}[htp] {\small
\begin{center}
\begin{tabular}{c|c|c|c|c}
\hline \hline
   ${\sG\uG}(8)$   &  $\gG_{531}$  & $\gG_{431}$  & $\gG_{331}$  &  $\gG_{\rm SM}$  \\
\hline \hline
 $\repb{ 8_F}^\Omega$   
 & $( \repb{5} \,, \rep{1}\,,  +\frac{1}{5} )_{ \mathbf{F} }^\Omega$  
 & $(\repb{4} \,, \rep{1} \,, +\frac{1}{4} )_{ \mathbf{F} }^\Omega$  
 & $(\repb{3} \,, \rep{1} \,, +\frac{1}{3} )_{ \mathbf{F} }^\Omega$  
 &  $( \repb{3} \,, \rep{ 1}  \,, +\frac{1}{3} )_{ \mathbf{F} }^{\Omega}~:~ { \Dc_R^\Omega}^c$  \\
 &  &  &  $( \rep{1} \,, \rep{1} \,, 0)_{ \mathbf{F} }^{\Omega}$ 
 &  $( \rep{1} \,, \rep{1} \,, 0)_{ \mathbf{F} }^{\Omega} ~:~ \check \Nc_L^{\Omega }$  \\
 &  & $( \rep{1} \,, \rep{1} \,, 0)_{ \mathbf{F} }^{\Omega^\prime}$
 &  $( \rep{1} \,, \rep{1} \,, 0)_{ \mathbf{F} }^{\Omega^\prime}$ 
 &  $( \rep{1} \,, \rep{1} \,, 0)_{ \mathbf{F} }^{\Omega^\prime} ~:~ \check \Nc_L^{\Omega^\prime }$  \\[1mm]
 & $(\rep{1}\,, \repb{3}  \,,  -\frac{1}{3})_{ \mathbf{F} }^\Omega$  
 &  $(\rep{1}\,, \repb{3}  \,,  -\frac{1}{3})_{ \mathbf{F} }^\Omega$  
 &  $( \rep{1} \,, \repb{3} \,,  -\frac{1}{3})_{ \mathbf{F} }^{\Omega}$  
 &  $( \rep{1} \,, \repb{2} \,,  -\frac{1}{2})_{ \mathbf{F} }^{\Omega } ~:~\Lc_L^\Omega =( \Ec_L^\Omega \,, - \Nc_L^\Omega )^T$   \\
 &   &  &  &  $( \rep{1} \,, \rep{1} \,,  0)_{ \mathbf{F} }^{\Omega^{\prime\prime} } ~:~ \check \Nc_L^{\Omega^{\prime\prime} }$  \\
\hline\hline
\end{tabular}
\caption{The ${\sG\uG}(8)$ fermion representation of $\repb{8_F}^\Omega$ under the $\gG_{531}\,,\gG_{431}\,, \gG_{331}\,, \gG_{\rm SM}$ subalgebras.}
\label{tab:SU8_8ferm_SSW}
\end{center} 
}
\end{table}%

\begin{table}[htp] {\small
\begin{center}
\begin{tabular}{c|c|c|c|c}
\hline \hline
   ${\sG\uG}(8)$   &  $\gG_{531}$  & $\gG_{431}$  & $\gG_{331}$  &  $\gG_{\rm SM}$  \\
\hline \hline 
 $\rep{28_F}$   
 & $( \rep{1}\,, \repb{ 3} \,, + \frac{2}{3})_{ \mathbf{F}}$ 
 & $( \rep{1}\,, \repb{ 3} \,, + \frac{2}{3})_{ \mathbf{F}}$
 & $( \rep{1}\,, \repb{ 3} \,, + \frac{2}{3})_{ \mathbf{F}}$
 & $( \rep{1}\,, \repb{ 2} \,, + \frac{1}{2})_{ \mathbf{F}} ~:~( { \nG_R }^c\,, - {\eG_R }^c  )^T$  \\
&   &  &   & $( \rep{1}\,, \rep{ 1} \,, +1)_{ \mathbf{F}}~:~ {\tau_R }^c $   \\[1mm]
& $( \rep{5}\,, \rep{3} \,,  +\frac{2}{15})_{ \mathbf{F}}$ 
& $( \rep{1}\,, \rep{3} \,,  +\frac{1}{3})_{ \mathbf{F}}^{\prime}$   
& $( \rep{1}\,, \rep{3} \,,  +\frac{1}{3})_{ \mathbf{F}}^{\prime}$  
& $( \rep{1}\,, \rep{2} \,,  +\frac{1}{2})_{ \mathbf{F}}^{\prime}~:~ ( {\eG_R^{\prime} }^c \,, { \nG_R^{\prime} }^c)^T$  \\
&   &   &   & $( \rep{1}\,, \rep{1} \,, 0 )_{ \mathbf{F}} ~:~ \check \nG_R^c $ \\
&   &  $( \rep{4}\,, \rep{3} \,,  + \frac{1}{12} )_{ \mathbf{F}}$ & $( \rep{1}\,, \rep{3} \,,  +\frac{1}{3})_{ \mathbf{F}}^{\prime\prime}$  & $( \rep{1}\,, \rep{2} \,,  +\frac{1}{2})_{ \mathbf{F}}^{\prime\prime} ~:~( {\eG_R^{\prime\prime} }^c \,, { \nG_R^{\prime\prime}}^c )^T$ \\
&   &   &   & $( \rep{1}\,, \rep{1}\,, 0)_{ \mathbf{F}}^{\prime} ~:~ \check \nG_R^{\prime\,c}$ \\
&   &  & $( \rep{3}\,, \rep{3} \,,  0 )_{ \mathbf{F}}$  
& $( \rep{3}\,, \rep{1} \,,  -\frac{1}{3} )_{ \mathbf{F}} ~:~\DG_L$  \\
&   &   &   & $( \rep{3}\,, \rep{2}\,, + \frac{1}{6})_{ \mathbf{F}} ~:~(t_L\,, b_L)^T$ \\[1mm]
& $( \rep{10}\,, \rep{ 1} \,, -\frac{2}{5})_{ \mathbf{F}}$ 
&$( \rep{4}\,, \rep{ 1} \,, -\frac{1}{4})_{ \mathbf{F}}$
&  $( \rep{3}\,, \rep{ 1} \,, -\frac{1}{3})_{ \mathbf{F}}^{\prime}$
& $( \rep{3}\,, \rep{1} \,, -\frac{1}{3})_{ \mathbf{F}}^{\prime} ~:~\DG_L^\prime$  \\
&   &   &  $( \rep{1}\,, \rep{1}\,, 0)_{ \mathbf{F}}^{\prime\prime}$ & $( \rep{1}\,, \rep{1} \,,  0)_{ \mathbf{F}}^{\prime\prime} ~:~\check \nG_R^{\prime \prime \,c}$  \\
&   &  $( \rep{6}\,, \rep{ 1} \,, -\frac{1}{2})_{ \mathbf{F}}$
& $( \rep{3}\,, \rep{ 1} \,, -\frac{1}{3})_{ \mathbf{F}}^{\prime\prime}$  
& $( \rep{3}\,, \rep{1} \,, -\frac{1}{3})_{ \mathbf{F}}^{\prime\prime} ~:~\DG_L^{\prime \prime}$   \\
&   &   & $( \repb{3}\,, \rep{1} \,, -\frac{2}{3} )_{ \mathbf{F}}$  & $ ( \repb{3}\,, \rep{1} \,, -\frac{2}{3} )_{ \mathbf{F}} ~:~ {t_R}^c$ \\[1mm]  
\hline\hline
\end{tabular}
\caption{
The ${\sG\uG}(8)$ fermion representation of $\rep{28_F}$ under the $\gG_{531}\,,\gG_{431}\,, \gG_{331}\,, \gG_{\rm SM}$ subalgebras.
}
\label{tab:SU8_28ferm_SSW}
\end{center}
}
\end{table}

\begin{table}[htp] {\small
\begin{center}
\begin{tabular}{c|c|c|c|c}
\hline \hline
   ${\sG\uG}(8)$   &  $\gG_{531}$  & $\gG_{431}$  & $\gG_{331}$  &  $\gG_{\rm SM}$  \\
\hline \hline
     $\rep{56_F}$   
     & $( \rep{ 1}\,, \rep{1} \,, +1)_{ \mathbf{F}}^{\prime}$  
     & $( \rep{ 1}\,, \rep{1} \,, +1)_{ \mathbf{F}}^{\prime}$ 
     & $( \rep{ 1}\,, \rep{1} \,, +1)_{ \mathbf{F}}^{\prime}$   
     & $ ( \rep{ 1}\,, \rep{1} \,, +1)_{ \mathbf{F}}^{\prime} ~:~ {\EG_R}^c $  \\[1mm]
                       & $( \rep{ 5}\,, \repb{3} \,, +\frac{7}{15})_{ \mathbf{F}}$ 
                       & $( \rep{ 1}\,, \repb{3} \,, +\frac{2}{3})_{ \mathbf{F}}^{\prime} $  
                       & $( \rep{ 1}\,, \repb{3} \,, +\frac{2}{3})_{ \mathbf{F}}^{\prime} $  
                       & $( \rep{ 1}\,, \repb{2} \,, +\frac{1}{2})_{ \mathbf{F}}^{\prime\prime\prime} ~:~ ( {\nG_R^{\prime\prime\prime }}^c \,, -{\eG_R^{\prime\prime\prime } }^c )^T $ \\
                       &   &   &   & $( \rep{ 1}\,, \rep{1} \,, +1)_{ \mathbf{F}}^{\prime \prime} ~:~ {e_R}^c $   \\
                       &   &  $( \rep{ 4}\,, \repb{3} \,, +\frac{5}{12})_{ \mathbf{F}}$ 
                       & $( \rep{1}\,, \repb{3} \,, +\frac{2}{3})_{ \mathbf{F}}^{\prime\prime} $  
                       & $( \rep{1}\,, \repb{2} \,, +\frac{1}{2})_{ \mathbf{F}}^{\prime\prime\prime\prime} ~:~ ( {\nG_R^{\prime\prime\prime\prime }}^c \,, -{\eG_R^{\prime\prime\prime\prime } }^c )^T $ \\
                       &   &  &  &  $( \rep{1}\,, \rep{1} \,, +1)_{ \mathbf{F}}^{\prime\prime\prime} ~:~ {\mu_R}^c $  \\
                       &   &   & $( \rep{3}\,, \repb{3} \,, +\frac{1}{3})_{ \mathbf{F}} $  
                       & $( \rep{3}\,, \rep{1} \,, +\frac{2}{3})_{ \mathbf{F}} ~:~ \UG_L $ \\
                       &   &  &  &  $( \rep{3}\,, \repb{2} \,, +\frac{1}{6})_{ \mathbf{F}} ~:~ (\dG_L \,, - \uG_L )^T$  \\[1mm]
                       & $( \repb{10}\,, \rep{1} \,, -\frac{3}{5})_{ \mathbf{F}}$  
                       & $( \repb{4}\,, \rep{1} \,, -\frac{3}{4})_{ \mathbf{F}} $ 
                       & $( \rep{1}\,, \rep{1} \,, -1)_{ \mathbf{F}} $ 
                       & $( \rep{1}\,, \rep{1} \,, -1)_{ \mathbf{F}}^{  } ~:~ \EG_L $ \\
                       &   &   & $( \repb{3}\,, \rep{1} \,, -\frac{2}{3})_{ \mathbf{F}}^{\prime}$  &  $( \repb{3}\,, \rep{1} \,, -\frac{2}{3})_{ \mathbf{F}}^{\prime} ~:~ {c_R}^c$ \\
                       &   &  $( \rep{6}\,, \rep{1} \,, -\frac{1}{2})_{ \mathbf{F}}^{\prime}$ & $( \rep{3}\,, \rep{1} \,, -\frac{1}{3})_{ \mathbf{F}}^{\prime \prime \prime}$ & $( \rep{3}\,, \rep{1} \,, -\frac{1}{3})_{ \mathbf{F}}^{\prime\prime\prime} ~:~ \DG_L^{\prime \prime\prime}$  \\
                       &   &   & $( \repb{3}\,, \rep{1} \,, -\frac{2}{3})_{ \mathbf{F}}^{\prime\prime} $  & $( \repb{3}\,, \rep{1} \,, -\frac{2}{3})_{ \mathbf{F}}^{\prime \prime} ~:~ {u_R}^c $  \\[1mm]
                       & $( \rep{10}\,, \rep{3} \,, -\frac{1}{15})_{ \mathbf{F}}$  
                       &  $( \rep{4}\,, \rep{3} \,, +\frac{1}{12})_{ \mathbf{F}}^{\prime}$ 
                       & $( \rep{1}\,, \rep{3} \,, +\frac{1}{3})_{ \mathbf{F}}^{\prime\prime\prime} $ 
                       & $( \rep{1}\,, \rep{2} \,, +\frac{1}{2})_{ \mathbf{F}}^{\prime\prime\prime \prime \prime} ~:~ ( {\eG_R^{\prime\prime\prime\prime\prime }}^c \,, {\nG_R^{\prime\prime\prime\prime\prime } }^c )^T$ \\
                       &   &   &   & $( \rep{1}\,, \rep{1} \,, 0 )_{ \mathbf{F}}^{\prime\prime \prime} ~:~ {\check \nG_R}^{\prime \prime\prime \,c}$ \\
                       &   &  & $( \rep{3}\,, \rep{3} \,, 0 )_{ \mathbf{F}}^{\prime} $ & $( \rep{3}\,, \rep{1} \,, -\frac{1}{3})_{ \mathbf{F}}^{\prime\prime\prime \prime } ~:~\DG_L^{\prime\prime \prime \prime}$  \\
                       &   &   &   & $ ( \rep{3}\,, \rep{2} \,, +\frac{1}{6} )_{ \mathbf{F}}^{\prime} ~:~ (u_L\,,d_L)^T $ \\
                        &   &  $( \rep{6}\,, \rep{3} \,, -\frac{1}{6})_{ \mathbf{F}}$  & $( \rep{3}\,, \rep{3} \,, 0)_{ \mathbf{F}}^{\prime\prime}$ & $( \rep{3}\,, \rep{1} \,, -\frac{1}{3})_{ \mathbf{F}}^{\prime\prime\prime\prime\prime} ~:~ \DG_L^{\prime\prime\prime\prime\prime} $   \\
                       &   &   &   & $( \rep{3}\,, \rep{2} \,, +\frac{1}{6})_{ \mathbf{F}}^{\prime\prime} ~:~ (c_L\,,s_L)^T $  \\
                       &   &   & $( \repb{3}\,, \rep{3} \,, -\frac{1}{3})_{ \mathbf{F}} $  & $ ( \repb{3}\,, \rep{1} \,, -\frac{2}{3})_{ \mathbf{F}}^{\prime\prime\prime} ~:~ {\UG_R}^c $  \\
                       &   &   &   & $ ( \repb{3}\,, \rep{2} \,, -\frac{1}{6} )_{ \mathbf{F}}  ~:~ ( {\dG_R}^c \,,{\uG_R}^c )^T $ \\[1mm]
\hline\hline
\end{tabular}
\caption{
The ${\sG\uG}(8)$ fermion representation of $\rep{56_F}$ under the $\gG_{531}\,,\gG_{431}\,, \gG_{331}\,, \gG_{\rm SM}$ subalgebras. 
}
\label{tab:SU8_56ferm_SSW}
\end{center}
}
\end{table}%

\para
By following the SSW symmetry breaking pattern in Eq.~\eqref{eq:SUSYSU8_531_SSWPattern}, we tabulate the fermion representations at various stages in Tables~\ref{tab:SU8_8ferm_SSW}, \ref{tab:SU8_28ferm_SSW}, and \ref{tab:SU8_56ferm_SSW}. 
Along this pattern, the global $\widetilde{ {\rm U}}(1)_T$ symmetry evolves to the global $\widetilde{ {\rm U}}(1)_{B-L}$ at the EW scale according to the following sequence:
\beqn\label{eq:U1T_SSW_def}
&& \gG_{531}~:~ \Tc^\prime \equiv \Tc - 5t \Xc_0 \,, \quad  \gG_{431}~:~ \Tc^{ \prime \prime} \equiv \Tc^\prime  \,, \non
&& \gG_{331}~:~   \Tc^{ \prime \prime \prime} \equiv  \Tc^{ \prime \prime} + 8 t \Xc_2 \,, \quad  \gG_{\rm SM}~:~  \Bc- \Lc \equiv  \Tc^{ \prime \prime \prime} \,.
\eeqn

\subsection{Decompositions of the ${\sG \uG}$(8) Higgs fields}

\para
We decompose the Higgs fields in Eq.~\eqref{eq:SUSYSU8_Yukawa} into components that can develop VEVs for the sequential symmetry breaking stages in Eq.~\eqref{eq:SUSYSU8_531_SSWPattern}. 
For Higgs fields of $\repb{ 8_H}_{\,, \omega}$, they read
\begin{eqnarray}\label{eq:SU8_SSW_Higgs_Br01}
\repb{8_H}_{\,,\omega }  &\supset&  \underline{  ( \rep{1} \,, \repb{3} \,, -\frac{1}{3} )_{\mathbf{H}\,, \omega }  } \oplus  \langle ( \repb{5} \,, \rep{1} \,, +\frac{1}{5} )_{\mathbf{H}\,, \omega } \rangle \non
&\supset&  \underline{ ( \rep{1} \,, \repb{3} \,, -\frac{1}{3} )_{\mathbf{H}\,, \omega } }  \oplus  \langle ( \repb{4} \,, \rep{1} \,, +\frac{1}{4} )_{\mathbf{H}\,, \omega }   \rangle \non
&\supset&  \langle ( \rep{1} \,, \repb{3} \,, -\frac{1}{3} )_{\mathbf{H}\,, \omega }  \rangle \,.
\end{eqnarray}
For Higgs fields of $\repb{28_H}_{\,,\dot \omega } $, they read
\begin{eqnarray}\label{eq:SU8_SSW_Higgs_Br02}
\repb{28_H}_{\,,\dot \omega } &\supset& \underline{( \rep{1} \,, \rep{3} \,,  -\frac{2}{3} )_{\mathbf{H}\,, \dot\omega } } \oplus  \underline{ ( \repb{5} \,, \repb{3} \,, -\frac{2}{15} )_{\mathbf{H}\,, \dot\omega } } \oplus  \underline{ ( \repb{10} \,, \rep{1} \,, +\frac{2}{5} )_{\mathbf{H}\,, \dot\omega } }  \non
&\supset &  \underline{  ( \rep{1} \,, \rep{3} \,, -\frac{2}{3} )_{\mathbf{H}\,, \dot\omega }  } \oplus \Big[  \underline{ ( \rep{1} \,, \repb{3} \,, -\frac{1}{3} )_{\mathbf{H}\,, \dot\omega }  }  \oplus \underline{  ( \repb{4} \,, \repb{3} \,, -\frac{1}{12} )_{\mathbf{H}\,, \dot\omega }  } \Big] \oplus \langle ( \repb{4} \,, \rep{1} \,, +\frac{1}{4} )_{\mathbf{H}\,, \dot\omega }  \rangle \non
&\supset&  \underline{ ( \rep{1} \,, \rep{3} \,, -\frac{2}{3} )_{\mathbf{H}\,, \dot\omega }  }  \oplus \Big[ \langle ( \rep{1} \,, \repb{3} \,, -\frac{1}{3} )_{\mathbf{H}\,, \dot\omega }  \rangle \oplus   \langle ( \rep{1} \,, \repb{3} \,, -\frac{1}{3} )_{\mathbf{H}\,, \dot\omega }^{\prime}  \rangle  \Big]  \,.
%
\end{eqnarray}
For Higgs field of $\rep{70_H}$, it reads
\begin{eqnarray}\label{eq:SU8_SSW_Higgs_Br03} 
\rep{70_H} &\supset& ( \rep{5} \,, \rep{1 } \,, +\frac{4}{5} )_{\mathbf{H}}^{  }  \oplus  ( \repb{5} \,, \rep{1} \,, -\frac{4}{5} )_{\mathbf{H}}  \oplus \underline{ ( \rep{10} \,, \repb{3} \,, +\frac{4}{15} )_{\mathbf{H}} }   \oplus ( \repb{10} \,, \rep{3 } \,, -\frac{4}{15} )_{\mathbf{H}}^{ } \non
&\supset& \underline{ ( \rep{4} \,, \repb{3} \,, +\frac{5}{12} )_{\mathbf{H}} } \supset  \underline{ ( \rep{1} \,, \repb{3} \,, +\frac{2}{3} )_{\mathbf{H}}^{  } }  \supset  \langle ( \rep{1} \,, \repb{2} \,, +\frac{1}{2} )_{\mathbf{H}}^{  }  \rangle \,. 
\end{eqnarray}
The decompositions of the Higgs fields of $\rep{8_H}^\omega$ and $\rep{28_H}^{\dot \omega}$ can be made by taking the conjugate IRs in Eqs.~\eqref{eq:SU8_SSW_Higgs_Br01} and \eqref{eq:SU8_SSW_Higgs_Br02}, and we will neglect the corresponding expressions throughout the text.

\subsection{RGEs of the SSW symmetry breaking pattern}


\para
Between the $ v_{531} \leq \mu \leq v_{U} $, the massless Higgs fields are summarized as follows:
%
%
\beqn
&&( \repb{5}\,, \rep{1}\,, +\frac{1}{5})_{ \mathbf{H}\,, \omega} \oplus  ( \rep{1}\,, \repb{3}\,, -\frac{1}{3})_{ \mathbf{H}\,, \omega} \subset \repb{8_H}_{, \omega } \,, \non
&&(\rep{1}\,,\rep{3}\,, -\frac{2}{3})_{\mathbf{H}\,,\dot\omega} \oplus
(\repb{5}\,,\repb{3}\,, -\frac{2}{15})_{\mathbf{H}\,,\dot\omega} \oplus ( \repb{10}\,, \rep{1}\,, +\frac{2}{5})_{\mathbf{H}\,,\dot \omega}  \subset \repb{28_H}_{,\dot\omega} \,, \non 
&&(\rep{5}\,,\rep{1}\,, +\frac{4}{5})_\mathbf{H} \oplus (\repb{5}\,,\rep{1}\,, -\frac{4}{5})_\mathbf{H} \oplus (\rep{10}\,,\repb{3}\,, + \frac{4}{15})_\mathbf{H} \oplus (\repb{10}\,,\rep{3}\,, - \frac{4}{15})_\mathbf{H}  \subset \rep{70_H} \,, \non
&&(\repb{5}\,,\rep{1}\,, -\frac{4}{5})_\mathbf{H} \oplus (\rep{5}\,,\rep{1}\,, +\frac{4}{5})_\mathbf{H} \oplus (\repb{10}\,,\rep{3}\,, - \frac{4}{15})_\mathbf{H} \oplus (\rep{10}\,,\repb{3}\,, + \frac{4}{15})_\mathbf{H}  \subset \repb{70_H}\,, \non
&&( \rep{5}\,, \rep{1}\,, -\frac{1}{5})_{ \mathbf{H}}^{ \omega } \oplus ( \rep{1}\,, \rep{3}\,, +\frac{1}{3})_{ \mathbf{H}}^{ \omega } \subset \rep{8_H}^{ \omega } \,, \non
&&(\rep{1}\,,\repb{3}\,, +\frac{2}{3})_{\mathbf{H}}^{\dot\omega} \oplus (\rep{5}\,,\rep{3}\,, +\frac{2}{15})_{\mathbf{H}}^{\dot\omega} \oplus ( \rep{10}\,, \rep{1}\,, -\frac{2}{5})_{\mathbf{H}}^{\dot\omega} \subset \rep{28_H}^{\dot\omega} \,.
\eeqn
All fermionic components of $\repb{8_F}^\Omega \oplus \rep{28_F} \oplus \rep{56_F}$ remain massless after the decomposition into the $\gG_{531} $ IRs.
Consequently, we have the two-loop $\gG_{531} $ $ \beta $ coefficients of
\beqn\label{eq:HiggsA_531}
&& (b^{(1)}_{{\sG \uG}(5)_{s}}\,,b^{(1)}_{{\sG \uG}(3)_{W}}\,,b^{(1)}_{{\uG}(1)_{X_0}})   = (+54 \,, +60 \,, +\frac{368}{5}) \,, \non
&& b^{(2)}_{\gG_{531}}  = \begin{pmatrix}
6984/5 & 216 & 672/25\\
648 & 728 & 352/15\\
16128/25 & 2816/15 & 98816/1125
\end{pmatrix} \,.
\eeqn
%
%


\para
Between the $ v_{431} \leq \mu \leq v_{531} $, the massless Higgs fields are the following:
\beqn
&& ( \repb{4}\,, \rep{1}\,, +\frac{1}{4})_{ \mathbf{H}\,, 3 , \rm V ,\rm VI} \oplus ( \rep{1}\,, \repb{3}\,, -\frac{1}{3})_{\mathbf{H}\,, 3, \rm V,\rm VI } \subset ... \subset \repb{8_H}_{\,, 3 , \rm V ,\rm VI } \,,\non
&& (\rep{1}\,,\rep{3}\,, -\frac{2}{3})_{\mathbf{H}\,,\dot{\omega}} \oplus \Big[ (\rep{1}\,,\repb{3}\,, -\frac{1}{3})_{\mathbf{H}\,,\dot{\omega}} \oplus (\repb{4}\,,\repb{3}\,, -\frac{1}{12})_{\mathbf{H}\,,\dot{\omega}} \Big]  \non
& \oplus &  \Big[ (\repb{4}\,,\rep{1}\,, +\frac{1}{4})_{\mathbf{H}\,,\dot{\omega}} \oplus (\rep{6}\,,\rep{1}\,, +\frac{1}{2})_{\mathbf{H}\,,\dot{\omega}} \Big]  \subset ... \subset \repb{28_H}_{\,, \dot{\omega}} \,,\quad    \dot{\omega}=( \dot{1},\dot{2},\dot{\rm VII},\dot{\rm VIII},\dot{\rm IX} )\,, \non
&&(\rep{1}\,,\repb{3}\,, +\frac{2}{3})_{\mathbf{H}}^{\dot{\rm VII}} \oplus \Big[ (\rep{1}\,,\rep{3}\,, +\frac{1}{3})_{\mathbf{H}}^{\dot{\rm VII}} \oplus (\rep{4}\,,\rep{3}\,, +\frac{1}{12})_{\mathbf{H}}^{\dot{\rm VII}} \Big]    \non
& \oplus &  \Big[ (\rep{4}\,,\rep{1}\,, -\frac{1}{4})_{\mathbf{H}}^{\dot{\rm VII}} \oplus (\rep{6}\,,\rep{1}\,, -\frac{1}{2})_{\mathbf{H}}^{\dot{\rm VII}} \Big]  \subset ... \subset \rep{28_H}^{\dot{\rm VII}} \,, \non
&& (\rep{4}\,,\repb{3}\,, +\frac{5}{12})_\mathbf{H}  \subset ... \subset \rep{70_H} \,.
\eeqn
All ${\sG \uG}$(8) fermions that remain massless after the decomposition into the $\gG_{431}$ IRs can be found in Ref.~\cite{Wang:2023gut}.
Consequently, we have the two-loop $\gG_{431}$ $ \beta $ coefficients of
\beqn\label{eq:HiggsA_431}
&&(b^{(1)}_{{\sG \uG}(4)_{s}}\,,b^{(1)}_{{\sG \uG}(3)_{W}}\,,b^{(1)}_{{\uG}(1)_{X_1}})  = ( + \frac{5}{3} \,, +\frac{11}{2} \,, +\frac{115}{6}) \,,\non
&& b^{(2)}_{\gG_{431}}  = \begin{pmatrix}
3511/12 & 76 & 269/24\\
285/2 & 233 & 47/4\\
1345/8 & 94 & 5267/144
\end{pmatrix}\,.
\eeqn
The ${\sG \uG}(5)_s \oplus {\uG}(1)_{X_0}$ gauge couplings match with the ${\sG \uG}(4 )_s \oplus {\uG}(1)_{X_1}$ gauge couplings according to the following conditions:
\beqn\label{eq:531_coupMatch}
&& \alpha_{5s }^{-1} (v_{531} ) =  \alpha_{4s }^{-1} (v_{531} )  \,,~ \alpha_{X_1}^{-1} (v_{531} )  = \frac{1}{10} \alpha_{5s }^{-1} (v_{ 531} )  +  \alpha_{X_0}^{-1} (v_{ 531} ) \,.
\eeqn
%
%


\para
Between the $ v_{331} \leq \mu \leq v_{431} $, the massless Higgs fields are
\beqn
&& ( \repb{3}\,, \rep{1}\,, +\frac{1}{3})_{ \mathbf{H}\,, 3 ,\rm VI} \oplus ( \rep{1}\,, \repb{3}\,, -\frac{1}{3})_{ \mathbf{H}\,, 3 ,\rm VI}   \subset ... \subset \repb{8_H}_{, 3 ,\rm VI} \,, \non
&& (\rep{1}\,,\rep{3}\,, -\frac{2}{3})_{\mathbf{H}\,,\dot{\omega}} \oplus \Big[ (\rep{1}\,,\repb{3}\,, -\frac{1}{3})_{\mathbf{H}\,,\dot{\omega}} \oplus (\rep{1}\,,\repb{3}\,, -\frac{1}{3})_{\mathbf{H}\,,\dot{\omega}}^{\prime} \oplus (\repb{3}\,,\repb{3}\,, 0)_{\mathbf{H}\,,\dot{\omega}} \Big]   \non
& \oplus &  \Big[ (\repb{3}\,,\rep{1}\,, +\frac{1}{3})_{\mathbf{H}\,,\dot{\omega}} \oplus (\repb{3}\,,\rep{1}\,, +\frac{1}{3})_{\mathbf{H} \,,\dot{\omega}}^{\prime} \oplus (\rep{3}\,,\rep{1}\,, +\frac{2}{3})_{\mathbf{H}\,,\dot{\omega}} \Big] \subset ... \subset \repb{28_H}_{\,, \dot{\omega}} \,, \quad  \dot{\omega}=( \dot{2},\dot{\rm VIII},\dot{\rm IX} ) \,, \non
&& (\rep{1}\,,\repb{3}\,, +\frac{2}{3})_{\mathbf{H}} \subset ... \subset \rep{70_H} \,.
\eeqn
All ${\sG \uG}$(8) fermions that remain massless after the decomposition into the $\gG_{331}$ IRs can be found in Ref.~\cite{Wang:2023gut}. 
Consequently, we have the $\gG_{331}$ $\beta$ coefficients of
\beqn\label{eq:HiggsA_331}
&&  (b^{(1)}_{{\sG \uG}(3)_{c}}\,,b^{(1)}_{{\sG \uG}(3)_{W}}\,,b^{(1)}_{{\uG}(1)_{X_2}}) 
 = (-\frac{5}{3} \,, -\frac{3}{2} \,, +\frac{116}{9})\,,\non
&& b^{(2)}_{\gG_{331}}
 = \begin{pmatrix}
256/3 & 36 & 58/9 \\
36 & 89 & 22/3\\
464/9 & 176/3 & 728/27
\end{pmatrix} \,.
\eeqn
The ${\sG \uG}(4)_s \oplus {\uG}(1)_{X_1}$ gauge couplings match with the ${\sG \uG}(3 )_c \oplus {\uG}(1)_{X_2}$ gauge couplings according to the following conditions:
\beqn\label{eq:431_coupMatch}
&& \alpha_{4s }^{-1} (v_{431} ) =  \alpha_{3c }^{-1} (v_{431} )  \,,~ \alpha_{X_2}^{-1} (v_{ 431} ) = \frac{1}{6} \alpha_{4s }^{-1} (v_{431} )  +  \alpha_{X_1}^{-1} (v_{431} ) \,.
\eeqn
%
%


\para
Between the $ v_{\rm EW} \leq \mu \leq v_{331} $, the massless Higgs field is
\beq
\begin{aligned}
(\rep{1}\,,\repb{2}\,, +\frac{1}{2})_{\mathbf{H}} \subset
(\rep{1}\,,\repb{3}\,, +\frac{2}{3})_{\mathbf{H}} \subset
(\rep{4}\,,\repb{3}\,, +\frac{5}{12})_\mathbf{H} \subset
(\rep{10}\,,\repb{3}\,, +\frac{4}{15})_\mathbf{H}  
&\subset \rep{70_H} \,.
\end{aligned}
\eeq
All ${\sG \uG}$(8) fermions that remain massless after the decomposition into the $\gG_{\rm SM}$ IRs can be found in Ref.~\cite{Wang:2023gut}. 
Consequently, we have the $\gG_{\rm SM}$ $\beta$ coefficients of
\beqn\label{eq:HiggsA_SMto331}
&& (b^{(1)}_{{\sG \uG}(3)_{c}}\,,b^{(1)}_{{\sG \uG}(2)_{W}}\,,b^{(1)}_{{\uG}(1)_{Y}})   = (-7\,,-\frac{19}{6}\,,+\frac{41}{6}) \,, \non
&& b^{(2)}_{\gG_{\rm SM}} = \begin{pmatrix}
-26&9/2&11/6\\
12&35/6&3/2\\
44/3&9/2&199/18
\end{pmatrix} \,.
\eeqn
The ${\sG \uG}(3)_W \oplus {\uG}(1)_{X_2}$ gauge couplings of $(\alpha_{3W}\,, \alpha_{X_2} )$ match with the ${\sG \uG}(2 )_W \oplus {\uG}(1)_{\rm Y}$ gauge couplings according to the following conditions:
\beqn\label{eq:331_coupMatch}
&& \alpha_{3W }^{-1} (v_{331} ) =  \alpha_{2W }^{-1} (v_{331} )  \,,~ \alpha_{Y}^{-1} (v_{ 331} )  = \frac{1}{3} \alpha_{3W }^{-1} (v_{331} )  +  \alpha_{X_2}^{-1} (v_{331} ) \,.
\eeqn
Based on constructing the SM quark/lepton mass matrices to reproduce the observed hierarchical masses and the CKM mixing pattern, we find the following benchmark point:
\begin{align}\label{eq:benchmark SSW}
	v_{531}  \simeq 1.4 \times 10^{17} \, \mathrm{GeV} \,, \quad
	v_{431}  \simeq 4.8 \times 10^{15}\, \mathrm{GeV} \,, \quad
	v_{331}  \simeq 4.8 \times 10^{13} \, \mathrm{GeV} \,.
\end{align}
%
%

\begin{figure}[htb]
\centering
\includegraphics[height=4.5cm]{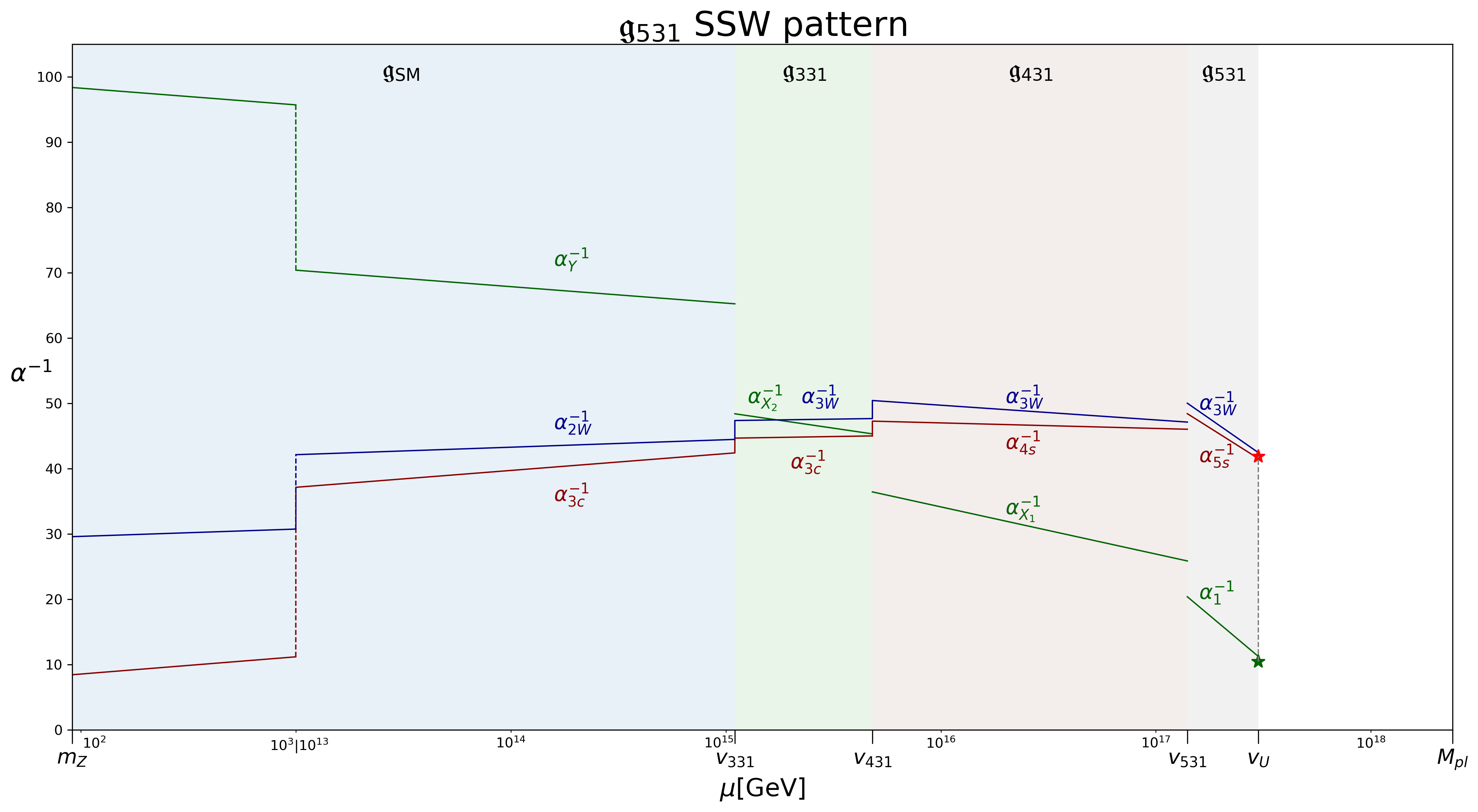}
\caption{The two-loop RGEs of the ${\sG \uG}(8)$ setup according to the SSW symmetry breaking pattern.
The RG evolutions within $10^{3}\,{\rm GeV} \lesssim \mu \lesssim 10^{13}\,{\rm GeV}$ are hidden in order to highlight the behaviors in three intermediate symmetry breaking scales given by the benchmark point in Eq.~\eqref{eq:benchmark SSW}.
}
\label{fig:SU8_RGE_SSW} 
\end{figure}

\para
With the one- and two-loop $\beta$ coefficients defined in Eqs.~\eqref{eq:HiggsA_531}, \eqref{eq:HiggsA_431}, \eqref{eq:HiggsA_331}, and \eqref{eq:HiggsA_SMto331}, we plot the RGEs of the ${\sG \uG}(8)$ setup along the SSW symmetry breaking pattern in Fig.~\ref{fig:SU8_RGE_SSW}.
The discontinuities of the ${\uG}(1)$ gauge couplings at three intermediate scales follow from their definitions in Eqs.~\eqref{eq:531_coupMatch}, \eqref{eq:431_coupMatch}, \eqref{eq:331_coupMatch}, respectively.
The benchmark point in Fig.~\ref{fig:SU8_RGE_SSW} is marked by $ \star $ and reads
\begin{gather}
c_{\rm HSW} \approx 0.24  \,,~v_{U} \approx 2.28 \times 10^{17} \, {\rm GeV} \,, \non
\alpha_{5c }^{-1} (v_{U} ) \approx 44.02 \,,~\alpha_{3W }^{-1} (v_{U} ) \approx 45.18\,,~\alpha_{1 }^{-1} (v_{U} ) \approx 10.50 \,.
\end{gather}

\section{THE SWS SYMMETRY BREAKING PATTERN}
\label{section:SWS_pattern}

\subsection{Decompositions of the ${\sG \uG}$(8) fermions}

\begin{table}[htp] {\small
\begin{center}
\begin{tabular}{c|c|c|c|c}
\hline \hline
   ${\sG\uG}(8)$   &  $\gG_{531}$  & $\gG_{431}$  & $\gG_{421}$  &  $\gG_{\rm SM}$  \\
\hline \hline
 $\repb{ 8_F}^\Omega$   
 & $( \repb{5} \,, \rep{1}\,,  +\frac{1}{5} )_{ \mathbf{F} }^\Omega$  
 & $(\repb{4} \,, \rep{1} \,, +\frac{1}{4} )_{ \mathbf{F} }^\Omega$  
 & $(\repb{4} \,, \rep{1} \,, +\frac{1}{4} )_{ \mathbf{F} }^\Omega$  
 &  $( \repb{3} \,, \rep{ 1}  \,, +\frac{1}{3} )_{ \mathbf{F} }^{\Omega}~:~ { \Dc_R^\Omega}^c$  \\
 &  &  &  &  $( \rep{1} \,, \rep{1} \,, 0)_{ \mathbf{F} }^{\Omega} ~:~ \check \Nc_L^{\Omega }$  \\
 &  & $( \rep{1} \,, \rep{1} \,, 0)_{ \mathbf{F} }^{\Omega^\prime}$
 &  $( \rep{1} \,, \rep{1} \,, 0)_{ \mathbf{F} }^{\Omega^\prime}$ 
 &  $( \rep{1} \,, \rep{1} \,, 0)_{ \mathbf{F} }^{\Omega^\prime} ~:~ \check \Nc_L^{\Omega^\prime }$  \\[1mm]
 & $(\rep{1}\,, \repb{3}  \,,  -\frac{1}{3})_{ \mathbf{F} }^\Omega$  
 &  $(\rep{1}\,, \repb{3}  \,,  -\frac{1}{3})_{ \mathbf{F} }^\Omega$  
 &  $( \rep{1} \,, \repb{2} \,,  -\frac{1}{2})_{ \mathbf{F} }^{\Omega}$  
 &  $( \rep{1} \,, \repb{2} \,,  -\frac{1}{2})_{ \mathbf{F} }^{\Omega } ~:~\Lc_L^\Omega =( \Ec_L^\Omega \,, - \Nc_L^\Omega )^T$   \\
 &   &  & $( \rep{1} \,, \rep{1} \,,  0)_{ \mathbf{F} }^{\Omega^{\prime\prime} }$ &  $( \rep{1} \,, \rep{1} \,,  0)_{ \mathbf{F} }^{\Omega^{\prime\prime} } ~:~ \check \Nc_L^{\Omega^{\prime\prime} }$  \\
\hline\hline
\end{tabular}
\caption{The ${\sG\uG}(8)$ fermion representation of $\repb{8_F}^\Omega$ under the $\gG_{531}\,,\gG_{431}\,, \gG_{421}\,, \gG_{\rm SM}$ subalgebras.}
\label{tab:SU8_8ferm_SWS}
\end{center} 
}
\end{table}%

\begin{table}[htp] {\small
\begin{center}
\begin{tabular}{c|c|c|c|c}
\hline \hline
   ${\sG\uG}(8)$   &  $\gG_{531}$  & $\gG_{431}$  & $\gG_{421}$  &  $\gG_{\rm SM}$  \\
\hline \hline 
 $\rep{28_F}$   
 & $( \rep{1}\,, \repb{ 3} \,, + \frac{2}{3})_{ \mathbf{F}}$ 
 & $( \rep{1}\,, \repb{ 3} \,, + \frac{2}{3})_{ \mathbf{F}}$
 & $( \rep{1}\,, \repb{ 2} \,, + \frac{1}{2})_{ \mathbf{F}}$
 & $( \rep{1}\,, \repb{ 2} \,, + \frac{1}{2})_{ \mathbf{F}} ~:~( { \nG_R }^c\,, - {\eG_R }^c  )^T$  \\
&   &  & $( \rep{1}\,, \rep{ 1} \,, +1)_{ \mathbf{F}}$  & $( \rep{1}\,, \rep{ 1} \,, +1)_{ \mathbf{F}}~:~ {\tau_R }^c $   \\[1mm]
& $( \rep{5}\,, \rep{3} \,,  +\frac{2}{15})_{ \mathbf{F}}$ 
& $( \rep{1}\,, \rep{3} \,,  +\frac{1}{3})_{ \mathbf{F}}^{\prime} $   
& $( \rep{1}\,, \rep{2} \,,  +\frac{1}{2})_{ \mathbf{F}}^{\prime} $  
& $( \rep{1}\,, \rep{2} \,,  +\frac{1}{2})_{ \mathbf{F}}^{\prime}~:~ ( {\eG_R^{\prime} }^c \,, { \nG_R^{\prime} }^c)^T$  \\
&   &   &  $( \rep{1}\,, \rep{1} \,, 0 )_{ \mathbf{F}}$  & $( \rep{1}\,, \rep{1} \,, 0 )_{ \mathbf{F}} ~:~ \check \nG_R^c $ \\
&   &  $( \rep{4}\,, \rep{3} \,,  + \frac{1}{12} )_{ \mathbf{F}}$ & $( \rep{4}\,, \rep{1} \,,  -\frac{1}{4})_{ \mathbf{F}}$  & $( \rep{3}\,, \rep{1} \,,  -\frac{1}{3})_{ \mathbf{F}} ~:~\DG_L$ \\
&   &   &   & $( \rep{1}\,, \rep{1}\,, 0)_{ \mathbf{F}}^{\prime} ~:~ \check \nG_R^{\prime\,c}$ \\
&   &  & $( \rep{4}\,, \rep{2} \,,  +\frac{1}{4} )_{ \mathbf{F}}$  
& $( \rep{1}\,, \rep{2} \,,  +\frac{1}{2} )_{ \mathbf{F}}^{\prime\prime} ~:~ ( {\eG_R^{\prime\prime} }^c \,, { \nG_R^{\prime\prime}}^c )^T$  \\
&   &   &   & $( \rep{3}\,, \rep{2}\,, + \frac{1}{6})_{ \mathbf{F}} ~:~(t_L\,, b_L)^T$ \\[1mm]
& $( \rep{10}\,, \rep{ 1} \,, -\frac{2}{5})_{ \mathbf{F}}$ 
&$( \rep{4}\,, \rep{ 1} \,, -\frac{1}{4})_{ \mathbf{F}}^{\prime} $
&  $( \rep{4}\,, \rep{ 1} \,, -\frac{1}{4})_{ \mathbf{F}}^{\prime} $
& $( \rep{3}\,, \rep{1} \,, -\frac{1}{3})_{ \mathbf{F}}^{\prime} ~:~\DG_L^\prime$  \\
&   &   &   & $( \rep{1}\,, \rep{1} \,,  0)_{ \mathbf{F}}^{\prime\prime} ~:~\check \nG_R^{\prime \prime \,c}$  \\
&   &  $( \rep{6}\,, \rep{ 1} \,, -\frac{1}{2})_{ \mathbf{F}}$
& $( \rep{6}\,, \rep{ 1} \,, -\frac{1}{2})_{ \mathbf{F}}$  
& $( \rep{3}\,, \rep{1} \,, -\frac{1}{3})_{ \mathbf{F}}^{\prime\prime} ~:~\DG_L^{\prime \prime}$   \\
&   &   &   & $ ( \repb{3}\,, \rep{1} \,, -\frac{2}{3} )_{ \mathbf{F}} ~:~ {t_R}^c$ \\[1mm]  
\hline\hline
\end{tabular}
\caption{
The ${\sG\uG}(8)$ fermion representation of $\rep{28_F}$ under the $\gG_{531}\,,\gG_{431}\,, \gG_{421}\,, \gG_{\rm SM}$ subalgebras.
}
\label{tab:SU8_28ferm_SWS}
\end{center}
}
\end{table}

\begin{table}[htp] {\small
\begin{center}
\begin{tabular}{c|c|c|c|c}
\hline \hline
   ${\sG\uG}(8)$   &  $\gG_{531}$  & $\gG_{431}$  & $\gG_{421}$  &  $\gG_{\rm SM}$  \\
\hline \hline
     $\rep{56_F}$   
     & $( \rep{ 1}\,, \rep{1} \,, +1)_{ \mathbf{F}}^{\prime} $  
     & $( \rep{ 1}\,, \rep{1} \,, +1)_{ \mathbf{F}}^{\prime} $ 
     & $( \rep{ 1}\,, \rep{1} \,, +1)_{ \mathbf{F}}^{\prime} $   
     & $ ( \rep{ 1}\,, \rep{1} \,, +1)_{ \mathbf{F}}^{\prime} ~:~ {\mu_R}^c $  \\[1mm]
                       & $( \rep{ 5}\,, \repb{3} \,, +\frac{7}{15})_{ \mathbf{F}} $ 
                       & $( \rep{ 1}\,, \repb{3} \,, +\frac{2}{3})_{ \mathbf{F}}^{\prime}  $  
                       & $( \rep{ 1}\,, \repb{2} \,, +\frac{1}{2})_{ \mathbf{F}}^{\prime \prime \prime}  $  
                       & $( \rep{ 1}\,, \repb{2} \,, +\frac{1}{2})_{ \mathbf{F}}^{\prime\prime\prime} ~:~ ( {\nG_R^{\prime\prime\prime }}^c \,, -{\eG_R^{\prime\prime\prime } }^c )^T $ \\
                       &   &   &  $( \rep{ 1}\,, \rep{1} \,, +1)_{ \mathbf{F}}^{\prime \prime}$ & $( \rep{ 1}\,, \rep{1} \,, +1)_{ \mathbf{F}}^{\prime \prime} ~:~{e_R}^c$   \\
                       &   &  $( \rep{ 4}\,, \repb{3} \,, +\frac{5}{12})_{ \mathbf{F}}$ 
                       & $( \rep{4}\,, \rep{1} \,, +\frac{3}{4})_{ \mathbf{F}} $  
                       & $( \rep{3}\,, \rep{1} \,, +\frac{2}{3})_{ \mathbf{F}} ~:~ \UG_L $ \\
                       &   &  &  &  $ ( \rep{1}\,, \rep{1} \,, +1)_{ \mathbf{F}}^{\prime\prime\prime} ~:~ {\EG_R}^c $  \\
                       &   &   & $( \rep{4}\,, \repb{2} \,, +\frac{1}{4})_{ \mathbf{F}} $  
                       & $( \rep{3}\,, \repb{2} \,, +\frac{1}{6})_{ \mathbf{F}} ~:~ (\dG_L \,, - \uG_L )^T$ \\
                       &   &  &  &  $( \rep{1}\,, \repb{2} \,, +\frac{1}{2})_{ \mathbf{F}}^{\prime\prime\prime\prime} ~:~ ( {\nG_R^{\prime\prime\prime\prime }}^c \,, -{\eG_R^{\prime\prime\prime\prime } }^c )^T $  \\[1mm]
                       & $( \repb{10}\,, \rep{1} \,, -\frac{3}{5})_{ \mathbf{F}}$  
                       & $( \repb{4}\,, \rep{1} \,, -\frac{3}{4})_{ \mathbf{F}} $ 
                       & $( \repb{4}\,, \rep{1} \,, -\frac{3}{4})_{ \mathbf{F}} $ 
                       & $( \rep{1}\,, \rep{1} \,, -1)_{ \mathbf{F}}^{  } ~:~ \EG_L $ \\
                       &   &   &   &  $ ( \repb{3}\,, \rep{1} \,, -\frac{2}{3})_{ \mathbf{F}}^{\prime} ~:~ {\UG_R}^c$ \\
                       &   &  $( \rep{6}\,, \rep{1} \,, -\frac{1}{2})_{ \mathbf{F}}^{\prime} $ & $( \rep{6}\,, \rep{1} \,, -\frac{1}{2})_{ \mathbf{F}}^{\prime} $ & $( \rep{3}\,, \rep{1} \,, -\frac{1}{3})_{ \mathbf{F}}^{\prime\prime\prime} ~:~ \DG_L^{\prime \prime\prime}$  \\
                       &   &   &   & $( \repb{3}\,, \rep{1} \,, -\frac{2}{3})_{ \mathbf{F}}^{\prime \prime} ~:~ {u_R}^c$  \\[1mm]
                       & $( \rep{10}\,, \rep{3} \,, -\frac{1}{15})_{ \mathbf{F}}$  
                       &  $( \rep{4}\,, \rep{3} \,, +\frac{1}{12})_{ \mathbf{F}}^{\prime} $ 
                       & $( \rep{4}\,, \rep{1} \,, -\frac{1}{4})_{ \mathbf{F}}^{\prime\prime}  $ 
                       & $( \rep{3}\,, \rep{1} \,, -\frac{1}{3})_{ \mathbf{F}}^{\prime\prime\prime \prime } ~:~\DG_L^{\prime\prime \prime \prime}$ \\
                       &   &   &   & $( \rep{1}\,, \rep{1} \,, 0 )_{ \mathbf{F}}^{\prime\prime \prime} ~:~ {\check \nG_R}^{\prime \prime\prime \,c}$ \\
                       &   &  & $( \rep{4}\,, \rep{2} \,, +\frac{1}{4} )_{ \mathbf{F}}^{\prime}  $ & $ ( \rep{3}\,, \rep{2} \,, +\frac{1}{6} )_{ \mathbf{F}}^{\prime} ~:~ (u_L\,,d_L)^T $  \\
                       &   &   &   & $( \rep{1}\,, \rep{2} \,, +\frac{1}{2})_{ \mathbf{F}}^{\prime\prime\prime \prime \prime} ~:~ ( {\eG_R^{\prime\prime\prime\prime\prime }}^c \,, {\nG_R^{\prime\prime\prime\prime\prime } }^c )^T$ \\
                        &   &  $( \rep{6}\,, \rep{3} \,, -\frac{1}{6})_{ \mathbf{F}}$  & $( \rep{6}\,, \rep{1} \,, -\frac{1}{2})_{ \mathbf{F}}^{\prime\prime} $ & $( \rep{3}\,, \rep{1} \,, -\frac{1}{3})_{ \mathbf{F}}^{\prime\prime\prime\prime\prime} ~:~ \DG_L^{\prime\prime\prime\prime\prime} $   \\
                       &   &   &   & $( \repb{3}\,, \rep{1} \,, -\frac{2}{3})_{ \mathbf{F}}^{\prime\prime\prime} ~:~ {c_R}^c $  \\
                       &   &   & $( \rep{6}\,, \rep{2} \,, 0)_{ \mathbf{F}} $  & $( \rep{3}\,, \rep{2} \,, +\frac{1}{6})_{ \mathbf{F}}^{\prime\prime} ~:~ (c_L\,,s_L)^T  $  \\
                       &   &   &   & $ ( \repb{3}\,, \rep{2} \,, -\frac{1}{6} )_{ \mathbf{F}}  ~:~ ( {\dG_R}^c \,,{\uG_R}^c )^T $ \\[1mm]
\hline\hline
\end{tabular}
\caption{
The ${\sG\uG}(8)$ fermion representation of $\rep{56_F}$ under the $\gG_{531}\,,\gG_{431}\,, \gG_{421}\,, \gG_{\rm SM}$ subalgebras. 
}
\label{tab:SU8_56ferm_SWS}
\end{center}
}
\end{table}%

\para
By following the SWS symmetry breaking pattern in Eq.~\eqref{eq:SUSYSU8_531_SWSPattern}, we tabulate the fermion representations at various stages in Tables~\ref{tab:SU8_8ferm_SWS}, \ref{tab:SU8_28ferm_SWS}, and \ref{tab:SU8_56ferm_SWS}. 
Along this pattern, the global $\widetilde{ {\rm U}}(1)_T$ symmetry evolves to the global $\widetilde{ {\rm U}}(1)_{B-L}$ at the EW scale according to the following sequence:
\beqn\label{eq:U1T_SWS_def}
&& \gG_{531}~:~ \Tc^\prime \equiv \Tc - 5t \Xc_0 \,, \quad  \gG_{431}~:~ \Tc^{ \prime \prime} \equiv \Tc^\prime + 8 t \Xc_1 \,, \non
&& \gG_{421}~:~   \Tc^{ \prime \prime \prime} \equiv  \Tc^{ \prime \prime} - 8 t \Xc_2 \,, \quad  \gG_{\rm SM}~:~  \Bc- \Lc \equiv  \Tc^{ \prime \prime \prime} + 8 t \Yc \,.
\eeqn

\subsection{Decompositions of the ${\sG \uG}$(8) Higgs fields}

\para
We decompose the Higgs fields in Eq.~\eqref{eq:SUSYSU8_Yukawa} into components that can be responsible for the sequential symmetry breaking stages in Eq.~\eqref{eq:SUSYSU8_531_SWSPattern}. 
For Higgs fields of $\repb{ 8_H}_{\,, \omega}$, they read
\begin{eqnarray}\label{eq:SU8_SWS_Higgs_Br01}
\repb{8_H}_{\,,\omega }  &\supset&  \underline{  ( \rep{1} \,, \repb{3} \,, -\frac{1}{3} )_{\mathbf{H}\,, \omega }  } \oplus  \langle ( \repb{5} \,, \rep{1} \,, +\frac{1}{5} )_{\mathbf{H}\,, \omega } \rangle \non
&\supset&  \langle ( \rep{1} \,, \repb{3} \,, -\frac{1}{3} )_{\mathbf{H}\,, \omega } \rangle  \oplus  \underline{ ( \repb{4} \,, \rep{1} \,, +\frac{1}{4} )_{\mathbf{H}\,, \omega }  }\non
&\supset&  \langle ( \repb{4} \,, \rep{1} \,, +\frac{1}{4} )_{\mathbf{H}\,, \omega } \rangle \,.
\end{eqnarray}
For Higgs fields of $\repb{28_H}_{\,,\dot \omega } $, they read
\begin{eqnarray}\label{eq:SU8_SWS_Higgs_Br02}
\repb{28_H}_{\,,\dot \omega } &\supset& \underline{( \rep{1} \,, \rep{3} \,,  -\frac{2}{3} )_{\mathbf{H}\,, \dot\omega } } \oplus  \underline{ ( \repb{5} \,, \repb{3} \,, -\frac{2}{15} )_{\mathbf{H}\,, \dot\omega } } \oplus  \underline{ ( \repb{10} \,, \rep{1} \,, +\frac{2}{5} )_{\mathbf{H}\,, \dot\omega } }  \non
&\supset &  \underline{  ( \rep{1} \,, \rep{3} \,, -\frac{2}{3} )_{\mathbf{H}\,, \dot\omega }  } \oplus \Big[  \langle ( \rep{1} \,, \repb{3} \,, -\frac{1}{3} )_{\mathbf{H}\,, \dot\omega }  \rangle  \oplus \underline{  ( \repb{4} \,, \repb{3} \,, -\frac{1}{12} )_{\mathbf{H}\,, \dot\omega }  } \Big] \oplus \underline{ ( \repb{4} \,, \rep{1} \,, +\frac{1}{4} )_{\mathbf{H}\,, \dot\omega }  } \non
&\supset&  \underline{ ( \rep{1} \,, \rep{2} \,, -\frac{1}{2} )_{\mathbf{H}\,, \dot\omega }  }  \oplus \Big[ \underline{ ( \rep{1} \,, \repb{2} \,, -\frac{1}{2} )_{\mathbf{H}\,, \dot\omega } } \oplus \langle ( \repb{4} \,, \rep{1} \,, +\frac{1}{4} )_{\mathbf{H}\,, \dot{\omega} }  \rangle \oplus   \underline {( \repb{4} \,, \repb{2} \,, -\frac{1}{4} )_{\mathbf{H}\,, \dot\omega }^{} }  \Big]  \non
&\oplus& \langle ( \repb{4} \,, \rep{1} \,, +\frac{1}{4} )_{\mathbf{H}\,, \dot\omega }^{\prime} \rangle \,.
%
\end{eqnarray}
For Higgs field of $\rep{70_H}$, it reads
\begin{eqnarray}\label{eq:SU8_SWS_Higgs_Br05} 
\rep{70_H} &\supset& ( \rep{5} \,, \rep{1 } \,, +\frac{4}{5} )_{\mathbf{H}}^{  }  \oplus  ( \repb{5} \,, \rep{1} \,, -\frac{4}{5} )_{\mathbf{H}}  \oplus \underline{ ( \rep{10} \,, \repb{3} \,, +\frac{4}{15} )_{\mathbf{H}} }   \oplus ( \repb{10} \,, \rep{3 } \,, -\frac{4}{15} )_{\mathbf{H}}^{ } \non
&\supset& \underline{ ( \rep{4} \,, \repb{3} \,, +\frac{5}{12} )_{\mathbf{H}} } \supset  \underline{ ( \rep{4} \,, \repb{2} \,, +\frac{1}{4} )_{\mathbf{H}}^{  } }  \supset  \langle ( \rep{1} \,, \repb{2} \,, +\frac{1}{2} )_{\mathbf{H}}^{  } \rangle \,. 
\end{eqnarray}
%
%

\subsection{RGEs of the SWS symmetry breaking pattern}


\para
Between the $ v_{531} \leq \mu \leq v_{U} $, almost all massless Higgs fields are in the first lines of Eq.~\eqref{eq:SU8_SWS_Higgs_Br01}, \eqref{eq:SU8_SWS_Higgs_Br02}, and Eq.~\eqref{eq:SU8_SWS_Higgs_Br05}, which are
\beqn
&& ( \repb{5}\,, \rep{1}\,, +\frac{1}{5})_{ \mathbf{H}\,, \omega} \oplus  ( \rep{1}\,, \repb{3}\,, -\frac{1}{3})_{ \mathbf{H}\,, \omega} \subset \repb{8_H}_{\,, \omega } \,, \non
&& (\rep{1}\,,\rep{3}\,, -\frac{2}{3})_{\mathbf{H}\,,\dot\omega} \oplus (\repb{5}\,,\repb{3}\,, -\frac{2}{15})_{\mathbf{H}\,,\dot\omega} \oplus ( \repb{10}\,, \rep{1}\,, +\frac{2}{5})_{\mathbf{H}\,,\dot \omega}  \subset \repb{28_H}_{\,,\dot\omega} \,, \non
&& (\rep{5}\,,\rep{1}\,, +\frac{4}{5})_\mathbf{H} \oplus  (\repb{5}\,,\rep{1}\,, -\frac{4}{5})_\mathbf{H} \oplus  (\rep{10}\,,\repb{3}\,, + \frac{4}{15})_\mathbf{H} \oplus (\repb{10}\,,\rep{3}\,, - \frac{4}{15})_\mathbf{H}  \subset \rep{70_H}  \,, \non
&& (\repb{5}\,,\rep{1}\,, -\frac{4}{5})_\mathbf{H} \oplus (\rep{5}\,,\rep{1}\,, +\frac{4}{5})_\mathbf{H} \oplus (\repb{10}\,,\rep{3}\,, - \frac{4}{15})_\mathbf{H} \oplus (\rep{10}\,,\repb{3}\,, + \frac{4}{15})_\mathbf{H}  \subset \repb{70_H} \,,\non
&& ( \rep{5}\,, \rep{1}\,, -\frac{1}{5})_{ \mathbf{H} }^{\omega} \oplus  ( \rep{1}\,, \rep{3}\,, +\frac{1}{3})_{ \mathbf{H} }^{\omega} \subset \rep{8_H}^{ \omega }  \,,\non
&& (\rep{1}\,,\repb{3}\,, +\frac{2}{3})_{\mathbf{H}}^{\dot\omega} \oplus (\rep{5}\,,\rep{3}\,, +\frac{2}{15})_{\mathbf{H}}^{\dot\omega} \oplus ( \rep{10}\,, \rep{1}\,, -\frac{2}{5})_{\mathbf{H}}^{\dot\omega}   \subset \rep{28_H}^{\dot\omega} \,.
\eeqn
All fermionic components of $\repb{8_F}^\Omega \oplus \rep{28_F} \oplus \rep{56_F}$ remain massless after the decomposition into the $\gG_{531} $ IRs. 
Consequently, we have the $ \gG_{531}$ $\beta$ coefficients of
\beqn\label{eq:HiggsB_531}
&& (b^{(1)}_{{\sG \uG}(5)_{s}}\,,b^{(1)}_{{\sG \uG}(3)_{W}}\,,b^{(1)}_{{\uG}(1)_{X_0}})  = (+54 \,, +60 \,, +\frac{368}{5}) \,, \non
&& b^{(2)}_{\gG_{531}} = \begin{pmatrix}
6984/5 & 216 & 672/25\\
648 & 728 & 352/15\\
16128/25 & 2816/15 & 98816/1125
\end{pmatrix}\,.
\eeqn
%
%


\para
Between the $ v_{431} \leq \mu \leq v_{531} $, the massless Higgs fields are the following:
\beqn
&& ( \repb{4}\,, \rep{1}\,, +\frac{1}{4})_{ \mathbf{H}\,, 3 , \rm V ,\rm VI} \oplus ( \rep{1}\,, \repb{3}\,, -\frac{1}{3})_{\mathbf{H}\,, 3, \rm V,\rm VI } \subset ... \subset \repb{8_H}_{\,, 3 , \rm V ,\rm VI }\,, \non
&& (\rep{1}\,,\rep{3}\,, -\frac{2}{3})_{\mathbf{H}\,,\dot{\omega}} \oplus  \Big[ (\repb{4}\,,\rep{1}\,, +\frac{1}{4})_{\mathbf{H}\,,\dot{\omega}} \oplus (\rep{6}\,,\rep{1}\,, +\frac{1}{2})_{\mathbf{H}\,,\dot{\omega}} \Big] \non
& \oplus &  \Big[ (\rep{1}\,,\repb{3}\,, -\frac{1}{3})_{\mathbf{H}\,,\dot{\omega}} \oplus (\repb{4}\,,\repb{3}\,, -\frac{1}{12})_{\mathbf{H}\,,\dot{\omega}} \Big]  \subset ... \subset \repb{28_H}_{\,, \dot{\omega}} \,, \quad  \dot{\omega}=( \dot{1},\dot{2},\dot{\rm VII},\dot{\rm VIII},\dot{\rm IX} ) \,, \non
&&  (\rep{1}\,,\repb{3}\,, +\frac{2}{3})_{\mathbf{H}}^{\dot{\rm \kappa}} \oplus \Big[ (\rep{1}\,,\rep{3}\,, +\frac{1}{3})_{\mathbf{H}}^{\dot{\rm \kappa}} \oplus (\rep{4}\,,\rep{3}\,, +\frac{1}{12})_{\mathbf{H}}^{\dot{\rm \kappa}} \Big]   \non
&\oplus&  \Big[ (\rep{4}\,,\rep{1}\,, -\frac{1}{4})_{\mathbf{H}}^{\dot{\rm \kappa}} \oplus (\rep{6}\,,\rep{1}\,, -\frac{1}{2})_{\mathbf{H}}^{\dot{\rm \kappa}} \Big] \subset ... \subset \rep{28_H}^{\dot{\rm \kappa}} \,, \quad  \dot{\kappa}=( \dot{\rm VII},\dot{\rm IX} ) \,,  \non
&& (\rep{4}\,,\repb{3}\,, +\frac{5}{12})_\mathbf{H} \subset ... \subset \rep{70_H} \,.
\eeqn
All ${\sG \uG}$(8) fermions that remain massless after the decomposition into the $\gG_{431}$ IRs can be found in Ref.~\cite{Wang:2023gut}. 
Consequently, we have the $\gG_{431}$ $\beta$ coefficients of
\beqn\label{eq:HiggsB_431}
&& (b^{(1)}_{{\sG \uG}(4)_{s}}\,,b^{(1)}_{{\sG \uG}(3)_{W}}\,,b^{(1)}_{{\uG}(1)_{X_1}})  = (+ \frac{8}{3}\,,+\frac{13}{2}\,,+\frac{61}{3}) \,, \non
&& b^{(2)}_{\gG_{431}} = \begin{pmatrix}
3907/12 & 84 & 99/8 \\
315/2 & 255 & 155/12 \\
1485/8 & 310/3 & 5855/144
\end{pmatrix} \,.
\eeqn
The ${\sG \uG}(5)_s \oplus {\uG}(1)_{X_0}$ gauge couplings match with the ${\sG \uG}(4 )_s \oplus {\uG}(1)_{X_1}$ gauge couplings as follows:
\beqn\label{eq:531_BcoupMatch}
&& \alpha_{5s }^{-1} (v_{531} ) =  \alpha_{4s }^{-1} (v_{531} )  \,,~ \alpha_{X_1}^{-1} (v_{ 531} )  = \frac{1}{10} \alpha_{ 5s }^{-1} (v_{531} )  +  \alpha_{X_0}^{-1} (v_{531} ) \,.
\eeqn
%
%


\para
Between the $ v_{421} \leq \mu \leq v_{431} $, the massless Higgs fields are
\beqn
&& ( \repb{4}\,, \rep{1}\,, +\frac{1}{4})_{ \mathbf{H}\,, 3 ,\rm VI} \oplus ( \rep{1}\,, \repb{2}\,, -\frac{1}{2})_{ \mathbf{H}\,, 3 ,\rm VI}   \subset ... \subset \repb{8_H}_{, 3 ,\rm VI} \,, \non
&& \Big[ (\rep{1}\,,\rep{2}\,, -\frac{1}{2})_{\mathbf{H}\,,\dot{\omega}} \oplus (\rep{1}\,,\rep{1}\,, -1)_{\mathbf{H}\,,\dot{\omega}}  \Big] \oplus \Big[ (\repb{4}\,,\rep{1}\,, +\frac{1}{4})_{\mathbf{H}\,,\dot{\omega}}^{\prime} \oplus (\rep{6}\,,\rep{1}\,, +\frac{1}{2})_{\mathbf{H}\,,\dot{\omega}}^{}  \Big]   \non
& \oplus &  \Big[ (\rep{1}\,,\repb{2}\,, -\frac{1}{2})_{\mathbf{H}\,,\dot{\omega}} \oplus (\repb{4}\,,\rep{1}\,, +\frac{1}{4})_{\mathbf{H} \,,\dot{\omega}}^{} \oplus (\repb{4}\,,\repb{2}\,, -\frac{1}{4})_{\mathbf{H}\,,\dot{\omega}} \Big] \subset ... \subset \repb{28_H}_{\,, \dot{\omega}}  \,, \quad \dot{\omega}=( \dot{2},\dot{\rm VIII},\dot{\rm IX} ) \,,\non
&&  \Big[ (\rep{1}\,,\repb{2}\,, +\frac{1}{2})_{\mathbf{H}}^{ \dot{\rm IX}} \oplus (\rep{1}\,,\rep{1}\,, +1)_{\mathbf{H}}^{ \dot{\rm IX}}  \Big] \oplus \Big[ (\rep{4}\,,\rep{1}\,, -\frac{1}{4})_{\mathbf{H}}^{ \dot{\rm IX}^{\prime}} \oplus (\rep{6}\,,\rep{1}\,, -\frac{1}{2})_{\mathbf{H}}^{ \dot{\rm IX}}  \Big]   \non
& \oplus &  \Big[ (\rep{1}\,,\rep{2}\,, +\frac{1}{2})_{\mathbf{H}}^{ \dot{\rm IX}} \oplus (\rep{4}\,,\rep{1}\,, -\frac{1}{4})_{\mathbf{H}}^{ \dot{\rm IX}} \oplus (\rep{4}\,,\rep{2}\,, +\frac{1}{4})_{\mathbf{H}}^{ \dot{\rm IX}} \Big] \subset ... \subset \rep{28_H}^{ \dot{\rm IX}} \,,   \non
&& (\rep{4}\,,\repb{2}\,, +\frac{1}{4})_{\mathbf{H}} \subset ... \subset \rep{70_H} \,.
\eeqn
All ${\sG \uG}$(8) fermions that remain massless after the decomposition into the $\gG_{421}$ IRs can be found in Ref.~\cite{Wang:2023gut}.  
Consequently, we have the $ \gG_{421} $ $ \beta $ coefficients of
\beqn\label{eq:HiggsB_421}
&& (b^{(1)}_{{\sG \uG}(4)_{s}}\,,b^{(1)}_{{\sG \uG}(2)_{W}}\,,b^{(1)}_{{\uG}(1)_{X_2}})  = (-\frac{8}{3}\,,+\frac{17}{3}\,,+\frac{47}{3}) \,, \non
&& b^{(2)}_{\gG_{421}}  = \begin{pmatrix}
991/6 & 45/2 & 35/4\\
225/2 & 353/3 & 39/4\\
525/4 & 117/4 & 307/8
\end{pmatrix} \,.
\eeqn
The ${\sG \uG}(3)_W \oplus {\uG}(1)_{X_1}$ gauge couplings match with the ${\sG \uG}(2 )_W \oplus {\uG}(1)_{X_2}$ gauge couplings as follows:
\beqn\label{eq:431_BcoupMatch}
&& \alpha_{3W }^{-1} (v_{431} ) =  \alpha_{2W }^{-1} (v_{431} )  \,,~ \alpha_{X_2}^{-1} (v_{431} )  = \frac{1}{3} \alpha_{3W }^{-1} (v_{431} )  +  \alpha_{X_1}^{-1} (v_{431} )  \,.
\eeqn
%
%


\para
Between the $ v_{\rm EW} \leq \mu \leq v_{421} $, we have the same $\gG_{\rm SM} $ $ \beta $ coefficients as Eq.~\eqref{eq:HiggsA_SMto331}.
The ${\sG \uG}(4)_s \oplus {\uG}(1)_{X_2}$ gauge couplings match with the ${\sG \uG}(3 )_c \oplus {\uG}(1)_{Y}$ gauge couplings as follows:
\beqn\label{eq:421_BcoupMatch}
&& \alpha_{4s }^{-1} (v_{421} ) =  \alpha_{3c }^{-1} (v_{421} )  \,,~ \alpha_{\rm Y}^{-1} (v_{ 421} ) = \frac{1}{6} \alpha_{4s }^{-1} (v_{421} )  +  \alpha_{X_2}^{-1} (v_{421} )  \,.
\eeqn
Based on constructing the SM quark/lepton mass matrices to reproduce the observed hierarchical masses and the CKM mixing pattern, we find the following benchmark point:
\begin{align}\label{eq:benchmark SWS}
	v_{531}  \simeq 1.4 \times 10^{17} \, \mathrm{GeV} \,, \quad
	v_{431}  \simeq 4.8 \times 10^{15}\, \mathrm{GeV} \,, \quad
	v_{421}  \simeq 1.1 \times 10^{15} \, \mathrm{GeV} \,.
\end{align}
%
%

\begin{figure}[htb]
\centering
\includegraphics[height=4.5cm]{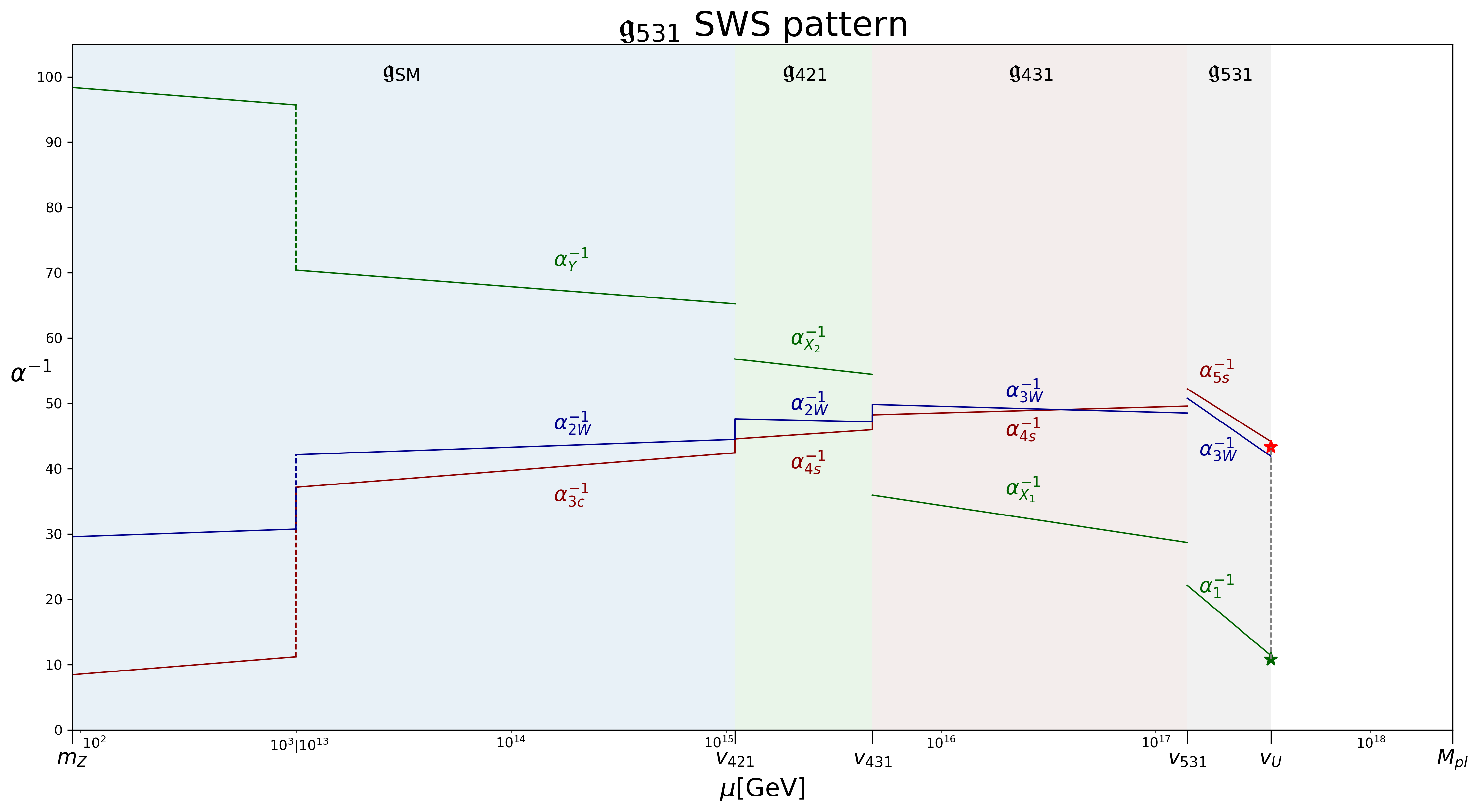}
\caption{The two-loop RGEs of the ${\sG \uG}(8)$ setup according to the SWS symmetry breaking pattern.
The RG evolutions within $10^{3}\,{\rm GeV} \lesssim \mu \lesssim 10^{13}\,{\rm GeV}$ are hidden in order to highlight the behaviors in three intermediate symmetry breaking scales given by the benchmark point in Eq.~\eqref{eq:benchmark SWS}.
}
\label{fig:SU8_RGE_SWS} 
\end{figure}

\para
With the one- and two-loop $\beta$ coefficients defined in Eqs.~\eqref{eq:HiggsB_531}, \eqref{eq:HiggsB_431}, \eqref{eq:HiggsB_421}, and \eqref{eq:HiggsA_SMto331}, we plot the RGEs of the ${\sG \uG}(8)$ setup along the SWS symmetry breaking pattern in Fig.~\ref{fig:SU8_RGE_SWS}.
The discontinuities of the ${\uG}(1)$ gauge couplings at three intermediate scales follow from their definitions in Eqs.~\eqref{eq:531_BcoupMatch}, \eqref{eq:431_BcoupMatch}, \eqref{eq:421_BcoupMatch}, respectively.
The benchmark point in Fig.~\ref{fig:SU8_RGE_SWS} is marked by $ \star $ and reads
\begin{gather}
c_{\rm HSW} \approx -0.31  \,,~v_{U} \approx 2.59 \times 10^{17}\, {\rm GeV} \,, \non
\alpha_{5c }^{-1} (v_{U} ) \approx 46.70 \,,~\alpha_{3W }^{-1} (v_{U} ) \approx 44.70\,,~\alpha_{1 }^{-1} (v_{U} ) \approx 10.82 \,.
\end{gather}

\section{THE WSS SYMMETRY BREAKING PATTERN}
\label{section:WSS_pattern}

\subsection{Decompositions of the ${\sG \uG}$(8) fermions}

\begin{table}[htp] {\small
\begin{center}
\begin{tabular}{c|c|c|c|c}
\hline \hline
   ${\sG\uG}(8)$   &  $\gG_{531}$  & $\gG_{521}$  & $\gG_{421}$  &  $\gG_{\rm SM}$  \\
\hline \hline
 $\repb{ 8_F}^\Omega$   
 & $( \repb{5} \,, \rep{1}\,,  +\frac{1}{5} )_{ \mathbf{F} }^\Omega$  
 & $(\repb{5} \,, \rep{1} \,, +\frac{1}{5} )_{ \mathbf{F} }^\Omega$  
 & $(\repb{4} \,, \rep{1} \,, +\frac{1}{4} )_{ \mathbf{F} }^\Omega$  
 &  $( \repb{3} \,, \rep{ 1}  \,, +\frac{1}{3} )_{ \mathbf{F} }^{\Omega}~:~ { \Dc_R^\Omega}^c$  \\
 &  &  &  &  $( \rep{1} \,, \rep{1} \,, 0)_{ \mathbf{F} }^{\Omega} ~:~ \check \Nc_L^{\Omega }$  \\
 &  &  &  $( \rep{1} \,, \rep{1} \,, 0)_{ \mathbf{F} }^{\Omega^\prime}$ 
 &  $( \rep{1} \,, \rep{1} \,, 0)_{ \mathbf{F} }^{\Omega^\prime} ~:~ \check \Nc_L^{\Omega^\prime }$  \\[1mm]
 & $(\rep{1}\,, \repb{3}  \,,  -\frac{1}{3})_{ \mathbf{F} }^\Omega$  
 &  $(\rep{1}\,, \repb{2}  \,,  -\frac{1}{2})_{ \mathbf{F} }^\Omega$  
 &  $( \rep{1} \,, \repb{2} \,,  -\frac{1}{2})_{ \mathbf{F} }^{\Omega}$  
 &  $( \rep{1} \,, \repb{2} \,,  -\frac{1}{2})_{ \mathbf{F} }^{\Omega } ~:~\Lc_L^\Omega =( \Ec_L^\Omega \,, - \Nc_L^\Omega )^T$   \\
 &   & $( \rep{1} \,, \rep{1} \,,  0)_{ \mathbf{F} }^{\Omega^{\prime\prime} }$ & $( \rep{1} \,, \rep{1} \,,  0)_{ \mathbf{F} }^{\Omega^{\prime\prime} }$ &  $( \rep{1} \,, \rep{1} \,,  0)_{ \mathbf{F} }^{\Omega^{\prime\prime} } ~:~ \check \Nc_L^{\Omega^{\prime\prime} }$  \\
\hline\hline
\end{tabular}
\caption{The ${\sG\uG}(8)$ fermion representation of $\repb{8_F}^\Omega$ under the $\gG_{531}\,,\gG_{521}\,, \gG_{421}\,, \gG_{\rm SM}$ subalgebras.}
\label{tab:SU8_8ferm_WSS}
\end{center} 
}
\end{table}%

\begin{table}[htp] {\small
\begin{center}
\begin{tabular}{c|c|c|c|c}
\hline \hline
   ${\sG\uG}(8)$   &  $\gG_{531}$  & $\gG_{521}$  & $\gG_{421}$  &  $\gG_{\rm SM}$  \\
\hline \hline 
 $\rep{28_F}$   
 & $( \rep{1}\,, \repb{ 3} \,, + \frac{2}{3})_{ \mathbf{F}}$ 
 & $( \rep{1}\,, \repb{ 2} \,, + \frac{1}{2})_{ \mathbf{F}}$
 & $( \rep{1}\,, \repb{ 2} \,, + \frac{1}{2})_{ \mathbf{F}}$
 & $( \rep{1}\,, \repb{ 2} \,, + \frac{1}{2})_{ \mathbf{F}} ~:~( { \nG_R }^c\,, - {\eG_R }^c  )^T$  \\
&   & $( \rep{1}\,, \rep{ 1} \,, +1)_{ \mathbf{F}}$  & $( \rep{1}\,, \rep{ 1} \,, +1)_{ \mathbf{F}}$  & $( \rep{1}\,, \rep{ 1} \,, +1)_{ \mathbf{F}}~:~ {\tau_R }^c $   \\[1mm]
& $( \rep{5}\,, \rep{3} \,,  +\frac{2}{15})_{ \mathbf{F}}$ 
& $( \rep{5}\,, \rep{1} \,,  -\frac{1}{5})_{ \mathbf{F}}$   
& $( \rep{4}\,, \rep{1} \,,  -\frac{1}{4})_{ \mathbf{F}}$  
& $( \rep{3}\,, \rep{1} \,,  -\frac{1}{3})_{ \mathbf{F}} ~:~\DG_L$  \\
&   &   &   & $( \rep{1}\,, \rep{1} \,, 0 )_{ \mathbf{F}} ~:~ \check \nG_R^c $ \\
&   &   & $( \rep{1}\,, \rep{1} \,, 0 )_{ \mathbf{F}}^{\prime} $  & $( \rep{1}\,, \rep{1}\,, 0)_{ \mathbf{F}}^{\prime} ~:~ \check \nG_R^{\prime\,c}$ \\
&   &  $( \rep{5}\,, \rep{2} \,,  + \frac{3}{10} )_{ \mathbf{F}}$ &  $( \rep{1}\,, \rep{2} \,,  +\frac{1}{2})_{ \mathbf{F}}^{\prime} $ & $( \rep{1}\,, \rep{2} \,,  +\frac{1}{2})_{ \mathbf{F}}^{\prime}~:~ ( {\eG_R^{\prime} }^c \,, { \nG_R^{\prime} }^c)^T$ \\
&   &  & $( \rep{4}\,, \rep{2} \,,  +\frac{1}{4} )_{ \mathbf{F}}$  
& $( \rep{1}\,, \rep{2} \,,  +\frac{1}{2} )_{ \mathbf{F}}^{\prime\prime} ~:~ ( {\eG_R^{\prime\prime} }^c \,, { \nG_R^{\prime\prime}}^c )^T$  \\
&   &   &   & $( \rep{3}\,, \rep{2}\,, + \frac{1}{6})_{ \mathbf{F}} ~:~(t_L\,, b_L)^T$ \\[1mm]
& $( \rep{10}\,, \rep{ 1} \,, -\frac{2}{5})_{ \mathbf{F}}$ 
& $( \rep{10}\,, \rep{ 1} \,, -\frac{2}{5})_{ \mathbf{F}}$
& $( \rep{4}\,, \rep{ 1} \,, -\frac{1}{4})_{ \mathbf{F}}^{\prime} $
& $( \rep{3}\,, \rep{1} \,, -\frac{1}{3})_{ \mathbf{F}}^{\prime} ~:~\DG_L^\prime$  \\
&   &   &   & $( \rep{1}\,, \rep{1} \,,  0)_{ \mathbf{F}}^{\prime\prime} ~:~\check \nG_R^{\prime \prime \,c}$  \\
&   &   & $( \rep{6}\,, \rep{ 1} \,, -\frac{1}{2})_{ \mathbf{F}}$  
& $( \rep{3}\,, \rep{1} \,, -\frac{1}{3})_{ \mathbf{F}}^{\prime\prime} ~:~\DG_L^{\prime \prime}$   \\
&   &   &   & $ ( \repb{3}\,, \rep{1} \,, -\frac{2}{3} )_{ \mathbf{F}} ~:~ {t_R}^c$ \\[1mm]  
\hline\hline
\end{tabular}
\caption{
The ${\sG\uG}(8)$ fermion representation of $\rep{28_F}$ under the $\gG_{531}\,,\gG_{521}\,, \gG_{421}\,, \gG_{\rm SM}$ subalgebras.
}
\label{tab:SU8_28ferm_WSS}
\end{center}
}
\end{table}

\begin{table}[htp] {\small
\begin{center}
\begin{tabular}{c|c|c|c|c}
\hline \hline
   ${\sG\uG}(8)$   &  $\gG_{531}$  & $\gG_{521}$  & $\gG_{421}$  &  $\gG_{\rm SM}$  \\
\hline \hline
     $\rep{56_F}$   
     & $( \rep{ 1}\,, \rep{1} \,, +1)_{ \mathbf{F}}^{\prime} $  
     & $( \rep{ 1}\,, \rep{1} \,, +1)_{ \mathbf{F}}^{\prime} $ 
     & $( \rep{ 1}\,, \rep{1} \,, +1)_{ \mathbf{F}}^{\prime} $   
     & $ ( \rep{ 1}\,, \rep{1} \,, +1)_{ \mathbf{F}}^{\prime} ~:~{ \mu_R }^c $  \\[1mm]
                       & $( \rep{ 5}\,, \repb{3} \,, +\frac{7}{15})_{ \mathbf{F}}$ 
                       & $( \rep{ 5}\,, \rep{1} \,, +\frac{4}{5})_{ \mathbf{F}} $  
                       & $( \rep{ 1}\,, \rep{1} \,, +1)_{ \mathbf{F}}^{\prime\prime}  $  
                       & $ ( \rep{ 1}\,, \rep{1} \,, +1)_{ \mathbf{F}}^{\prime \prime} ~:~ {e_R }^c  $ \\
                       &   &   &  $( \rep{ 4}\,, \rep{1} \,, +\frac{3}{4})_{ \mathbf{F}} $ & $ ( \rep{1}\,, \rep{1} \,, +1)_{ \mathbf{F}}^{\prime\prime\prime} ~:~ {\EG_R}^c $   \\
                       &   &   &   & $( \rep{3}\,, \rep{1} \,, +\frac{2}{3})_{ \mathbf{F}} ~:~ \UG_L $ \\
                       &   & $( \rep{ 5}\,, \repb{2} \,, +\frac{3}{10})_{ \mathbf{F}}$ & $( \rep{1}\,, \repb{2} \,, +\frac{1}{2})_{ \mathbf{F}}^{\prime\prime\prime}  $ &  $( \rep{ 1}\,, \repb{2} \,, +\frac{1}{2})_{ \mathbf{F}}^{\prime\prime\prime} ~:~ ( {\nG_R^{\prime\prime\prime }}^c \,, -{\eG_R^{\prime\prime\prime } }^c )^T $  \\
                       &   &   & $( \rep{4}\,, \repb{2} \,, +\frac{1}{4})_{ \mathbf{F}} $  
                       & $( \rep{1}\,, \repb{2} \,, +\frac{1}{2})_{ \mathbf{F}}^{\prime\prime\prime\prime} ~:~ ( {\nG_R^{\prime\prime\prime\prime }}^c \,, -{\eG_R^{\prime\prime\prime\prime } }^c )^T $ \\
                       &   &  &  &  $( \rep{3}\,, \repb{2} \,, +\frac{1}{6})_{ \mathbf{F}} ~:~ (\dG_L \,, - \uG_L )^T$  \\[1mm]
                       & $( \repb{10}\,, \rep{1} \,, -\frac{3}{5})_{ \mathbf{F}}$  
                       & $( \repb{10}\,, \rep{1} \,, -\frac{3}{5})_{ \mathbf{F}} $ 
                       & $( \repb{4}\,, \rep{1} \,, -\frac{3}{4})_{ \mathbf{F}} $ 
                       & $( \rep{1}\,, \rep{1} \,, -1)_{ \mathbf{F}}^{  } ~:~ \EG_L $ \\
                       &   &   &   &  $( \repb{3}\,, \rep{1} \,, -\frac{2}{3})_{ \mathbf{F}}^{\prime} ~:~ {\UG_R}^c $ \\
                       &   &   & $( \rep{6}\,, \rep{1} \,, -\frac{1}{2})_{ \mathbf{F}}^{\prime} $ & $( \rep{3}\,, \rep{1} \,, -\frac{1}{3})_{ \mathbf{F}}^{\prime\prime\prime} ~:~ \DG_L^{\prime \prime\prime}$  \\
                       &   &   &   & $( \repb{3}\,, \rep{1} \,, -\frac{2}{3})_{ \mathbf{F}}^{\prime \prime} ~:~ {u_R}^c $  \\[1mm]
                       & $( \rep{10}\,, \rep{3} \,, -\frac{1}{15})_{ \mathbf{F}}$  
                       &  $( \rep{10}\,, \rep{1} \,, -\frac{2}{5})_{ \mathbf{F}}^{\prime} $ 
                       & $( \rep{4}\,, \rep{1} \,, -\frac{1}{4})_{ \mathbf{F}}^{\prime\prime}  $ 
                       & $( \rep{3}\,, \rep{1} \,, -\frac{1}{3})_{ \mathbf{F}}^{\prime\prime\prime \prime } ~:~\DG_L^{\prime\prime \prime \prime}$ \\
                       &   &   &   & $( \rep{1}\,, \rep{1} \,, 0 )_{ \mathbf{F}}^{\prime\prime \prime} ~:~ {\check \nG_R}^{\prime \prime\prime \,c}$ \\
                       &   &   & $( \rep{6}\,, \rep{1} \,, -\frac{1}{2})_{ \mathbf{F}}^{\prime\prime} $ & $( \rep{3}\,, \rep{1} \,, -\frac{1}{3})_{ \mathbf{F}}^{\prime\prime\prime\prime\prime} ~:~ \DG_L^{\prime\prime\prime\prime\prime} $   \\
                       &   &   &   & $( \repb{3}\,, \rep{1} \,, -\frac{2}{3})_{ \mathbf{F}}^{\prime\prime\prime} ~:~ {c_R}^c $  \\
                       &   &  $( \rep{10}\,, \rep{2} \,, +\frac{1}{10})_{ \mathbf{F}}$  & $( \rep{4}\,, \rep{2} \,, +\frac{1}{4} )_{ \mathbf{F}}^{\prime}  $ & $ ( \rep{3}\,, \rep{2} \,, +\frac{1}{6} )_{ \mathbf{F}}^{\prime} ~:~ (u_L\,,d_L)^T $  \\
                       &   &   &   & $( \rep{1}\,, \rep{2} \,, +\frac{1}{2})_{ \mathbf{F}}^{\prime\prime\prime \prime \prime} ~:~ ( {\eG_R^{\prime\prime\prime\prime\prime }}^c \,, {\nG_R^{\prime\prime\prime\prime\prime } }^c )^T$ \\
                       &   &   & $( \rep{6}\,, \rep{2} \,, 0)_{ \mathbf{F}} $  & $( \rep{3}\,, \rep{2} \,, +\frac{1}{6})_{ \mathbf{F}}^{\prime\prime} ~:~ (c_L\,,s_L)^T  $  \\
                       &   &   &   & $ ( \repb{3}\,, \rep{2} \,, -\frac{1}{6} )_{ \mathbf{F}}  ~:~ ( {\dG_R}^c \,,{\uG_R}^c )^T $ \\[1mm]
\hline\hline
\end{tabular}
\caption{
The ${\sG\uG}(8)$ fermion representation of $\rep{56_F}$ under the $\gG_{531}\,,\gG_{521}\,, \gG_{421}\,, \gG_{\rm SM}$ subalgebras. 
}
\label{tab:SU8_56ferm_WSS}
\end{center}
}
\end{table}%

\para
By following the WSS symmetry breaking pattern in Eq.~\eqref{eq:SUSYSU8_531_WSSPattern}, we tabulate the fermion representations at various stages in Tables~\ref{tab:SU8_8ferm_WSS}, \ref{tab:SU8_28ferm_WSS}, and \ref{tab:SU8_56ferm_WSS}. 
Along this pattern, the global $\widetilde{ {\rm U}}(1)_T$ symmetry evolves to the global $\widetilde{ {\rm U}}(1)_{B-L}$ at the EW scale according to the following sequence:
\beqn\label{eq:U1T_WSS_def}
&& \gG_{531}~:~ \Tc^\prime \equiv \Tc + 3t \Xc_0 \,, \quad  \gG_{521}~:~ \Tc^{ \prime \prime} \equiv \Tc^\prime - 8 t \Xc_1  \,, \non
&& \gG_{421}~:~   \Tc^{ \prime \prime \prime} \equiv  \Tc^{ \prime \prime}  \,, \quad  \gG_{\rm SM}~:~  \Bc- \Lc \equiv  \Tc^{ \prime \prime \prime} + 8 t \Yc \,.
\eeqn

\subsection{Decompositions of the ${\sG \uG}$(8) Higgs fields}

\para
We decompose the Higgs fields in Eq.~\eqref{eq:SUSYSU8_Yukawa} into components that can be responsible for the sequential symmetry breaking stages in Eq.~\eqref{eq:SUSYSU8_531_WSSPattern}. 
For Higgs fields of $\repb{ 8_H}_{\,, \omega}$, they read
\begin{eqnarray}\label{eq:SU8_WSS_Higgs_Br01}
\repb{8_H}_{\,,\omega }  &\supset&  \langle ( \rep{1} \,, \repb{3} \,, -\frac{1}{3} )_{\mathbf{H}\,, \omega }  \rangle \oplus  \underline{ ( \repb{5} \,, \rep{1} \,, +\frac{1}{5} )_{\mathbf{H}\,, \omega } }\non
&\supset&  \underline{ ( \rep{1} \,, \repb{2} \,, -\frac{1}{2} )_{\mathbf{H}\,, \omega } }  \oplus  \langle ( \repb{5} \,, \rep{1} \,, +\frac{1}{5} )_{\mathbf{H}\,, \omega }  \rangle  \non
&\supset&  \underline{ ( \rep{1} \,, \repb{2} \,, -\frac{1}{2} )_{\mathbf{H}\,, \omega } }  \oplus  \langle ( \repb{4} \,, \rep{1} \,, +\frac{1}{4} )_{\mathbf{H}\,, \omega }  \rangle  \,.
\end{eqnarray}
For Higgs fields of $\repb{28_H}_{\,,\dot \omega } $, they read
\begin{eqnarray}\label{eq:SU8_WSS_Higgs_Br02}
\repb{28_H}_{\,,\dot \omega } &\supset& \underline{( \rep{1} \,, \rep{3} \,,  -\frac{2}{3} )_{\mathbf{H}\,, \dot\omega } } \oplus  \underline{ ( \repb{5} \,, \repb{3} \,, -\frac{2}{15} )_{\mathbf{H}\,, \dot\omega } } \oplus  \underline{ ( \repb{10} \,, \rep{1} \,, +\frac{2}{5} )_{\mathbf{H}\,, \dot\omega } }  \non
&\supset &  \underline{  ( \rep{1} \,, \rep{2} \,, -\frac{1}{2} )_{\mathbf{H}\,, \dot\omega }  } \oplus \Big[  \underline{ ( \repb{5} \,, \repb{2} \,, -\frac{3}{10} )_{\mathbf{H}\,, \dot\omega }  }  \oplus \langle  ( \repb{5} \,, \rep{1} \,, +\frac{1}{5} )_{\mathbf{H}\,, \dot\omega }  \rangle  \Big] \oplus \underline{ ( \repb{10} \,, \rep{1} \,, +\frac{2}{5} )_{\mathbf{H}\,, \dot\omega }  } \non
&\supset&  \underline{ ( \rep{1} \,, \rep{2} \,, -\frac{1}{2} )_{\mathbf{H}\,, \dot\omega }  }  \oplus \Big[ \underline{ ( \rep{1} \,, \repb{2} \,, -\frac{1}{2} )_{\mathbf{H}\,, \dot\omega } } \oplus \underline{ ( \repb{4} \,, \repb{2} \,, -\frac{1}{4} )_{\mathbf{H}\,, \dot{\omega} }  } \oplus   \langle  ( \repb{4} \,, \rep{1} \,, +\frac{1}{4} )_{\mathbf{H}\,, \dot\omega }  \rangle  \Big]  \non
&\oplus& \langle ( \repb{4} \,, \rep{1} \,, +\frac{1}{4} )_{\mathbf{H}\,, \dot\omega }^{\prime}  \rangle \,.
%
\end{eqnarray}
For Higgs field of $\rep{70_H}$, it reads
\begin{eqnarray}\label{eq:SU8_WSS_Higgs_Br05} 
\rep{70_H} &\supset& ( \rep{5} \,, \rep{1 } \,, +\frac{4}{5} )_{\mathbf{H}}^{  }  \oplus  ( \repb{5} \,, \rep{1} \,, -\frac{4}{5} )_{\mathbf{H}}  \oplus \underline{ ( \rep{10} \,, \repb{3} \,, +\frac{4}{15} )_{\mathbf{H}} }   \oplus ( \repb{10} \,, \rep{3 } \,, -\frac{4}{15} )_{\mathbf{H}}^{ } \non
&\supset& \underline{ ( \rep{10} \,, \repb{2} \,, +\frac{1}{10} )_{\mathbf{H}} } \supset  \underline{ ( \rep{4} \,, \repb{2} \,, +\frac{1}{4} )_{\mathbf{H}}^{  } }  \supset  \langle ( \rep{1} \,, \repb{2} \,, +\frac{1}{2} )_{\mathbf{H}}  \rangle  \,. 
\end{eqnarray}
%
%

\subsection{RGEs of the WSS symmetry breaking pattern}


\para
Between the $ v_{531} \leq \mu \leq v_{U} $, almost all massless Higgs fields are summarized as follows:
\beqn
&& ( \repb{5}\,, \rep{1}\,, +\frac{1}{5})_{ \mathbf{H}\,, \omega} \oplus  ( \rep{1}\,, \repb{3}\,, -\frac{1}{3})_{ \mathbf{H}\,, \omega} \subset \repb{8_H}_{, \omega } \,, \non
&& (\rep{1}\,,\rep{3}\,, -\frac{2}{3})_{\mathbf{H}\,,\dot\omega} \oplus (\repb{5}\,,\repb{3}\,, -\frac{2}{15})_{\mathbf{H}\,,\dot\omega} \oplus ( \repb{10}\,, \rep{1}\,, +\frac{2}{5})_{\mathbf{H}\,,\dot \omega} \subset \repb{28_H}_{,\dot\omega} \,, \non
&& (\rep{5}\,,\rep{1}\,, +\frac{4}{5})_\mathbf{H} \oplus (\repb{5}\,,\rep{1}\,, -\frac{4}{5})_\mathbf{H} \oplus (\rep{10}\,,\repb{3}\,, + \frac{4}{15})_\mathbf{H} \oplus (\repb{10}\,,\rep{3}\,, - \frac{4}{15})_\mathbf{H} \subset \rep{70_H} \,, \non
&& (\repb{5}\,,\rep{1}\,, -\frac{4}{5})_\mathbf{H} \oplus  (\rep{5}\,,\rep{1}\,, +\frac{4}{5})_\mathbf{H} \oplus (\repb{10}\,,\rep{3}\,, - \frac{4}{15})_\mathbf{H} \oplus (\rep{10}\,,\repb{3}\,, + \frac{4}{15})_\mathbf{H} \subset \repb{70_H} \,, \non
&& ( \rep{5}\,, \rep{1}\,, -\frac{1}{5})_{ \mathbf{H}}^{\omega} \oplus  ( \rep{1}\,, \rep{3}\,, +\frac{1}{3})_{ \mathbf{H}}^{\omega} \subset \rep{8_H}^{\omega} \,,\non
&& (\rep{1}\,,\repb{3}\,, +\frac{2}{3})_{\mathbf{H}}^{\dot\omega} \oplus (\rep{5}\,,\rep{3}\,, +\frac{2}{15})_{\mathbf{H}}^{\dot\omega} \oplus ( \rep{10}\,, \rep{1}\,, -\frac{2}{5})_{\mathbf{H}}^{\dot\omega} \subset \rep{28_H}^{\dot\omega} \,.
\eeqn
All fermionic components of $\repb{8_F}^\Omega \oplus \rep{28_F} \oplus \rep{56_F}$ remain massless after the decomposition into the $\gG_{531} $ IRs.
Consequently, we have the $\gG_{531} $ $ \beta $ coefficients of
\beqn\label{eq:HiggsC_531}
&& (b^{(1)}_{{\sG \uG}(5)_{s}}\,,b^{(1)}_{{\sG \uG}(3)_{W}}\,,b^{(1)}_{{\uG}(1)_{X_0}})   = (+54 \,, +60 \,, +\frac{368}{5}) \,, \non
&& b^{(2)}_{\gG_{531}}  = \begin{pmatrix}
6984/5 & 216 & 672/25\\
648 & 728 & 352/15\\
16128/25 & 2816/15 & 98816/1125
\end{pmatrix} \,.
\eeqn


\para
Between the $ v_{521} \leq \mu \leq v_{531} $, the massless Higgs fields are
\beqn
&& ( \repb{5}\,, \rep{1}\,, +\frac{1}{5})_{ \mathbf{H}\,, 3 , \rm V ,\rm VI} \oplus ( \rep{1}\,, \repb{2}\,, -\frac{1}{2})_{\mathbf{H}\,, 3, \rm V,\rm VI } \subset ... \subset \repb{8_H}_{\,, 3 , \rm V ,\rm VI }\,, \non
&& (\repb{10}\,,\rep{1}\,, +\frac{2}{5})_{\mathbf{H}\,,\dot{\omega}} \oplus \Big[ (\rep{1}\,,\rep{2}\,, -\frac{1}{2})_{\mathbf{H}\,,\dot{\omega}} \oplus (\rep{1}\,,\rep{1}\,, -1)_{\mathbf{H}\,,\dot{\omega}} \Big]  \non
& \oplus &  \Big[ (\repb{5}\,,\rep{1}\,, +\frac{1}{5})_{\mathbf{H}\,,\dot{\omega}} \oplus (\repb{5}\,,\repb{2}\,, -\frac{3}{10})_{\mathbf{H}\,,\dot{\omega}} \Big]  \subset ... \subset \repb{28_H}_{\,, \dot{\omega}}  \,, \quad  \dot{\omega}=( \dot{1},\dot{2},\dot{\rm VII},\dot{\rm VIII},\dot{\rm IX} ) \,,  \non
&&  (\rep{10}\,,\rep{1}\,, -\frac{2}{5})_{\mathbf{H}}^{\dot{\rm \kappa}} \oplus \Big[ (\rep{1}\,,\repb{2}\,, +\frac{1}{2})_{\mathbf{H}}^{\dot{\rm \kappa}} \oplus (\rep{1}\,,\rep{1}\,, +1)_{\mathbf{H}}^{\dot{\rm \kappa}} \Big]    \non
& \oplus &  \Big[ (\rep{5}\,,\rep{1}\,, -\frac{1}{5})_{\mathbf{H}}^{\dot{\rm \kappa}} \oplus (\rep{5}\,,\rep{2}\,, +\frac{3}{10})_{\mathbf{H}}^{\dot{\rm \kappa}} \Big]  \subset ... \subset \rep{28_H}^{\dot{\rm \kappa}} \,, \quad  \dot{\kappa}=(\dot{\rm VII},\dot{\rm IX} ) \,,   \non 
&& (\rep{10}\,,\repb{2}\,, +\frac{1}{10})_\mathbf{H}  \subset ... \subset \rep{70_H} \,.
\eeqn
All ${\sG \uG}$(8) fermions that remain massless after the decomposition into the $\gG_{521}$ IRs can be found in Ref.~\cite{Wang:2023gut}.  
Consequently, we have the $\gG_{521} $ $ \beta $ coefficients of
\beqn\label{eq:HiggsC_521}
&& (b^{(1)}_{{\sG \uG}(5)_{s}}\,,b^{(1)}_{{\sG \uG}(2)_{W}}\,,b^{(1)}_{{\uG}(1)_{X_1}})  = (-\frac{1}{2}\,,+\frac{67}{6}\,,+\frac{709}{30}) \,, \non
&& b^{(2)}_{\gG_{521}}  = \begin{pmatrix}
2149/5 & 75/2 & 679/50 \\
300 & 1129/6 & 29/2 \\
8148/25 & 87/2 & 14233/250
\end{pmatrix} \,.
\eeqn
The ${\sG \uG}(3)_W \oplus {\uG}(1)_{X_0}$ gauge couplings match with the ${\sG \uG}(2 )_W \oplus {\uG}(1)_{X_1}$ gauge couplings as follows:
\beqn\label{eq:531_CcoupMatch}
&& \alpha_{3W }^{-1} (v_{531} ) =  \alpha_{2W }^{-1} (v_{531} )  \,,~ \alpha_{X_1}^{-1} (v_{531} )  = \frac{1}{3} \alpha_{3W }^{-1} (v_{531} )  +  \alpha_{X_0}^{-1} (v_{531} )  \,.
\eeqn


\para
Between the $ v_{421} \leq \mu \leq v_{521} $, the massless Higgs fields are
\beqn
&& ( \repb{4}\,, \rep{1}\,, +\frac{1}{4})_{ \mathbf{H}\,, 3 ,\rm VI} \oplus ( \rep{1}\,, \repb{2}\,, -\frac{1}{2})_{ \mathbf{H}\,, 3 ,\rm VI}   \subset ... \subset \repb{8_H}_{, 3 ,\rm VI} \,,\non
&& \Big[ (\repb{4}\,,\rep{1}\,, +\frac{1}{4})_{\mathbf{H}\,,\dot{\omega}}^{\prime} \oplus  (\rep{6}\,,\rep{1}\,, +\frac{1}{2})_{\mathbf{H}\,,\dot{\omega}} \Big] \oplus \Big[(\rep{1}\,,\rep{2}\,, -\frac{1}{2})_{\mathbf{H}\,,\dot{\omega}}^{} \oplus (\rep{1}\,,\rep{1}\,, -1)_{\mathbf{H}\,,\dot{\omega}} \Big]   \non
& \oplus &  \Big[ (\repb{4}\,,\rep{1}\,, +\frac{1}{4})_{\mathbf{H}\,,\dot{\omega}} \oplus (\rep{1}\,,\repb{2}\,, -\frac{1}{2})_{\mathbf{H} \,,\dot{\omega}}^{ } \oplus (\repb{4}\,,\repb{2}\,, -\frac{1}{4})_{\mathbf{H}\,,\dot{\omega}} \Big] \subset ... \subset \repb{28_H}_{\,, \dot{\omega}} \,, \quad  \dot{\omega}=( \dot{2},\dot{\rm VIII},\dot{\rm IX} ) \,,\non
&& \Big[ (\rep{4}\,,\rep{1}\,, -\frac{1}{4})_{\mathbf{H}}^{\dot{\rm IX}^\prime} \oplus  (\rep{6}\,,\rep{1}\,, -\frac{1}{2})_{\mathbf{H}}^{\dot{\rm IX}} \Big] \oplus \Big[(\rep{1}\,,\repb{2}\,, +\frac{1}{2})_{\mathbf{H}}^{\dot{\rm IX}} \oplus (\rep{1}\,,\rep{1}\,, +1)_{\mathbf{H}}^{\dot{\rm IX}} \Big]   \non
&\oplus&  \Big[ (\rep{4}\,,\rep{1}\,, -\frac{1}{4})_{\mathbf{H}}^{\dot{\rm IX}} \oplus (\rep{1}\,,\rep{2}\,, +\frac{1}{2})_{\mathbf{H}}^{\dot{\rm IX}} \oplus (\rep{4}\,,\rep{2}\,, +\frac{1}{4})_{\mathbf{H}}^{\dot{\rm IX}} \Big] \subset ... \subset \rep{28_H}^{\dot{\rm IX}} \,,   \non
&& (\rep{4}\,,\repb{2}\,, +\frac{1}{4})_{\mathbf{H}} \subset ... \subset \rep{70_H} \,.
\eeqn
All ${\sG \uG}$(8) fermions that remain massless after the decomposition into the $\gG_{421}$ IRs can be found in Ref.~\cite{Wang:2023gut}.  
Consequently, we have the $\gG_{421}$ $\beta$ coefficients of
\beqn\label{eq:HiggsC_421}
&& (b^{(1)}_{{\sG \uG}(4)_{s}}\,,b^{(1)}_{{\sG \uG}(2)_{W}}\,,b^{(1)}_{{\uG}(1)_{X_2}})  = (-\frac{8}{3}\,,+\frac{17}{3} \,,+\frac{47}{3}) \,, \non
&& b^{(2)}_{\gG_{421}}  = \begin{pmatrix}
991/6 & 45/2 & 35/4 \\
225/2 & 353/3 & 39/4\\
525/4 & 117/4 & 307/8
\end{pmatrix}\,.
\eeqn
The ${\sG \uG}(5)_s \oplus {\uG}(1)_{X_1}$ gauge couplings match with the ${\sG \uG}(4 )_s \oplus {\uG}(1)_{X_2}$ gauge couplings as follows:
\beqn\label{eq:521_CcoupMatch}
&& \alpha_{5s }^{-1} (v_{521} ) =  \alpha_{4s }^{-1} (v_{521} )  \,,~ \alpha_{X_2}^{-1} (v_{521} )  = \frac{1}{10} \alpha_{5s }^{-1} (v_{521} )  +  \alpha_{X_1}^{-1} (v_{521} )  \,.
\eeqn


\para
Between the $ v_{\rm EW} \leq \mu \leq v_{421} $, we have the same $\gG_{\rm SM} $ $ \beta $ coefficients as Eq.~\eqref{eq:HiggsA_SMto331}.
The ${\sG \uG}(4)_s \oplus {\uG}(1)_{X_2}$ gauge couplings match with the ${\sG \uG}(3 )_c \oplus {\uG}(1)_{\rm Y}$ gauge couplings as follows:
\beqn\label{eq:421_CcoupMatch}
&& \alpha_{4s }^{-1} (v_{421} ) =  \alpha_{3c }^{-1} (v_{ 421 } )  \,,~ \alpha_{\rm Y}^{-1} (v_{ 421} ) = \frac{1}{6} \alpha_{4s }^{-1} (v_{421} )  +  \alpha_{X_2}^{-1} (v_{421} )  \,.
\eeqn
Based on constructing the SM quark/lepton mass matrices to reproduce the observed hierarchical masses and the CKM mixing pattern, we find the following benchmark point:
\begin{align}\label{eq:benchmark WSS}
	v_{531}  \simeq 1.4 \times 10^{17} \, \mathrm{GeV} \,, \quad
	v_{521}  \simeq 4.8 \times 10^{15}\, \mathrm{GeV} \,, \quad
	v_{421}  \simeq 1.1 \times 10^{15} \, \mathrm{GeV} \,.
\end{align}
%
%

%
%
\begin{figure}[htb]
\centering
\includegraphics[height=5cm]{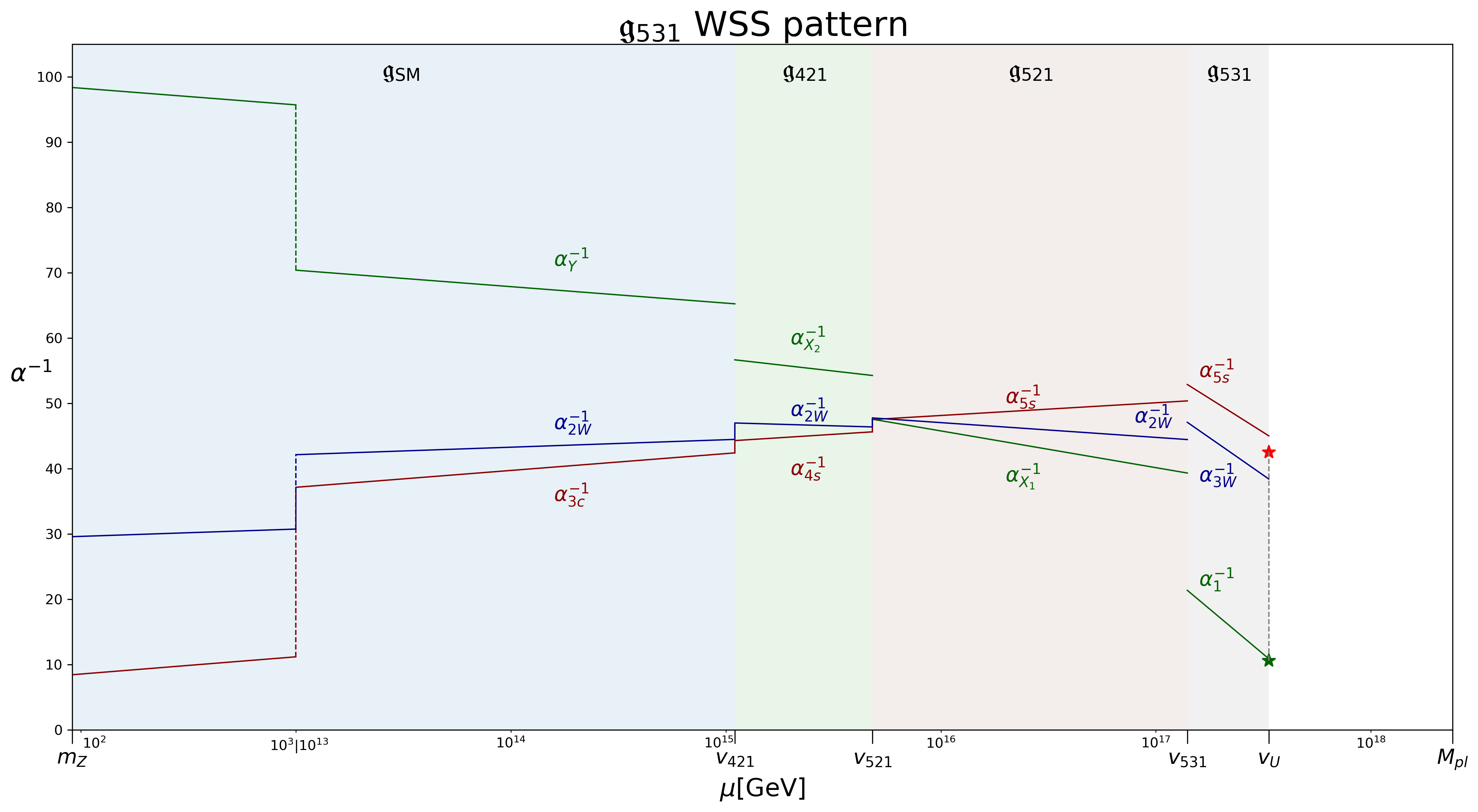}
\caption{
The two-loop RGEs of the ${\sG \uG}(8)$ setup according to the WSS symmetry breaking pattern.
The RG evolutions within $10^{3}\,{\rm GeV} \lesssim \mu \lesssim 10^{13}\,{\rm GeV}$ are hidden in order to highlight the behaviors in three intermediate symmetry breaking scales given by the benchmark point in Eq.~\eqref{eq:benchmark WSS}.
}
\label{fig:SU8_RGE_WSS} 
\end{figure}

\para
With the one- and two-loop $\beta$ coefficients defined in Eqs.~\eqref{eq:HiggsC_531}, \eqref{eq:HiggsC_521}, \eqref{eq:HiggsC_421}, and \eqref{eq:HiggsA_SMto331}, we plot the RGEs of the ${\sG \uG}(8)$ setup along the WSS symmetry breaking pattern in Fig.~\ref{fig:SU8_RGE_WSS}.
The discontinuities of the ${\uG}(1)$ gauge couplings at three intermediate scales follow from their definitions in Eqs.~\eqref{eq:531_CcoupMatch}, \eqref{eq:521_CcoupMatch}, \eqref{eq:421_CcoupMatch}, respectively.
The benchmark point in Fig.~\ref{fig:SU8_RGE_WSS} is marked by $ \star $ and reads
\begin{gather}
c_{\rm HSW} \approx -0.98  \,,~v_{U} \approx 2.56 \times 10^{17}\, {\rm GeV} \,, \non
\alpha_{5c }^{-1} (v_{U} ) \approx 47.45 \,,~\alpha_{3W }^{-1} (v_{U} ) \approx 41.09\,,~\alpha_{1 }^{-1} (v_{U} ) \approx 10.63\,.
\end{gather}
%
%

\section{THE WWW SYMMETRY BREAKING PATTERN}
\label{section:WWW_pattern}

\subsection{Decompositions of the ${\sG \uG}$(8) fermions}

\begin{table}[htp] {\small
\begin{center}
\begin{tabular}{c|c|c|c|c}
\hline \hline
   ${\sG \uG}(8)$   &  $\gG_{351}$  & $\gG_{341}$  & $\gG_{331}$  &  $\gG_{\rm SM}$  \\
\hline \hline
 $\repb{ 8_F}^\Omega$   
 & $( \repb{3} \,, \rep{1}\,,  +\frac{1}{3} )_{ \mathbf{F} }^\Omega$  
 & $(\repb{3} \,, \rep{1} \,, +\frac{1}{3} )_{ \mathbf{F} }^\Omega$  
 & $(\repb{3} \,, \rep{1} \,, +\frac{1}{3} )_{ \mathbf{F} }^\Omega$  
 &  $( \repb{3} \,, \rep{ 1}  \,, +\frac{1}{3} )_{ \mathbf{F} }^{\Omega}~:~ { \Dc_R^\Omega}^c$  \\[1mm]
&  $( \rep{1} \,, \repb{5} \,, -\frac{1}{5})_{ \mathbf{F} }^{\Omega}$   &  $( \rep{1} \,, \repb{4} \,, -\frac{1}{4})_{ \mathbf{F} }^{\Omega} $&  $( \rep{1} \,, \repb{3} \,, -\frac{1}{3})_{ \mathbf{F} }^{\Omega}$  &   $( \rep{1} \,, \repb{2} \,, -\frac{1}{2})_{ \mathbf{F} }^{\Omega}    ~:~ \Lc_L^\Omega =( \Ec_L^\Omega \,, - \Nc_L^\Omega )^T$  \\
&   &  &  &  $( \rep{1} \,, \rep{1} \,,  0)_{ \mathbf{F} }^{\Omega^{\prime} } ~:~ \check \Nc_L^{\Omega^{\prime} }$  \\[1mm]
 &  &
 &  $( \rep{1} \,, \rep{1} \,, 0)_{ \mathbf{F} }^{\Omega^{\prime\prime}}$ 
 &  $( \rep{1} \,, \rep{1} \,, 0)_{ \mathbf{F} }^{\Omega^{\prime\prime}} ~:~ \check \Nc_L^{\Omega^{\prime \prime}}$  \\[1mm]
 &  & $( \rep{1} \,, \rep{1} \,, 0)_{ \mathbf{F} }^{\Omega}$
 &  $( \rep{1} \,, \rep{1} \,, 0)_{ \mathbf{F} }^{\Omega}$ 
 &  $( \rep{1} \,, \rep{1} \,, 0)_{ \mathbf{F} }^{\Omega} ~:~ \check \Nc_L^{\Omega }$  \\[1mm]
\hline\hline
\end{tabular}
\caption{The ${\sG \uG}(8)$ fermion representation of $\repb{8_F}^\Omega$ under the $\gG_{351}\,,\gG_{341}\,, \gG_{331}\,, \gG_{\rm SM}$ subalgebras.}
\label{tab:SU8_8ferm_WWW}
\end{center} 
}
\end{table}%

\begin{table}[htp] {\small
\begin{center}
\begin{tabular}{c|c|c|c|c}
\hline \hline
   ${\sG \uG}(8)$   &  $\gG_{351}$  & $\gG_{341}$  & $\gG_{331}$  &  $\gG_{\rm SM}$  \\
\hline \hline 
 $\rep{28_F}$   
 & $( \repb{3}\,, \rep{ 1} \,, - \frac{2}{3})_{ \mathbf{F}}$ 
 & $( \repb{3}\,, \rep{ 1} \,, - \frac{2}{3})_{ \mathbf{F}}$
 & $( \repb{3}\,, \rep{ 1} \,, - \frac{2}{3})_{ \mathbf{F}}$
 & $( \repb{3}\,, \rep{ 1} \,, - \frac{2}{3})_{ \mathbf{F}} ~:~{t_R}^c$  \\[1mm]
& $( \rep{3}\,, \rep{5} \,,  -\frac{2}{15})_{ \mathbf{F}}$ 
& $( \rep{3}\,, \rep{4} \,,  -\frac{1}{12})_{ \mathbf{F}}$   
& $( \rep{3}\,, \rep{3} \,,  0)_{ \mathbf{F}}$  
& $( \rep{3}\,, \rep{2} \,,  +\frac{1}{6})_{ \mathbf{F}}~:~(t_L\,, b_L)^T$  \\
&   &   &   & $( \rep{3}\,, \rep{1} \,, -\frac{1}{3} )_{ \mathbf{F}}^\prime ~:~\DG_L^\prime  $ \\[1mm]
&   & &  $( \rep{3}\,, \rep{1} \,, -\frac{1}{3} )_{ \mathbf{F}} ^{\prime\prime} $ & $ ( \rep{3}\,, \rep{1} \,, -\frac{1}{3} )_{ \mathbf{F}}  ^{\prime\prime} ~:~\DG_L^{\prime\prime}$ \\[1mm]

&   & $( \rep{3}\,, \rep{1} \,,  -\frac{1}{3} )_{ \mathbf{F}}$   &  $( \rep{3}\,, \rep{1} \,,  -\frac{1}{3} )_{ \mathbf{F}}$  
& $( \rep{3}\,, \rep{1} \,,  -\frac{1}{3} )_{ \mathbf{F}} ~:~\DG_L$  \\[1mm]
&$( \rep{1}\,, \rep{10} \,,  +\frac{2}{5} )_{ \mathbf{F}}$     & $( \rep{1}\,, \rep{4} \,,  +\frac{1}{4} )_{ \mathbf{F}}$    & $( \rep{1}\,, \rep{3} \,,  +\frac{1}{3} )_{ \mathbf{F}}$    & $( \rep{1}\,, \rep{2}\,, + \frac{1}{2})_{ \mathbf{F}}^{\prime\prime} ~:~( {\eG_R^{\prime\prime} }^c \,, { \nG_R^{\prime\prime}}^c )^T$ \\
&   &   &  & $( \rep{1}\,, \rep{1} \,,  0)_{ \mathbf{F}}^{\prime} ~:~\check \nG_R^{\prime  \,c}$  \\[1mm]
&   &   & $( \rep{1}\,, \rep{1} \,,  0)_{ \mathbf{F}}^{\prime\prime} $ & $( \rep{1}\,, \rep{1} \,,  0)_{ \mathbf{F}}^{\prime\prime} ~:~\check \nG_R^{\prime\prime  \,c}$  \\[1mm]
&   &  $( \rep{1}\,, \rep{ 6} \,, +\frac{1}{2})_{ \mathbf{F}}$
& $( \rep{1}\,, \rep{ 3} \,, +\frac{1}{3})_{ \mathbf{F}}^{\prime}$  
& $( \rep{1}\,, \rep{1} \,, 0)_{ \mathbf{F}}^{\prime\prime} ~:~\check \nG_R^c$   \\
&   &   &   & $ ( \rep{1}\,, \rep{2} \,, +\frac{1}{2} )_{ \mathbf{F}} ~:~  ( {\eG_R }^c \,, { \nG_R }^c)^T$ \\[1mm]  
&   &   & $ ( \rep{1}\,, \repb{3} \,, +\frac{2}{3} )_{ \mathbf{F}} $  & $ ( \rep{1}\,, \rep{1} \,, +1 )_{ \mathbf{F}} ~:~ {\tau_R }^c $ \\
&   &   &   & $ ( \rep{1}\,, \repb{2} \,, +\frac{1}{2} )_{ \mathbf{F}}^{\prime} ~:~  (  \nG_R^{\prime \, c }\,, - \eG_R^{\prime \, c }  )^T$ \\[1mm]  
\hline\hline
\end{tabular}
\caption{
The ${\sG \uG}(8)$ fermion representation of $\rep{28_F}$ under the $\gG_{351}\,,\gG_{341}\,, \gG_{331}\,, \gG_{\rm SM}$ subalgebras.
}
\label{tab:SU8_28ferm_WWW}
\end{center}
}
\end{table}

\begin{table}[htp] {\small
\begin{center}
\begin{tabular}{c|c|c|c|c}
\hline \hline
   ${\sG \uG}(8)$   &  $\gG_{351}$  & $\gG_{341}$  & $\gG_{331}$  &  $\gG_{\rm SM}$  \\
\hline \hline
     $\rep{56_F}$   
  
                       & $( \repb{ 3}\,, \rep{5} \,, -\frac{7}{15})_{ \mathbf{F}}$ 
                       & $( \repb{ 3}\,, \rep{4} \,, -\frac{5}{12})_{ \mathbf{F}}$  
                       & $( \repb{ 3}\,, \rep{3} \,, -\frac{1}{3})_{ \mathbf{F}}$  
                       & $( \repb{ 3}\,, \rep{2} \,, -\frac{1}{6})_{ \mathbf{F}} ~:~ ( {\dG_R}^c \,,{\uG_R}^c )^T $ \\
                       &   &   &   & $( \repb{ 3}\,, \rep{1} \,, -\frac{2}{3})_{ \mathbf{F}}^{\prime } ~:~ {\UG_R}^c $   \\[1mm]
                       &   &   & $( \repb{ 3}\,, \rep{1} \,, -\frac{2}{3})_{ \mathbf{F}}^{\prime\prime } $ & $ ( \repb{ 3}\,, \rep{1} \,, -\frac{2}{3})_{ \mathbf{F}}^{\prime\prime } ~:~ {c_R}^c $\\[1mm]
                       &   & $( \repb{3}\,, \rep{1} \,, -\frac{2}{3})_{ \mathbf{F}}^{\prime\prime\prime}$  & $( \repb{3}\,, \rep{1} \,, -\frac{2}{3})_{ \mathbf{F}}^{\prime\prime\prime}$  &  $( \repb{3}\,, \rep{1} \,, -\frac{2}{3})_{ \mathbf{F}}^{\prime\prime\prime} ~:~ {u_R}^c$ \\[1mm]
                       &$( \rep{1}\,, \repb{10} \,, +\frac{3}{5})_{ \mathbf{F}}$   & $( \rep{1}\,, \repb{4} \,, +\frac{3}{4})_{ \mathbf{F}}$  & $( \rep{1}\,, \repb{3} \,, +\frac{2}{3})_{ \mathbf{F}}^{\prime}$  &  $( \rep{1}\,, \repb{2} \,, +\frac{1}{2})_{ \mathbf{F}} ~:~ ( {\nG_R^{\prime\prime\prime }}^c \,, -{\eG_R^{\prime\prime\prime } }^c )^T $ \\
                       &   &  &  &  $( \rep{1}\,, \rep{1} \,, +1)_{ \mathbf{F}}^{\prime} ~:~ {\mu_R}^c $  \\[1mm]
                       &   &  & $( \rep{1}\,, \rep{1} \,, +1)_{ \mathbf{F}}^{\prime\prime} $  &  $( \rep{1}\,, \rep{1} \,, +1)_{ \mathbf{F}}^{\prime\prime} ~:~ {\EG_R}^{\,c} $  \\[1mm]
                       &   &$( \rep{1}\,, \rep{6} \,, +\frac{1}{2})_{ \mathbf{F}}^{\prime} $   & $( \rep{1}\,, \rep{3} \,, +\frac{1}{3})_{ \mathbf{F}}^{\prime\prime} $  & $( \rep{1}\,, \rep{2} \,, +\frac{1}{2})_{ \mathbf{F}}^{\prime \prime\prime \prime} ~:~ ( {\eG_R^{\prime\prime\prime\prime }}^c \,, {\nG_R^{\prime\prime\prime\prime } }^c )^T $  \\
                        &   &   &   & $( \rep{1}\,, \rep{1} \,, 0 )_{ \mathbf{F}}^{\prime\prime \prime} ~:~ {\check \nG_R}^{\prime \prime\prime \,c}$ \\[1mm]
                      
                       &   
                       &  
                       & $( \rep{1}\,, \repb{3} \,, +\frac{2}{3})_{ \mathbf{F}}^{\prime\prime} $ 
                       & $( \rep{1}\,, \repb{2} \,, +\frac{1}{2})_{ \mathbf{F}}^{\prime\prime\prime\prime\prime  } ~:~  ( {\nG_R^{\prime\prime\prime\prime\prime }}^c \,, -{\eG_R^{\prime\prime\prime\prime\prime } }^c )^T  $ \\
                       && &&$( \rep{1}\,, \rep{1} \,, +1)_{ \mathbf{F}}^{\prime\prime\prime}~:~ {e_R}^c $\\[1mm]
                       
                       & $( \rep{3}\,, \rep{10} \,, +\frac{1}{15})_{ \mathbf{F}}$  
                       &  $( \rep{3}\,, \rep{4} \,, -\frac{1}{12})_{ \mathbf{F}}^{\prime}$ 
                       & $( \rep{3}\,, \rep{3} \,, 0)_{ \mathbf{F}}^{\prime} $ 
                       & $( \rep{3}\,, \rep{1} \,, -\frac{1}{3})_{ \mathbf{F}}^{\prime\prime\prime \prime } ~:~\DG_L^{\prime\prime \prime \prime}$ \\
                       &   &   &   & $ ( \rep{3}\,, \rep{2} \,, +\frac{1}{6} )_{ \mathbf{F}}^{\prime} ~:~ (u_L\,,d_L)^T $ \\[1mm]
                       &&   &$( \rep{3}\,, \rep{1} \,, -\frac{1}{3})_{ \mathbf{F}}^{\prime\prime\prime\prime\prime} $ & $ ( \rep{3}\,, \rep{1} \,, -\frac{1}{3})_{ \mathbf{F}}^{ \prime\prime\prime\prime\prime }~:~\DG_L^{\prime\prime \prime \prime\prime}$ \\[1mm]
                       
                       &   & $( \rep{3}\,, \rep{6} \,, +\frac{1}{6})_{ \mathbf{F}} $ & $( \rep{3}\,, \rep{3} \,, 0 )_{ \mathbf{F}}^{\prime\prime} $ & $( \rep{3}\,, \rep{1} \,, -\frac{1}{3})_{ \mathbf{F}}^{\prime\prime\prime } ~:~\DG_L^{\prime\prime\prime}$  \\

                       &   &   &   & $( \rep{3}\,, \rep{2} \,, +\frac{1}{6})_{ \mathbf{F}}^{\prime\prime} ~:~ (c_L\,,s_L)^T  $  \\[1mm]

                       &   &   & $( \rep{3}\,, \repb{3} \,, +\frac{1}{3})_{ \mathbf{F}} $  
                       & $( \rep{3}\,, \rep{1} \,, +\frac{2}{3})_{ \mathbf{F}} ~:~ \UG_L $ \\
                        &   &  &  &  $( \rep{3}\,, \repb{2} \,, +\frac{1}{6})_{ \mathbf{F}}^{\prime\prime\prime} ~:~ (\dG_L \,, - \uG_L )^T$  \\[1mm]
    & $( \rep{ 1}\,, \rep{1} \,, -1)_{ \mathbf{F}}$  
     & $( \rep{ 1}\,, \rep{1} \,, -1)_{ \mathbf{F}}$ 
     & $( \rep{ 1}\,, \rep{1} \,, -1)_{ \mathbf{F}}$   
     & $ ( \rep{ 1}\,, \rep{1} \,, -1)_{ \mathbf{F}} ~:~ {\EG_L} $  \\[1mm]
\hline\hline
\end{tabular}
\caption{
The ${\sG \uG}(8)$ fermion representation of $\rep{56_F}$ under the $\gG_{351}\,,\gG_{341}\,, \gG_{331}\,, \gG_{\rm SM}$ subalgebras. 
}
\label{tab:SU8_56ferm_WWW}
\end{center}
}
\end{table}%

\para
By following the WWW symmetry breaking pattern in Eq.~\eqref{eq:SUSYSU8_351_WWWPattern}, we tabulate the fermion representations at various stages in Tables~\ref{tab:SU8_8ferm_WWW}, \ref{tab:SU8_28ferm_WWW}, and \ref{tab:SU8_56ferm_WWW}. 
Along this pattern, the global $\widetilde{ {\rm U}}(1)_T$ symmetry evolves to the global $\widetilde{ {\rm U}}(1)_{B-L}$ at the EW scale according to the following sequence:
\beqn\label{eq:U1T_WWW_def}
&& \gG_{351}~:~ \Tc^\prime \equiv \Tc + 5 \Xc_0 \,, \quad  \gG_{341}~:~ \Tc^{ \prime \prime} \equiv \Tc^\prime   \,, \non
&& \gG_{331}~:~   \Tc^{ \prime \prime \prime} \equiv  \Tc^{ \prime \prime}  \,, \quad  \gG_{\rm SM}~:~  \Bc- \Lc \equiv  \Tc^{ \prime \prime \prime} \,.
\eeqn

\subsection{Decompositions of the ${\sG \uG}$(8) Higgs fields}

\para
We decompose the Higgs fields in Eq.~\eqref{eq:SUSYSU8_Yukawa} into components that can be responsible for the sequential symmetry breaking stages in Eq.~\eqref{eq:SUSYSU8_351_WWWPattern}. 
For Higgs fields of $\repb{ 8_H}_{\,, \omega}$, they read
\begin{eqnarray}\label{eq:SU8_WWW_Higgs_Br01}
\repb{8_H}_{\,,\omega }  &\supset&   \langle ( \rep{1} \,, \repb{5} \,, -\frac{1}{5} )_{\mathbf{H}\,, \omega } \rangle \supset    \langle ( \rep{1} \,, \repb{4} \,, -\frac{1}{4} )_{\mathbf{H}\,, \omega }  \rangle \supset  \langle ( \rep{1} \,, \repb{3} \,, -\frac{1}{3} )_{\mathbf{H}\,, \omega } \rangle \,. 
\end{eqnarray}
For Higgs fields of $\repb{28_H}_{\,,\dot \omega } $, they read
\begin{eqnarray}\label{eq:SU8_WWW_Higgs_Br02}
\repb{28_H}_{\,,\dot \omega } &\supset& \underline{( \rep{1} \,, \repb{10} \,,  -\frac{2}{5} )_{\mathbf{H}\,, \dot\omega } }  \non
&\supset &  \underline{  ( \rep{1} \,, \rep{6} \,, -\frac{1}{2} )_{\mathbf{H}\,, \dot\omega }  } \oplus \langle ( \rep{1} \,, \repb{4} \,, -\frac{1}{4} )_{\mathbf{H}\,, \dot\omega }  \rangle \non
&\supset& \Big[ \underline{ ( \rep{1} \,, \rep{3} \,, -\frac{2}{3} )_{\mathbf{H}\,, \dot\omega }  }  \oplus  \langle ( \rep{1} \,, \repb{3} \,, -\frac{1}{3} )_{\mathbf{H}\,, \dot\omega } \rangle  \Big] \oplus   \langle ( \rep{1} \,, \repb{3} \,, -\frac{1}{3} )_{\mathbf{H}\,, \dot\omega }^{\prime} \rangle   \,.
\end{eqnarray}
For Higgs field of $\rep{70_H}$, it reads
\begin{eqnarray}\label{eq:SU8_WWW_Higgs_Br05} 
\rep{70_H} &\supset& ( \rep{1} \,, \rep{5 } \,, -\frac{4}{5} )_{\mathbf{H}}^{  }  \oplus \underline{ ( \rep{1} \,, \repb{5} \,, +\frac{4}{5} )_{\mathbf{H}}}  \oplus  ( \rep{3} \,, \repb{10} \,, +\frac{4}{15} )_{\mathbf{H}}    \oplus ( \repb{3} \,, \rep{10 } \,, -\frac{4}{15} )_{\mathbf{H}}^{ } \non
&\supset& \underline{ ( \rep{1} \,, \repb{4} \,, +\frac{3}{4} )_{\mathbf{H}} } \supset  \underline{ ( \rep{1} \,, \repb{3} \,, +\frac{2}{3} )_{\mathbf{H}}^{  } }  \supset  \langle  ( \rep{1} \,, \repb{2} \,, +\frac{1}{2} )_{\mathbf{H}}^{  }  \rangle \,. 
\end{eqnarray}
%
%

\subsection{RGEs of the WWW symmetry breaking pattern}


\para
Between the $ v_{351} \leq \mu \leq v_{U} $, almost all massless Higgs fields are summarized as follows:
\beqn
&& ( \repb{3}\,, \rep{1}\,, +\frac{1}{3})_{ \mathbf{H}\,, \omega} \oplus  ( \rep{1}\,, \repb{5}\,, -\frac{1}{5})_{ \mathbf{H}\,, \omega} \subset \repb{8_H}_{, \omega } \,,\non
&& (\rep{3}\,,\rep{1}\,, +\frac{2}{3})_{\mathbf{H}\,,\dot\omega} \oplus (\repb{3}\,,\repb{5}\,, +\frac{2}{15})_{\mathbf{H}\,,\dot\omega} \oplus ( \rep{1}\,, \repb{10}\,, -\frac{2}{5})_{\mathbf{H}\,,\dot \omega} \subset \repb{28_H}_{,\dot\omega} \,, \non
&& (\rep{1}\,,\rep{5}\,, -\frac{4}{5})_\mathbf{H} \oplus (\rep{1}\,,\repb{5}\,, +\frac{4}{5})_\mathbf{H} \oplus
(\rep{3}\,,\repb{10}\,, + \frac{4}{15})_\mathbf{H} \oplus (\repb{3}\,,\rep{10}\,, - \frac{4}{15})_\mathbf{H} \subset \rep{70_H} \,, \non
&& (\rep{1}\,,\repb{5}\,, +\frac{4}{5})_\mathbf{H} \oplus (\rep{1}\,,\rep{5}\,, -\frac{4}{5})_\mathbf{H} \oplus (\repb{3}\,,\rep{10}\,, - \frac{4}{15})_\mathbf{H} \oplus (\rep{3}\,,\repb{10}\,, + \frac{4}{15})_\mathbf{H} \subset \repb{70_H} \,,\non
&& ( \rep{3}\,, \rep{1}\,, -\frac{1}{3})_{ \mathbf{H}}^{\omega} \oplus  ( \rep{1}\,, \rep{5}\,, +\frac{1}{5})_{ \mathbf{H}}^{\omega}\subset \rep{8_H}^{ \omega } \,, \non
&& (\repb{3}\,,\rep{1}\,, -\frac{2}{3})_{\mathbf{H}}^{\dot\omega} \oplus (\rep{3}\,,\rep{5}\,, -\frac{2}{15})_{\mathbf{H}}^{\dot\omega} \oplus ( \rep{1}\,, \rep{10}\,, +\frac{2}{5})_{\mathbf{H}}^{\dot\omega} \subset \rep{28_H}^{\dot\omega} \,.
\eeqn
All fermionic components of $\repb{8_F}^\Omega \oplus \rep{28_F} \oplus \rep{56_F}$ remain massless after the decomposition into the $\gG_{351}$ IRs.
Consequently, we have the $\gG_{351}$ $\beta$ coefficients of
\beqn\label{eq:HiggsD_351}
&& (b^{(1)}_{{\sG \uG}(3)_{c}}\,,b^{(1)}_{{\sG \uG}(5)_{W}}\,,b^{(1)}_{{\uG}(1)_{X_0}})  = (+60\,,+54\,,+\frac{368}{5} ) \,, \non 
&& b^{(2)}_{\Gc_{351}}  = \begin{pmatrix}
728&648&352/15\\
216&6984/5&672/25\\
2816/15&16128/25&98816/1125
\end{pmatrix}\,.
\eeqn


\para
Between the $ v_{341} \leq \mu \leq v_{351} $, the massless Higgs fields are
\beqn
&& ( \repb{3}\,, \rep{1}\,, +\frac{1}{3})_{ \mathbf{H}\,, 3 , \rm V ,\rm VI} \oplus ( \rep{1}\,, \repb{4}\,, -\frac{1}{4})_{\mathbf{H}\,, 3, \rm V,\rm VI } \subset ... \subset \repb{8_H}_{\,, 3 , \rm V ,\rm VI } \,,\non
&& (\rep{3}\,,\rep{1}\,, +\frac{2}{3})_{\mathbf{H}\,,\dot{\omega}} \oplus \Big[ (\repb{3}\,,\repb{4}\,, +\frac{1}{12})_{\mathbf{H}\,,\dot{\omega}} \oplus (\repb{3}\,,\rep{1}\,, +\frac{1}{3})_{\mathbf{H}\,,\dot{\omega}} \Big]  \non
& \oplus &  \Big[ (\rep{1}\,,\repb{4}\,, -\frac{1}{4})_{\mathbf{H}\,,\dot{\omega}} \oplus (\rep{1}\,,\rep{6}\,, -\frac{1}{2})_{\mathbf{H}\,,\dot{\omega}} \Big]  \subset ... \subset \repb{28_H}_{\,, \dot{\omega}} \,, \quad  \dot{\omega}=( \dot{1},\dot{2},\dot{\rm VII},\dot{\rm VIII},\dot{\rm IX} ) \,,  \non
&&(\repb{3}\,,\rep{1}\,, -\frac{2}{3})_{\mathbf{H}}^{\dot{\rm VII}} \oplus \Big[ (\rep{3}\,,\rep{4}\,, -\frac{1}{12})_{\mathbf{H}}^{\dot{\rm VII}} \oplus (\rep{3}\,,\rep{1}\,, -\frac{1}{3})_{\mathbf{H}}^{\dot{\rm VII}} \Big]    \non
& \oplus &  \Big[ (\rep{1}\,,\rep{4}\,, +\frac{1}{4})_{\mathbf{H}}^{\dot{\rm VII}} \oplus (\rep{1}\,,\rep{6}\,, +\frac{1}{2})_{\mathbf{H}}^{\dot{\rm VII}} \Big]  \subset ... \subset \rep{28_H}^{\dot{\rm VII}} \,, \non
&& (\rep{1}\,,\repb{4}\,, +\frac{3}{4})_\mathbf{H}  \subset ... \subset \rep{70_H} \,.
\eeqn
All ${\sG \uG}$(8) fermions that remain massless after the decomposition into the $\gG_{341}$ IRs can be found in Ref.~\cite{Wang:2023gut}.   
Consequently, we have the $\gG_{341} $ $\beta$ coefficients of

\beqn\label{eq:HiggsD_341}
&& (b^{(1)}_{{\sG \uG}(3)_{c}}\,,b^{(1)}_{{\sG \uG}(4)_{W}}\,,b^{(1)}_{{\uG}(1)_{X_1}})  = (+ \frac{29}{6}\,,+\frac{4}{3}\,,+\frac{173}{9})\,,\non
&& b^{(2)}_{\gG_{341}} = \begin{pmatrix}
655/3 & 255/2 & 373/36 \\
68 & 3389/12 & 271/24 \\
746/9 & 1355/8 & 17363/432
\end{pmatrix} \,.
\eeqn
The ${\sG \uG}(5)_W \oplus {\uG}(1)_{X_0}$ gauge couplings match with the ${\sG \uG}(4 )_W \oplus {\uG}(1)_{X_1}$ gauge couplings as follows:
\beqn\label{eq:351_DcoupMatch}
&& \alpha_{5W }^{-1} (v_{351} ) =  \alpha_{4W }^{-1} (v_{351} )  \,,~ \alpha_{X_1}^{-1} (v_{351} )  = \frac{1}{10} \alpha_{5W }^{-1} (v_{351} )  +  \alpha_{X_0}^{-1} (v_{351} )  \,.
\eeqn


\para
Between the $ v_{331} \leq \mu \leq v_{341} $, the massless Higgs fields are
\beqn
&& ( \repb{3}\,, \rep{1}\,, +\frac{1}{3})_{ \mathbf{H}\,, 3 ,\rm VI} \oplus ( \rep{1}\,, \repb{3}\,, -\frac{1}{3})_{ \mathbf{H}\,, 3 ,\rm VI}   \subset ... \subset \repb{8_H}_{, 3 ,\rm VI} \,, \non
&& (\rep{3}\,,\rep{1}\,, +\frac{2}{3})_{\mathbf{H}\,,\dot{\omega}} \oplus \Big[ (\repb{3}\,,\rep{1}\,, +\frac{1}{3})_{\mathbf{H}\,,\dot{\omega}} \oplus (\repb{3}\,,\rep{1}\,, +\frac{1}{3})_{\mathbf{H} \,,\dot{\omega}}^{\prime} \oplus (\repb{3}\,,\repb{3}\,, 0)_{\mathbf{H}\,,\dot{\omega}} \Big]   \non
& \oplus &  \Big[ (\rep{1}\,,\repb{3}\,, -\frac{1}{3})_{\mathbf{H}\,,\dot{\omega}} \oplus (\rep{1}\,,\repb{3}\,, -\frac{1}{3})_{\mathbf{H}\,,\dot{\omega}}^{\prime} \oplus (\rep{1}\,,\rep{3}\,, -\frac{2}{3})_{\mathbf{H}\,,\dot{\omega}} \Big] \subset ... \subset \repb{28_H}_{\,, \dot{\omega}} \,, \quad \dot{\omega}=( \dot{2},\dot{\rm VIII},\dot{\rm IX} ) \,, \non
&& (\rep{1}\,,\repb{3}\,, +\frac{2}{3})_{\mathbf{H}} \subset ... \subset \rep{70_H} \,.
\eeqn
All ${\sG \uG}$(8) fermions that remain massless after the decomposition into the $\gG_{331}$ IRs can be found in Ref.~\cite{Wang:2023gut}.  
Consequently, we have the $\gG_{331}$ $\beta$ coefficients of
\beqn\label{eq:HiggsD_331}
&& (b^{(1)}_{{\sG \uG}(3)_{c}}\,,b^{(1)}_{{\sG \uG}(3)_{W}}\,,b^{(1)}_{{\uG}(1)_{X_2}})   = (-\frac{5}{3}\,,-\frac{3}{2}\,,+\frac{122}{9})  \,,\non
&& b^{(2)}_{\gG_{331}} = \begin{pmatrix}
256/3 & 36 & 58/9 \\
36 & 89 & 22/3 \\
464/9 & 176/3 & 782/27
\end{pmatrix}\,.
\eeqn
The ${\sG \uG}(4)_W \oplus {\uG}(1)_{X_1}$ gauge couplings match with the ${\sG \uG}(3 )_W \oplus {\uG}(1)_{X_2}$ gauge couplings as follows:
\beqn\label{eq:341_DcoupMatch}
&& \alpha_{4W }^{-1} (v_{341} ) =  \alpha_{3W }^{-1} (v_{341} )  \,,~ \alpha_{X_2}^{-1} (v_{341} ) = \frac{1}{6} \alpha_{4W }^{-1} (v_{341} )  +  \alpha_{X_1}^{-1} (v_{341} )  \,.
\eeqn


\para
Between the $ v_{\rm EW} \leq \mu \leq v_{331} $, we have the same $\gG_{\rm SM} $ $ \beta $ coefficients as Eq.~\eqref{eq:HiggsA_SMto331}.
The ${\sG \uG}(3)_W \oplus {\uG}(1)_{X_2}$ gauge couplings match with the ${\sG \uG}(2 )_W \oplus {\uG}(1)_{Y}$ gauge couplings as follows:
\beqn\label{eq:331_DcoupMatch}
&& \alpha_{3W }^{-1} (v_{331} ) =  \alpha_{2W }^{-1} (v_{ 331} )  \,,~ \alpha_{\rm Y}^{-1} (v_{ 331} )  = \frac{1}{3} \alpha_{3W }^{-1} (v_{331} )  +  \alpha_{X_2}^{-1} (v_{331} )  \,.
\eeqn
Based on constructing the SM quark/lepton mass matrices to reproduce the observed hierarchical masses and the CKM mixing pattern, we find the following benchmark point:
\begin{align}\label{eq:benchmark WWW}
	v_{351}  \simeq 1.4 \times 10^{17} \, \mathrm{GeV} \,, \quad
	v_{341}  \simeq 4.8 \times 10^{15}\, \mathrm{GeV} \,, \quad
	v_{331}  \simeq 4.8 \times 10^{13} \, \mathrm{GeV} \,.
\end{align}
%
%

\begin{figure}[htb]
\centering
\includegraphics[height=5cm]{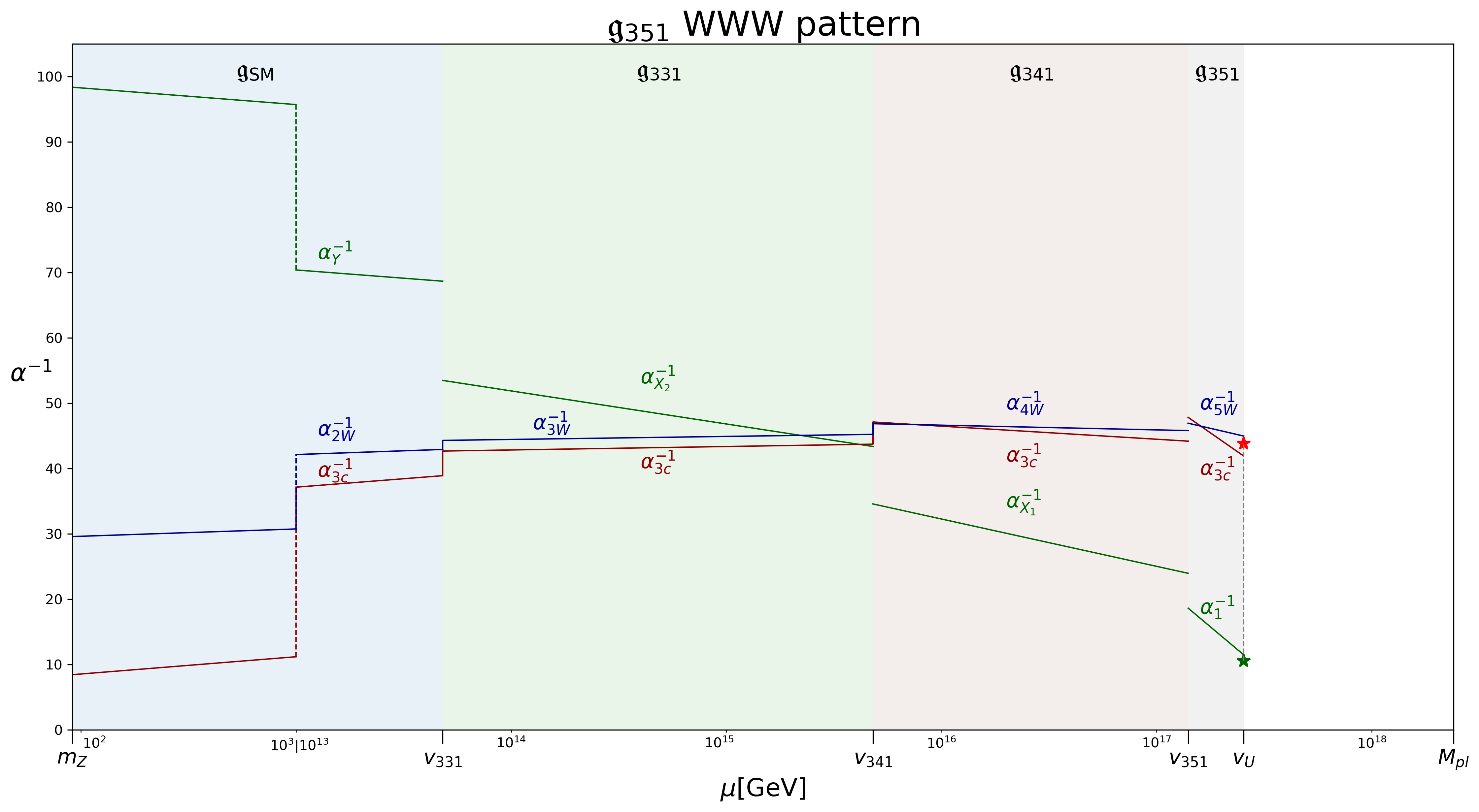}
\caption{
The two-loop RGEs of the ${\sG \uG}(8)$ setup according to the WWW symmetry breaking pattern.
The RG evolutions within $10^{3}\,{\rm GeV} \lesssim \mu \lesssim 10^{13}\,{\rm GeV}$ are hidden in order to highlight the behaviors in three intermediate symmetry breaking scales given by the benchmark point in Eq.~\eqref{eq:benchmark WWW}.
}
\label{SU8_RGE_WWW} 
\end{figure}

\para
With the one- and two-loop $\beta$ coefficients defined in Eqs.~\eqref{eq:HiggsD_351}, \eqref{eq:HiggsD_341}, \eqref{eq:HiggsD_331}, and \eqref{eq:HiggsA_SMto331}, we plot the RGEs of the ${\sG \uG}(8)$ setup along the WWW symmetry breaking pattern in Fig.~\ref{SU8_RGE_WWW}.
The discontinuities of the $\uG(1)$ gauge couplings at three intermediate scales follow from their definitions in Eqs.~\eqref{eq:351_DcoupMatch}, \eqref{eq:341_DcoupMatch}, and \eqref{eq:331_DcoupMatch}.
The benchmark point in Fig.~\ref{SU8_RGE_WWW} is marked by $\star$ and reads
\begin{gather}
c_{\rm HSW} \approx 0.51  \,,~v_{U} \approx 2.72 \times 10^{17} \, {\rm GeV} \,, \non
\alpha_{3c }^{-1} (v_{U} ) \approx 41.21 \,,~\alpha_{5W }^{-1} (v_{U} ) \approx 44.73\,,~\alpha_{1 }^{-1} (v_{U} ) \approx 10.57\,.
\end{gather}

\section{THE NO-GO PATTERN OF ${\sG\uG}(8)\to \gG_{621}$}
\label{section:SSS_pattern}

\para
In this section, we prove that the nonmaximal symmetry breaking pattern of ${\sG\uG}(8)\to \gG_{621}$ cannot be realistic, since the vectorlike quarks of $(\uG\,, \dG)$ from the $\rep{56_F}$ are massless at the EWSB scale.
For this purpose, we only tabulate the fermion representation of $\rep{56_F}$ at various stages in Table~\ref{tab:SU8_56ferm_SSS} by following the SSS symmetry breaking pattern in Eq.~\eqref{eq:SUSYSU8_621Pattern}.

\begin{table}[htp] {\small
		\begin{center}
			\begin{tabular}{c|c|c|c|c}
				\hline \hline
				${\sG \uG}(8)$   &  $\gG_{621}$  & $\gG_{521}$  & $\gG_{421}$  &  $\gG_{\rm SM}$  \\
				\hline \hline
				$\rep{56_F}$   
				& $( \rep{ 20}\,, \rep{1} \,, -\frac{1}{2})_{ \mathbf{F}}$  
				& $( \repb{ 10}\,, \rep{1} \,, -\frac{3}{5})_{ \mathbf{F}}$ 
				& $( \repb{ 4}\,, \rep{1} \,, -\frac{3}{4})_{ \mathbf{F}}$   
				& $( \repb{ 3}\,, \rep{1} \,, -\frac{2}{3})_{ \mathbf{F}}^{\prime} ~:~ {c_R}^c  $  \\
				&        &       &         & $ ( \rep{ 1}\,, \rep{1} \,, -1)_\mathbf{F} ~:~ \EG_L $  \\
				&        &       &    $ ( \rep{6}\,, \rep{1} \,, -\frac{1}{2})_{\mathbf{F}}^{\prime} $      & $ ( \repb{3}\,, \rep{1} \,, -\frac{2}{3})_{\mathbf{F}}^{\prime\prime}  ~:~{u_R}^c $  \\
				&        &       &         & $ ( \rep{3}\,, \rep{1} \,,  -\frac{1}{3})_{\mathbf{F}}^{\prime\prime\prime} ~:~ \DG_L^{\prime \prime\prime} $  \\[1mm]
				&
				& $( \rep{ 10}\,, \rep{1} \,, -\frac{2}{5})_{ \mathbf{F}}^{\prime} $  
				& $( \rep{6}\,, \rep{1} \,, -\frac{1}{2})_{\mathbf{F}}^{\prime\prime} $  
				& $ ( \repb{ 3}\,, \rep{1} \,, -\frac{2}{3})_{ \mathbf{F}}^{\prime\prime\prime}  ~:~ {\UG_R}^c $ \\
				&   &   &   & $ ( \rep{3}\,, \rep{1} \,,  -\frac{1}{3})_{\mathbf{F}}^{\prime \prime\prime\prime} ~:~ \DG_L^{\prime \prime\prime\prime} $  \\
				&   &   
				& $( \rep{4}\,, \rep{1} \,, -\frac{1}{4})_{ \mathbf{F}}^{\prime\prime} $  
				& $ ( \rep{3}\,, \rep{1} \,, -\frac{1}{3})_{\mathbf{F}}^{\prime\prime\prime\prime\prime} ~:~ \DG_L^{\prime\prime\prime\prime\prime} $ \\
				&   &  &  &  $( \rep{1}\,, \rep{1} \,, 0)_{ \mathbf{F}}^{\prime \prime\prime} ~:~{\check \nG_R}^{\prime \prime\prime \,c}$  \\[1mm]
				& $( \rep{15}\,, \rep{2} \,, +\frac{1}{6})_{ \mathbf{F}} $  &        $( \rep{10}\,, \rep{2} \,, +\frac{1}{10})_{ \mathbf{F}} $& $( \rep{6}\,, \rep{2} \,, 0)_{ \mathbf{F}} $  
				& $( \repb{3}\,, \rep{2} \,, -\frac{1}{6})_{ \mathbf{F}}  ~:~ ( {\dG_R}^c \,,{\uG_R}^c )^T $ \\
				&   &  &  &  $( \rep{3}\,, \rep{2} \,, +\frac{1}{6})_{ \mathbf{F}}^{\prime} ~:~ (\uG_L \,,  \dG_L )^T$  \\
				&  
				& 
				& $( \rep{4}\,, \rep{2} \,, +\frac{1}{4})_{ \mathbf{F}}^{\prime} $ 
				& $( \rep{3}\,, \rep{2} \,, +\frac{1}{6})_{ \mathbf{F}}^{\prime\prime} ~:~ (c_L\,,s_L)^T $ \\
				&   &   &   &  $( \rep{1}\,, \rep{2} \,, +\frac{1}{2})_{ \mathbf{F}}^{\prime\prime\prime} ~:~ ( {\eG_R^{\prime\prime\prime }}^c \,, {\nG_R^{\prime\prime\prime} }^c )^T$ \\
				&   &  $( \rep{5}\,, \rep{2} \,, +\frac{3}{10})_{ \mathbf{F}}^{\prime}$ & $( \rep{4}\,, \rep{2} \,, +\frac{1}{4})_{ \mathbf{F}}^{\prime\prime}$ & $( \rep{3}\,, \rep{2} \,, +\frac{1}{6})_{ \mathbf{F}}^{\prime\prime\prime} ~:~ (u_L\,,d_L)^T$  \\
				&   &   &   & $( \rep{1}\,, \rep{2} \,, +\frac{1}{2})_{\mathbf{F}}^{\prime\prime\prime\prime} ~:~ ({\eG_R^{\prime\prime\prime\prime}}^c \,, {\nG_R^{\prime\prime\prime\prime}}^c )^T $ \\
				&   &   &  $( \rep{1}\,, \rep{2} \,, +\frac{1}{2})_{\mathbf{F}}^{\prime\prime\prime\prime\prime} $ 
				& $( \rep{1}\,, \rep{2} \,, +\frac{1}{2})_{\mathbf{F}}^{\prime\prime\prime\prime\prime} ~:~ ({\eG_R^{\prime\prime\prime\prime\prime}}^c \,, {\nG_R^{\prime\prime\prime\prime\prime}}^c )^T $ \\[1mm]
				& $( \rep{6}\,, \rep{1} \,, +\frac{5}{6})_{ \mathbf{F}}$  
				&  $( \rep{5}\,, \rep{1} \,, +\frac{4}{5})_{ \mathbf{F}}$ 
				& $( \rep{4}\,, \rep{1} \,, +\frac{3}{4})_{ \mathbf{F}}$ 
				& $( \rep{3}\,, \rep{1} \,, +\frac{2}{3})_{ \mathbf{F}}~:~ \UG_L$ \\
				&   &   &   & $( \rep{1}\,, \rep{1} \,, +1 )_{ \mathbf{F}}^{\prime}~:~{e_R}^c$ \\
				&   &  & $( \rep{1}\,, \rep{1} \,, +1 )_{ \mathbf{F}}^{\prime\prime}$ & $( \rep{1}\,, \rep{1} \,, +1 )_{ \mathbf{F}}^{\prime\prime}~:~{\mu_R}^c$  \\
				&   & $( \rep{1}\,, \rep{1} \,, +1 )_{ \mathbf{F}}^{\prime\prime\prime}$ & $( \rep{1}\,, \rep{1} \,, +1 )_{ \mathbf{F}}^{\prime\prime\prime}$ & $( \rep{1}\,, \rep{1} \,, +1 )_{ \mathbf{F}}^{\prime\prime\prime} ~:~ {\EG_R}^c $  \\
				[1mm]
				\hline\hline
			\end{tabular}
			\caption{
				The ${\sG \uG}(8)$ fermion representation of $\rep{56_F}$ under the $\gG_{621}\,,\gG_{521}\,, \gG_{421}\,, \gG_{\rm SM}$ subalgebras. 
			}
			\label{tab:SU8_56ferm_SSS}
		\end{center}
	}
\end{table}

\para
Analogous the other symmetry breaking patterns analyzed before, the potential vectorlike fermion mass terms are due to the $d=5$ operator from the superpotential in Eq.~\eqref{eq:SUSYSU8_Yukawa}, where a Higgs field of $\rep{28_H}^{\dot \omega} $ is involved.
The corresponding decompositions are the following:
\begin{eqnarray}\label{eq:SU8_SSS_Higgs_Br02}
\rep{28_H}^{\dot \omega} &\supset& \underline{( \rep{15} \,, \rep{1} \,,  - \frac{1}{3} )_{\mathbf{H} }^{ \dot \omega } } \oplus   ( \rep{6} \,, \rep{2} \,, +\frac{1}{3} )_{\mathbf{H} }^{\dot \omega }   \oplus ( \rep{1} \,, \rep{1} \,,  +1 )_{\mathbf{H} }^{\dot \omega }  \non 
&\supset &   \underline{ ( \rep{10} \,, \rep{1} \,, - \frac{2}{5} )_{\mathbf{H} }^{\dot \omega } }  \oplus \langle ( \rep{5} \,, \rep{1} \,, - \frac{1}{5} )_{\mathbf{H} }^{ \dot \omega }  \rangle     \non
&\supset&  ( \rep{6 } \,, \rep{1} \,, -\frac{1}{2 } )_{\mathbf{H} }^{ \dot \omega }    \oplus \langle ( \rep{4} \,, \rep{1} \,, - \frac{1}{4} )_{\mathbf{H} }^{\dot \omega }  \rangle \oplus   \langle ( \rep{4} \,, \rep{1} \,, -\frac{1}{4} )_{\mathbf{H} }^{ \prime\, \dot \omega } \rangle  \,.
\end{eqnarray}
Accordingly, we decompose the Yukawa coupling along the SSS symmetry breaking pattern between two $\rep{56_F}$'s due to the following $d=5$ operator from the superpotential in Eq.~\eqref{eq:SUSYSU8_Yukawa}:
\beqn
&&\frac{1}{ M_{\rm pl} } \rep{56_F} \rep{56_F} \langle \rep{63_H} \rangle \rep{28_H}^{\dot \omega}  \non
&\supset& \zeta_0  (  \rep{15}\,, \rep{2}\,, +\frac{1}{6} )_{ \mathbf{F}}^{}  \otimes  ( \rep{15}\,, \rep{2}\,, +\frac{1}{6} )_{ \mathbf{F} } \otimes ( \rep{15}\,, \rep{1}\,, -\frac{1}{3})_{ \mathbf{H} }^{\dot \omega } \non
&\supset& \zeta_0 (  \rep{10}\,, \rep{2}\,, +\frac{1}{10} )_{ \mathbf{F}}^{}  \otimes  ( \rep{10}\,, \rep{2}\,,  +\frac{1}{10})_{ \mathbf{F} } \otimes \langle( \rep{5}\,, \rep{1}\,, - \frac{1}{5})_{ \mathbf{H} }^{\dot \omega } \rangle  \,,
%
\eeqn
where we use the decompositions of the Higgs fields of $\rep{28_H}^{\dot \omega}$ in Eq.~\eqref{eq:SU8_SSS_Higgs_Br02}.
To see why this term vanishes, let us denote relevant fields in terms of their components as $\Psi_\alpha^{ \[ \bar{ a }\, \bar{ b } \] \,,  i }  \equiv ( \rep{10 } \,, \rep{2 } \,, + \frac{1 }{10 } )_{ \rep{F}}$ and $\Phi^{ \bar e\,, \dot \omega } \equiv ( \rep{5 } \,, \rep{1 } \,, - \frac{1}{5 } )_{ \rep{H} }^{\dot \omega }$, with $\bar a=1\,,...\,,5$ being the ${\sG \uG}(5)_s$ indices, $i=1\,,2$ being the ${\sG \uG}(2)_W$ indices, and $\alpha =1\,,2$ being the two-component Weyl spinor indices.
The above gauge-invariant Yukawa coupling term can be explicitly expressed as
\beqn
&& \epsilon^{\alpha\beta} \epsilon_{ \bar a \bar b \bar c \bar d  \bar e } { \epsilon_{  i  j  } } \,  \Psi_\alpha^{ \[ \bar a \, \bar b \] \,,  i }  \Psi_\beta^{ \[ \bar c\, \bar d \] \,, j } \Phi^{ \bar e \,, \dot \omega }    = \epsilon^{\beta \alpha}  \epsilon_{ \bar c \bar d \bar a \bar b \bar e }  { \epsilon_{  j  i  } } \,  \Psi_\beta^{ \[ \bar c \, \bar d \] \,,  j }  \Psi_\alpha^{ \[ \bar a\, \bar b \] \,,  i } \Phi^{ \bar e  \,, \dot \omega }   \non
&=& + \epsilon^{ \alpha\beta}  \epsilon_{ \bar a \bar b \bar c \bar d  \bar e } { \epsilon_{  i  j  } } \,  \Psi_\beta^{ \[ \bar c \, \bar d \] \,,  j }  \Psi_\alpha^{ \[ \bar a\, \bar b \] \,, i }  \Phi^{ \bar e  \,, \dot \omega } = - \epsilon^{ \alpha\beta}  \epsilon_{ \bar a \bar b \bar c \bar d  \bar e }  { \epsilon_{  i  j  } } \,  \Psi_\alpha^{ \[ \bar a\, \bar b \] \,,  i } \Psi_\beta^{ \[ \bar c \, \bar d \] \,,  j }  \Phi^{ \bar e  \,, \dot \omega }   \non
&\Rightarrow&  ( \rep{ 10 }\,, \rep{2 }\,, + \frac{1}{ 10 } )_{ \mathbf{F}}  \otimes  ( \rep{ 10 }\,, \rep{2 }\,, + \frac{1}{ 10 } )_{ \mathbf{F}} \otimes ( \rep{5 } \,, \rep{1 } \,, -\frac{1 }{ 5} )_{ \rep{H}  }^{\dot \omega }   = 0 \,,
\eeqn
where we have swapped all gauge and spinor indices of two Weyl fermions in the second line.
By further looking for the potential mass mixing terms between the $(\uG\,, \dG)$ quarks and other flavors of quarks~\cite{Wang:2023gut}, we find that the $\dG$ quark remains massless since its left-handed component cannot form a gauge-invariant mixing term with any other down-type quarks.

\section{SUMMARY}
\label{section:conclusion}

\begin{table}[htp]
\begin{center}
\begin{tabular}{c|c|c|c|c|c|c|c}
\toprule\hline
  Patterns  &  $\Delta_{\rm IR}^s$   & $\Delta_{\rm IR}^W $  & $\Delta_{\rm IR}^1 $  & $c_{\rm HSW}$  & $v_{U}\, [{\rm GeV}]$   & $\alpha_{U }^{} (v_{U} )$ & $\tau_p\,[{\rm yrs}]$  \\
\toprule\hline
                 & $ 90.01 $  & $ 180.96 $ & $ -33.16 $ &  &  &  &   \\
                 \cline{2-4}
  ${\rm SSW}$    & $ 85.27 $  & $ 104.22 $ & $ -52.11 $ &  $ 0.24 $  &  $ 2.28 \times 10^{17} $  &  $ 0.021 $  &  $ \gtrsim 1.76 \times 10^{41} $  \\
                 \cline{2-4}
                 & $ 85.30 $  & $ 108.96 $ & $ -75.80 $ &  &  &  &   \\
                 \hline
                 &  $ 99.49 $ & $ 85.29 $ & $ -61.58 $ &  &  &  &   \\
                 \cline{2-4}
  ${\rm SWS}$    & $ 85.27 $ & $ 99.50 $ & $ -104.22 $ &  $-0.31 $  &  $ 2.59 \times 10^{17} $  &  $ 0.022 $  &  $ \gtrsim 2.67 \times 10^{41} $  \\
                 \cline{2-4}
				 & $ 80.53 $ & $ 118.43 $ & $ -52.10 $ &  &  &  &   \\
				 \hline
				 & $ 23.69 $  & $ 23.68 $ & $ -118.42 $ &  &  &  &   \\
			   	 \cline{2-4}
  ${\rm WSS}$    & $ 71.06 $  & $ 52.11 $ & $ -80.54 $ &  $ -0.98 $  &  $ 2.56 \times 10^{17}  $  &  $ 0.022 $  &  $ \gtrsim 2.55 \times 10^{41} $  \\
				 \cline{2-4}
				 & $ 71.10 $  & $ 94.75 $ & $ -56.85 $ &  &  &  &   \\
                 \hline\hline
                 & $ 137.39 $ & $ 42.63 $ & $ -28.42 $ &  &  &  &   \\
                 \cline{2-4}
  ${\rm WWW}$    & $ 127.91 $ & $ 61.58 $ & $ -47.37 $ &  $ 0.51 $  &  $ 2.72 \times 10^{17} $  &  $ 0.023 $  &  $ \gtrsim 2.97 \times 10^{41} $  \\
                 \cline{2-4}
                 & $ 142.12 $ & $ 52.11 $ & $ -33.16 $ &  &  &  &   \\
\hline\toprule
\end{tabular}
\caption{The benchmark points of the affine Lie algebra $\widehat {\sG \uG}(8)_{k_U=1}$ through four possible nonmaximal symmetry breaking patterns with the $\Nc=1$ SUSY extension. 
We use the $(\Delta_{\rm IR}^s\,, \Delta_{\rm IR}^W\,, \Delta_{\rm IR}^1)$ to represent the reasonable threshold effects estimated according to Eq.~\eqref{eq:threshold_match} at each symmetry breaking stage from the UV scale to the IR scale successively.}	
\label{tab:summary_BM}
\end{center} 
\end{table}%

\para
In this paper, we focus on the possible nonmaximal symmetry breaking patterns of the affine Lie algebra $\widehat {\sG \uG}(8)_{k_U=1}\to \gG_{531}/\gG_{351}$ allowed by the SUSY extension.
Based on the analyses to the two-loop RGEs of each pattern, we found that the gauge coupling unification can be achieved below the reduced Planck scale according to the relations given in Eqs.~\eqref{eqs:SU8_531351_GCU} in the context of the affine Lie algebra $\widehat {\sG \uG}(8)_{k_U=1}$.
The corresponding unification scales and the unified gauge couplings have been obtained and summarized in Table~\ref{tab:summary_BM}.
With the joint effects of the threshold effects and the  $d=5$ gravity-induced HSW operator in Eq.~\eqref{eq:HSW_op}, we find that the gauge coupling unification can be realized around $\Oc(10^{17})\,{\rm GeV}$ for four distinctive patterns.
By using the naive estimation~\cite{Langacker:1980js,Nath:2006ut} of 
\beqn
\tau [ p\to \pi^0 + e^+ ]&\approx& \frac{ v_U^4 }{\alpha_{U }^{2} m_p^5} \,, \quad m_p = 938\,{\rm MeV} \,,
\eeqn
purely from the gauge sector contribution, we have found significantly enhanced proton lifetime of $\tau[ p\to \pi^0 + e^+] \gtrsim  \Oc(10^{41})\,{\rm yrs}$, even without considering the mixing effects.
This means several ongoing experimental probes of the proton decay modes are unlikely to discover such a signal, with the typical exclusion limit of $\sim {\cal O}(10^{34})\,{\rm yrs}$ from the Super-Kamiokande~\cite{Super-Kamiokande:2020wjk} or DUNE Collaboration~\cite{DUNE:2016hlj}.
In addition, we have also proved that the nonmaximal symmetry breaking pattern of $\widehat {\sG \uG}(8)_{k_U=1}\to \gG_{621}$ is not realistic, since the vectorlike quarks of $\dG$ within the $\rep{56_F}$ remain massless.

\para
Besides the gauge coupling unifications, the flavor identifications from Tab.~\ref{tab:SU8_8ferm_SSW} through \ref{tab:SU8_56ferm_WWW} according to the SM fermion mass hierarchies for each symmetry breaking pattern have been analyzed in Ref.~\cite{Wang:2023gut}.
There are several differences in the corresponding mass spectra among these patterns, as well as compared to the maximal breaking patterns analyzed previously~\cite{Chen:2024cht,Chen:2024yhb}.
\begin{itemize}

\item[(i)] The vectorlike fermions of $(\EG\,, \UG\,, \uG\,, \dG)$ within the $\rep{56_F}$ may obtain their masses at different stages along different patterns, which are summarized in Ref.~\cite{Wang:2023gut}.

\item[(ii)] Within the $\rep{56_F}$, we found different SM flavor identifications of the first- and second-generational SM fermions along the SSW, SWS, and the WSS symmetry breaking patterns, which can be found in Tables~\ref{tab:SU8_56ferm_SSW}, \ref{tab:SU8_56ferm_SWS}, and \ref{tab:SU8_56ferm_WSS}.
The third-generational SM fermions of $(t_L\,, b_L\,, {t_R}^c\,, {\tau_R}^c)$ always live in the $\rep{28_F}$, regardless of the specific symmetry breaking patterns.

\item[(iii)] Given the SM flavor identifications along the SWS and WSS symmetry breaking patterns, we have already observed the tree-level mass splitting between the strange quark and the muon.
The origin of the Cabbibo angle is interpreted as the ratio of $\zeta_{23}$ between two different symmetry breaking scales along the SWS and WSS patterns.

\end{itemize}
Above all, to determine the realistic symmetry breaking patterns, one should rely on the RGEs of the corresponding SM fermion mass terms based on the current results together with the previous ones along the maximal symmetry breaking patterns in Refs.~\cite{Chen:2024cht,Chen:2024deo}.
We defer these to future work of performing the detailed analyses of the RGEs of the SM fermion masses.

\section*{ACKNOWLEDGMENTS}
%
%
\para
We thank Kaiwen Sun, Yuan Sun, and Yongchao Zhang for very enlightening discussions and communications. 
N.C. thanks the University of Science and Technology of China, Southeast University, and Shandong University for hospitality when preparing this work.
N.C. is partially supported by the National Natural Science Foundation of China (under Grants No. 12035008 and No. 12275140) and Nankai University.

\section*{DATA AVAILABILITY}
%
%
\para
The data are not publicly available. The data are available from the authors upon reasonable request.

\appendix

\section{APPENDIX: THE GAUGE COUPLING UNIFICATION IN THE $\widehat {\sG\uG}(8)$ THEORY}
\label{section:RGEs}

\para
Generically, the two-loop RGE of a gauge coupling of $\alpha^{}_{\Upsilon}$ in the $\overline{\rm MS}$ scheme for the Lie algebra $\gG_\Upsilon$ is given by~\cite{Machacek:1983tz}
\begin{equation}
\frac{\mathrm{d} \alpha^{-1}_{\Upsilon}\(\mu\)}{\mathrm{d} \log\mu} = - \frac{b^{\(1\)}_{\Upsilon}}{2\pi} - \sum_{\Upsilon^\prime} \frac{b^{\(2\)}_{\Upsilon \Upsilon^\prime}}{8\pi^2 \alpha^{ -1 }_{ \Upsilon^\prime } \(\mu\)} \,.
\end{equation}
The nonSUSY one- and two-loop $ \beta $ coefficients for the non-Abelian Lie algebras are given by
\begin{subequations}
\begin{eqnarray}
b^{(1)}_{\Upsilon} & = & -\frac{11}{3} C_2 ( \gG_\Upsilon ) + \frac{2}{3} \sum_{F}T ( \Rc^{F}_{ \Upsilon } ) + \frac{1}{3} \sum_{S} T (\Rc^{S}_{\Upsilon} ) \,, \\
b^{ (2)}_{ \Upsilon \Upsilon^\prime } & = & -\frac{34}{3} \delta_{ \Upsilon \Upsilon^\prime } C_2 ( \gG_\Upsilon )^2  + \sum_F \Big[ 2 \sum_{ \Upsilon^\prime }C_2 ( \Rc^{F}_{ \Upsilon^\prime } )  + \frac{10}{3} \delta_{ \Upsilon \Upsilon^\prime } C_2 ( \gG_\Upsilon )  \Big] T ( \Rc^F_\Upsilon ) \nonumber \\
& & + \sum_S \Big[ 4\sum_{ \Upsilon^\prime } C_2 (\Rc^{S}_{\Upsilon^\prime} )  + \frac{2}{3} \delta_{ \Upsilon \Upsilon^{\prime}} C_2 ( \gG_\Upsilon ) \Big] T  ( \Rc^S_\Upsilon ) \,,
\end{eqnarray}
\end{subequations}
with $\Rc^{F/S}_{ \Upsilon } \in \gG_\Upsilon$.
Here, $C_2( \gG_\Upsilon )$ is the quadratic Casimir of the Lie algebra $\gG_\Upsilon$, and $T(\Rc^{F/S}_{ \Upsilon })$ represent the trace invariants of the chiral fermions in the IRs of $\Rc^{F}_{ \Upsilon }$ and complex scalars in the IRs of $\Rc^{S}_{ \Upsilon }$.
For the Abelian Lie algebras with the corresponding charges denoted as $ \Xc^{F/S}_{\Upsilon} $, the one- and two-loop $\beta$ coefficients are 
\begin{subequations}
\begin{eqnarray}
b^{\(1\)}_{\Upsilon} & = & \frac{2}{3}\sum_{F} \(\Xc^{F}_{\Upsilon}\)^2 + \frac{1}{3} \sum_{S} (\Xc^{S}_{\Upsilon} )^2 \,, \\
b^{\( 2 \)}_{\Upsilon \Upsilon^\prime } & = & 2 \sum_{ F\,, \Upsilon^\prime } ( \Xc^{F}_{\Upsilon^\prime } )^2 \cdot  (\Xc^{F}_{\Upsilon} )^2 + 4 \sum_{ S \,, \Upsilon^\prime }  (\Xc^{S}_{\Upsilon^\prime } )^2 \cdot (\Xc^{S}_{\Upsilon} )^2 \,.
\end{eqnarray}
\end{subequations}
We will also assume a set of SUSY RGEs between the GUT scale and the first symmetry breaking stage, and the corresponding one- and two-loop $\beta$ coefficients~\cite{Martin:1993zk} are given by
\begin{subequations}
\begin{eqnarray}
b^{(1)}_{\Upsilon} & = & -3 C_2 ( \gG_\Upsilon ) + \sum_{\Phi} T ( \Rc^{\Phi}_{ \Upsilon } ) \,, \\
b^{ (2)}_{ \Upsilon \Upsilon^\prime } & = & -6 \delta_{ \Upsilon \Upsilon^{\prime} } C_2 ( \gG_\Upsilon )^2  + \sum_{\Phi} \Big[ 4 \sum_{ \Upsilon^\prime } C_2 ( \Rc^{\Phi}_{ \Upsilon^\prime } )  + 2 \delta_{ \Upsilon \Upsilon^{\prime} } C_2 ( \gG_\Upsilon )  \Big] T ( \Rc^{\Phi}_\Upsilon )  \,,
\end{eqnarray}
\end{subequations}
for the non-Abelian $\gG_\Upsilon$, and
\begin{subequations}
\begin{eqnarray}
b^{(1)}_{\Upsilon} & = &  \sum_{\Phi}  \(\Xc^{\Phi}_{\Upsilon}\)^2 \,, \\
b^{ (2)}_{ \Upsilon \Upsilon^\prime } & = & 4 \sum_{\Phi\,, \Upsilon^\prime }   \(\Xc^{\Phi}_{\Upsilon^\prime }\)^2  \cdot \(\Xc^{\Phi}_{\Upsilon}\)^2  \,,
\end{eqnarray}
\end{subequations}
for the Abelian $\gG_\Upsilon$.
In practice, we derive the two-loop RGEs numerically by using the PyR@TE code~\cite{Sartore:2020gou}.

\para
The one-loop threshold effect was usually considered as a major source to modify the RG behavior within the field theory~\cite{Weinberg:1980wa,Hall:1980kf,Ernst:2018bib}, and it can be generically expressed as
\beqn\label{eq:threshold_match}
\alpha_{ \rm UV }^{-1} ( \mu )  &=&  \alpha_{ \rm IR }^{-1} ( \mu ) +  \frac{ \Delta_{  \rm IR } ( \mu ) }{ 12\pi} \,, \non
\Delta_{  \rm IR } ( \mu ) &=& C_2 ( \Gc_{  \rm UV } ) - C_2 ( \Gc_{ \rm IR } ) - 21 \sum_{ V_k \in \Gc_{ \rm IR }   }  {\rm dim} ( V_k ) \log \frac{ M_{V_k } }{ \mu } \non
&&+ 2 \sum_{ \Sc_k \in \Gc_{ \rm IR }  } {\rm dim} ( \Sc_k) \log \frac{ M_{ \Sc_k } }{ \mu } + 4 \sum_{ \Fc_k \in \Gc_{ \rm IR }  } {\rm dim}( \Fc_k ) \log \frac{ M_{ \Fc_k} }{ \mu }  \,,
\eeqn
where $V_k$, $S_k$, and $\Fc_k$ represent the vector bosons, complex scalars, and Weyl fermions that are integrated out at the threshold scale $\mu$.
Their dimensions are always given in terms of the broken subgroup of $\Gc_{ \rm IR } $.
One may evaluate contributions due to the threshold effects in Eq.~\eqref{eq:threshold_match} to see if they can alter the gauge coupling evolutions significantly.
By taking the renormalization scale of $\mu = \overline M_{V_k}$ (with $\overline M_{V_k}$ being the average of the gauge boson masses) in Eq.~\eqref{eq:threshold_match}, we list the contributions from the massive scalars and fermions at each symmetry breaking stage.
Accordingly, the reasonable one-loop threshold effects in terms of $\Delta_{\rm IR}$ at each stage can be estimated.
For simplicity, we take a common mass for both massive fermions and massive scalar fields at each stage, and we estimate the largest possible threshold effects by assuming these masses within the range of $[\frac{1}{10}\,, 10]$ to the renormalization scale of $\mu = \overline M_{V_k}$.


\providecommand{\href}[2]{#2}\begingroup\raggedright\endgroup

\end{document}